\documentclass[reprint, prl,
 amsmath,amssymb,
 aps, 
 floatfix, 
]{revtex4-2}

\usepackage{graphicx}
\usepackage{dcolumn}
\usepackage{bm}
\usepackage{soul} 
\usepackage{xcolor}
\usepackage{hyperref}
\usepackage{subcaption}
\usepackage{ulem}
\begin{document}

\title{
High-Statistics Measurement of the Cosmic-Ray Electron Spectrum with H.E.S.S. }
\collaboration{H.E.S.S. Collaboration}
\author{  F.~Aharonian~$^{1,2,3 }$
}\author{  F.~Ait~Benkhali~$^{4 }$
}\author{  J.~Aschersleben~$^{ 5 }$
}\author{  H.~Ashkar~$^{  6}$
}\author{  M.~Backes~$^{ 7,8 }$
}\author{  V.~Barbosa~Martins~$^{ 9 }$
}\author{  R.~Batzofin~$^{ 10 }$
}\author{  Y.~Becherini~$^{ 11,12}$
}\author{  D.~Berge~$^{ 9,13 }$
}\author{  K.~Bernl\"ohr~$^{ 2}$
}
% ...

\vspace{9mm}

\affiliation{
$^{1}$
Dublin Institute for Advanced Studies, 31 Fitzwilliam Place, Dublin 2, Ireland}
\affiliation{$^2$
Max-Planck-Institut f\"ur Kernphysik, P.O. Box 103980, D 69029 Heidelberg, Germany}
\affiliation{$^3$
Yerevan State University,  1 Alek Manukyan St, Yerevan 0025, Armenia}
\affiliation{$^4$
Landessternwarte, Universit\"at Heidelberg, K\"onigstuhl, D 69117 Heidelberg, Germany}
\affiliation{$^5$
Kapteyn Astronomical Institute, University of Groningen, Landleven 12, 9747 AD Groningen, The Netherlands}
\affiliation{$^6$
Laboratoire Leprince-Ringuet, École Polytechnique, CNRS, Institut Polytechnique de Paris, F-91128 Palaiseau, France}
\affiliation{$^7$
University of Namibia, Department of Physics, Private Bag 13301, Windhoek 10005, Namibia}
\affiliation{$^8$
Centre for Space Research, North-West University, Potchefstroom 2520, South Africa}
\affiliation{$^9$
Deutsches Elektronen-Synchrotron DESY, Platanenallee 6, 15738 Zeuthen, Germany}
\affiliation{$^{10}$
Institut f\"ur Physik und Astronomie, Universit\"at Potsdam,  Karl-Liebknecht-Strasse 24/25, D 14476 Potsdam, Germany}
\affiliation{$^{11}$
Université de Paris, CNRS, Astroparticule et Cosmologie, F-75013 Paris, France}
\affiliation{$^{12}$
Department of Physics and Electrical Engineering, Linnaeus University,  351 95 V\"axj\"o, Sweden}
\affiliation{$^{13}$
Institut f\"ur Physik, Humboldt-Universit\"at zu Berlin, Newtonstr. 15, D 12489 Berlin, Germany}
% ...

\noaffiliation

\date{\today}

\begin{abstract}
Owing to their rapid cooling rate and hence loss-limited propagation distance, cosmic-ray electrons and positrons (CRe) at very high energies probe local cosmic-ray accelerators 
and provide constraints on exotic production mechanisms such as annihilation of dark matter particles. 
We present a high-statistics measurement of the spectrum of CRe candidate events from {0.3 to 
40}~TeV with the High Energy Stereoscopic System (H.E.S.S.), 
covering two orders of magnitude in energy and reaching a proton rejection power of better than $10^{4}$. The measured spectrum is well described by a broken power law, with a break 
around 1~TeV, where the spectral index increases  from $\Gamma_1 = 3.25$ $\pm$~0.02 (stat) $\pm$~0.2 (sys) to $\Gamma_2 = 4.49$ $\pm$ 0.04 (stat) $\pm$ 0.2 (sys). Apart from the break, the spectrum is featureless. The absence of distinct signatures at multi-TeV energies imposes constraints on the presence of nearby CRe accelerators and the local CRe propagation mechanisms.

\end{abstract}

\maketitle

\section{Introduction}

Cosmic-ray electrons and positrons (CRe) at very high energies ($E \gtrsim$ 100~GeV) undergo fast radiation losses while propagating in the Galaxy. Both inverse-Compton scattering and synchrotron radiation losses limit their cooling times and therefore propagation distance to typically kpc-scales or below~\cite{1970ApJ...162L.181S, 1979ApJ...228..297C} in diffusion-dominated Galactic cosmic-ray transport~\cite{1975SSRv...17...45D}. 
 Therefore, local CRe either indicate the existence of one (or several) primary CRe sources in the local vicinity~\cite{1980A&A....90..140M, 1995PhRvD..52.3265A}, or are the result of cosmic-ray nuclei interacting with interstellar gas producing secondary CRe~\cite{1998ApJ...493..694M}, which would suggest a more distributed origin of CRe. 
In particular, nearby pulsars and their environments as well as supernova remnants have been suggested as CRe sources~\cite{1995A&A...294L..41A,2004ApJ...601..340K,2010A&A...524A..51D, 2021PhRvD.103h3010E}.
However, no unambiguous imprint of such local CRe sources has been revealed in the spectrum, nor inferred through anisotropy~\cite{1971ApL.....9..169S, 1995ICRC....3...56P,2017PhRvL.118i1103A,2017JCAP...01..006M} of the CRe observed at Earth.
The measurements of the positron fraction rising with energy~\cite{2009Natur.458..607A,2014PhRvL.113l1102A} triggered interest in more exotic scenarios, such as the imprint of dark-matter annihilation, although conventional scenarios with pulsars as CRe sources seem to be favoured~\cite{2018PhRvD..97l3008P,2019MNRAS.484.3491T,2019PhRvL.122j1101A}.
The high-energy end of the CRe spectrum has been made accessible by indirect, ground-based measurement techniques.
H.E.S.S. measurements up to $\sim~4$ TeV~\cite{2008PhRvL.101z1104A,2009A&A...508..561A} revealed the existence of a break in the CRe spectrum at around 1~TeV, confirmed later by MAGIC and VERITAS~\cite{2011ICRC....6...47B, 2018PhRvD..98f2004A}. Direct measurements by  the  {\it Fermi}-LAT telescope and AMS-02 reached the onset of the break in the electron spectrum~\cite{2010PhRvD..82i2004A, AGUILAR20211}, allowed for discrimination between electrons and positrons~\cite{2012PhRvL.108a1103A}, and ultimately extended the energy range of direct measurements~\cite{2017PhRvD..95h2007A}. Further extension to 4.6~TeV and 7.5~TeV, respectively, and a first confirmation of the break by direct measurements was obtained by
DAMPE~\cite{2017Natur.552...63D} and CALET~\cite{2023PhRvL.131s1001A}. 

Using the vast statistics provided by 12 years of data, advanced particle discrimination schemes, improved instrument response functions, we now constrain the CRe measurement to an unequaled energy of
{40~TeV}, and resolve the spectral break with high statistics.

\section{\label{sec:Data}The Data Set}

H.E.S.S. is an array of imaging atmospheric Cherenkov telescopes (IACTs) situated in Namibia, with four  12~m diameter telescopes (CT1-4), operational since 2003, and one large-size telescope with a mirror diameter of 28~m in the center of the array (CT5), inaugurated in 2012. H.E.S.S. has proven its capability to measure CRe already in 2008~\cite{2008PhRvL.101z1104A, 2009A&A...508..561A}, becoming the first ground-based instrument to do so. Since leptonic charge separation cannot be achieved with H.E.S.S., measurements of the sum of electrons and positrons are performed, and the term CRe is used for the sum of both hereafter. 

The cameras of the CT1-4 telescopes were upgraded in 2017 with new electronics.
The analysis presented here uses all data taken with CT1-4 before the camera upgrade, spanning a period of almost 12 
years of H.E.S.S. observations, from December 2003 to June 2015.

Strict quality criteria have been applied to guarantee the best possible data quality. Besides standard quality cuts \cite{SM}, only observations with all four CT1-4 telescopes operational and with zenith angles smaller than $45^\circ$ were used.
The atmospheric transparency to Cherenkov radiation, as inferred from the trigger rate \cite{2014APh....54...25H}, is also used in the selection: only data with relative transparency larger than $60\%$ are used. 

Under the assumption of isotropy, all observations can be used for a measurement of CRe. However, since discrimination between electron and $\gamma$-ray induced air showers is challenging, the Galactic plane with its very extended $\gamma$-ray sources~\cite{2018A&A...612A...1H} and diffuse $\gamma$-ray emission~\cite{2014PhRvD..90l2007A} is excluded within $\pm 15^\circ$ of latitude. 
Runs with pointing positions within $5^\circ$ from the Small and Large Magellanic Clouds are also excluded.
This selection results in a total of $6830$ observation runs corresponding to $\sim 2728$ hours of data. Only events that are reconstructed to originate from within the central $4^\circ$ of the $5^\circ$ diameter field of view are retained.
Events within $0.25^\circ$ radius from known very-high-energy $\gamma$-ray sources are also rejected (the H.E.S.S. point spread function for these data is typically $0.06^\circ$).

Potential remaining contaminations are the high-latitude Galactic diffuse emission, and the diffuse extragalactic $\gamma$-ray background, which has a flux well below the CRe flux and shows an exponential cut-off at 250~GeV ~\cite{2015ApJ...799...86A}.

\begin{figure}[ht!]
\includegraphics[scale=0.43]{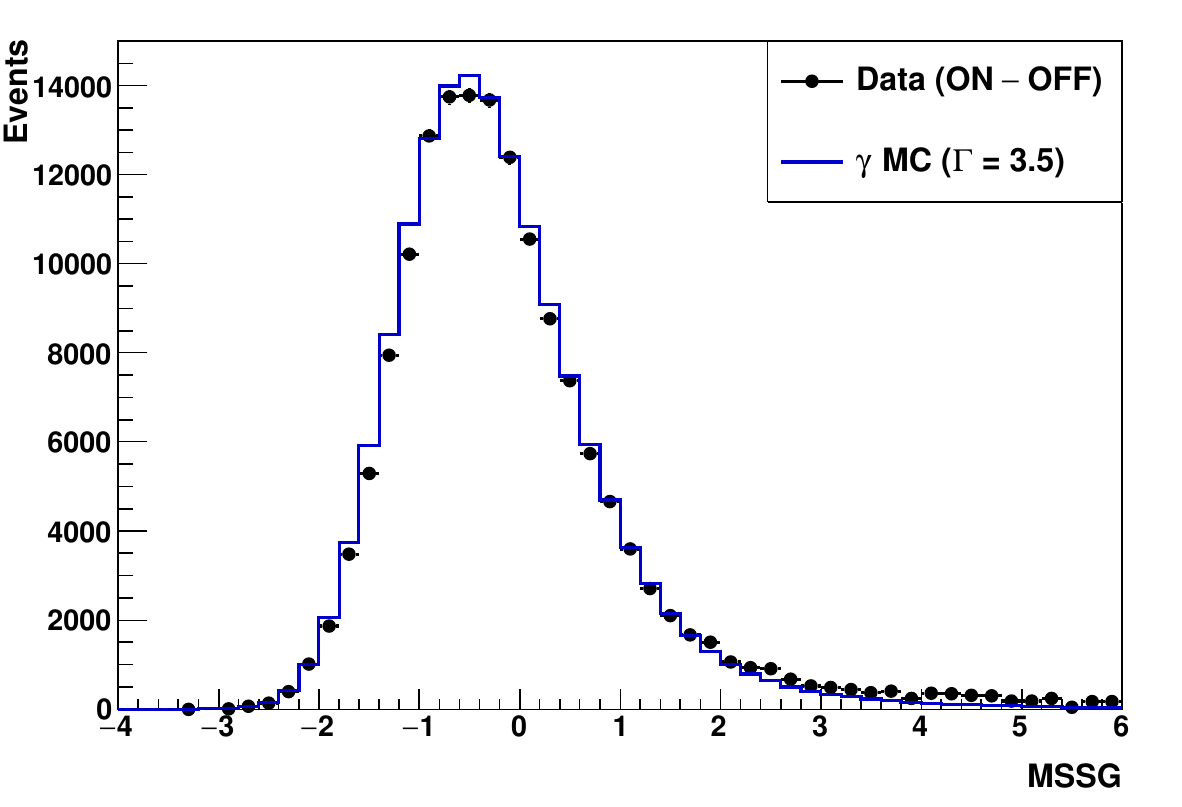} 
\caption{The distribution of the main classifier, the mean scaled shower goodness MSSG, for $\gamma$-rays from PKS~2155-304 in H.E.S.S. data after background subtraction (black), compared to $\gamma$-ray simulations generated with a spectral index $\Gamma = 3.5$ (blue). \label{fig:mssg_pks2155}}
\end{figure}
\begin{figure*}

  \subfloat{\includegraphics[width=.32\linewidth]{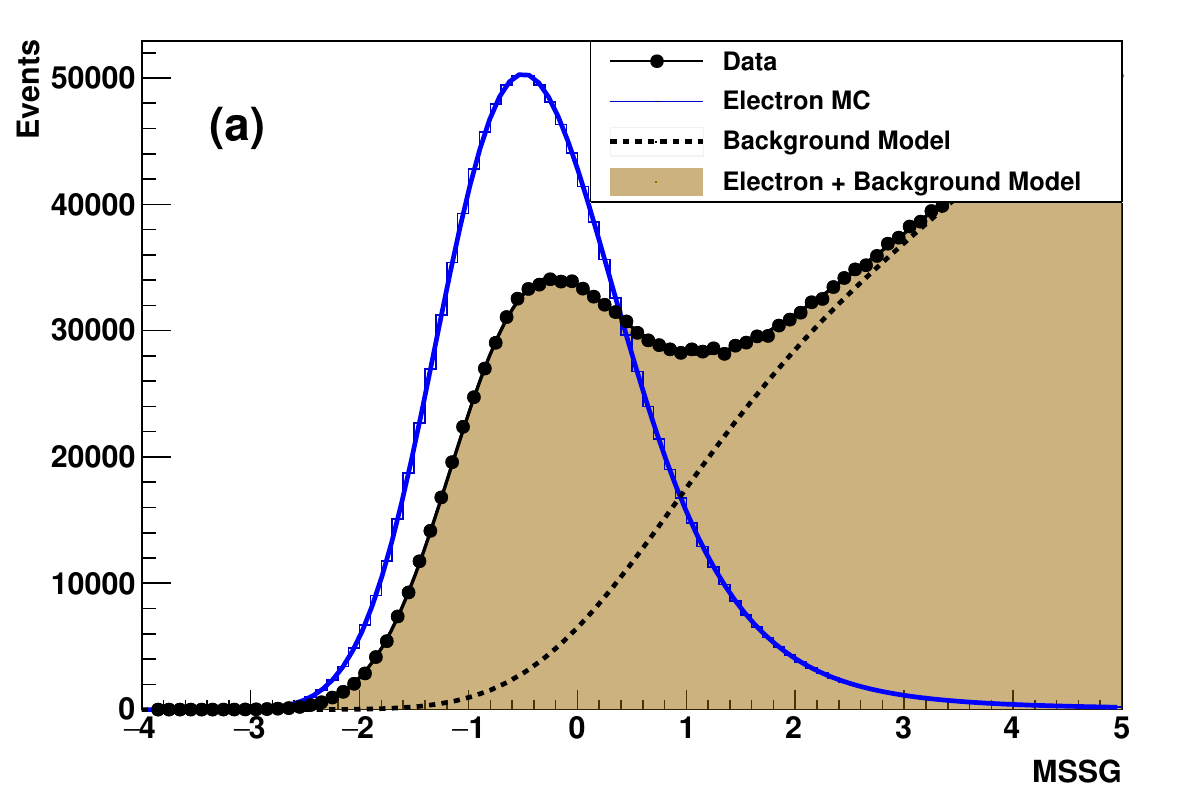} 
  }\hfill 
  \subfloat{\includegraphics[width=.32\linewidth]{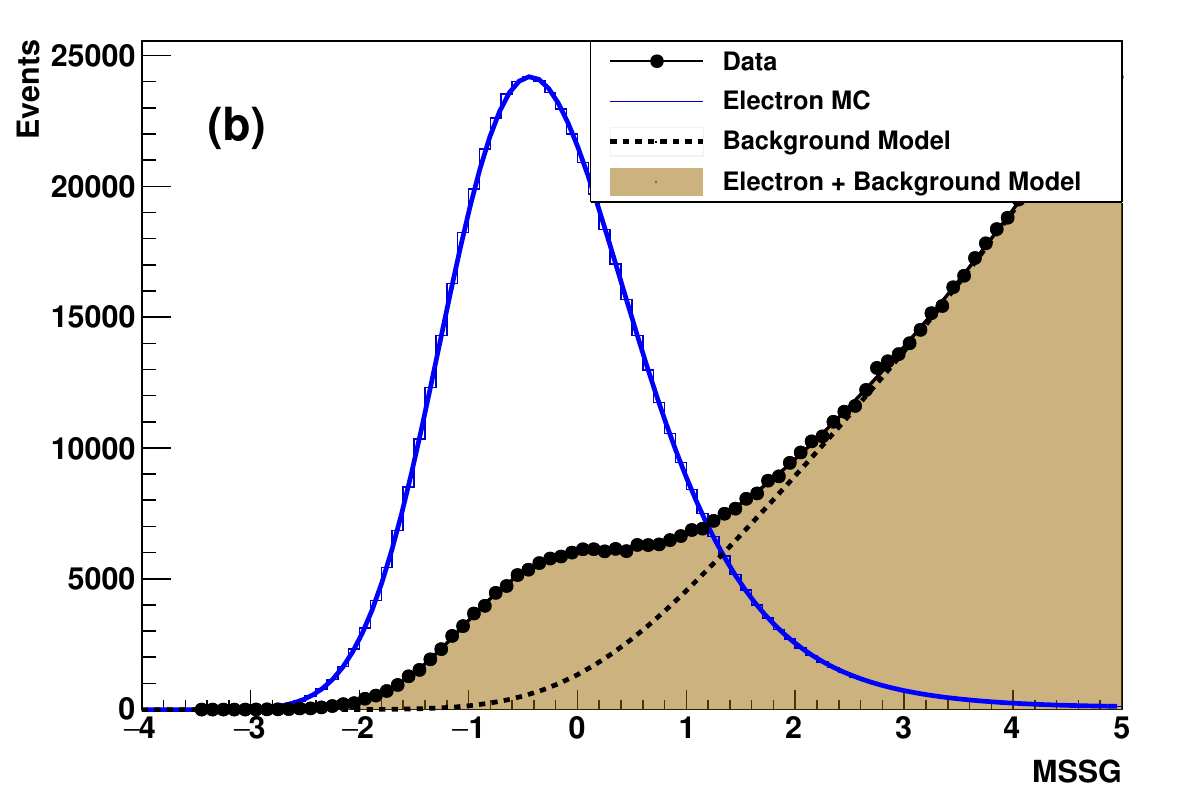}
  }\hfill 
  \subfloat{\includegraphics[width=.32\linewidth]{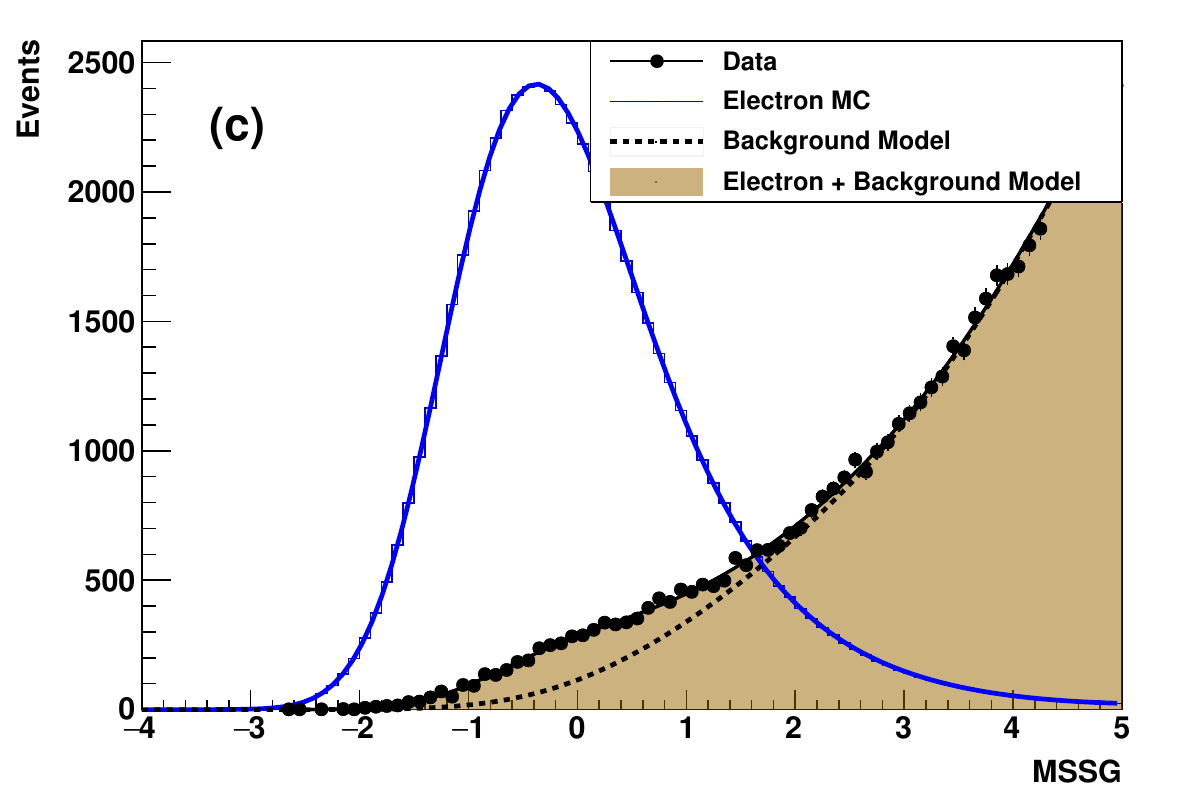}
  }

\caption{The MSSG distribution for H.E.S.S. data (black data points) in three energy bands: (a) $0.3-1$~TeV, (b) $1-3$~TeV, and (c) $>3$~TeV. The data are fitted with a parameterization of simulated electrons (blue, arb. norm.) and a parameterization of the hadronic cosmic rays 
(black dashed line). The combined model of electrons and hadronic background is shown as solid black line and ochre area. 
\label{mssg}}
\end{figure*}

\section{\label{sec:MC}Monte-Carlo Simulations}
Monte-Carlo (MC) simulations of IACTs are used at the reconstruction level to correct data for instrumental or environmental effects,
at the analysis level for validation,  and at the spectral
reconstruction level, where precise instrument response functions are needed.

Standard IACT analyses use interpolations between pre-generated MC data sets generated on a fixed grid of the parameter space. Here a more precise approach is used: In the run-wise simulation scheme~\cite{RunWise2020}, every run is simulated with the actual source trajectory and instrument parameters
of that run. This improves the agreement between data and simulations \cite{RunWise2020} and provides
more accurate response functions.

For each 
run of the data set, a large number of electron-induced showers (typically of the order of 200 000) are simulated with a hard spectrum ($\Gamma = 1.3$), ensuring 
sufficient statistics at high energies and reliable response functions up to more than $80\, \mathrm{TeV}$.
This MC data set is used at different levels of the analysis to check the quality and stability of the results (see \cite{SM}).

\section{\label{sec:Analysis}Shower Reconstruction and Primary Particle Identification}

For the event reconstruction an advanced reconstruction technique (Model++) based on comparison of the camera images with a semi-analytical model of the air showers is applied~\cite{2009APh....32..231D}. The shower image goodness-of-fit allows selection of relatively clean samples of $\gamma$-ray or CRe candiate events. The goodness of fit of each image is rescaled by computing its difference to its mean as derived from $\gamma$-ray simulations, and dividing the result by its rms. This allows us to combine the results of each telescope in a single variable, the \textit{mean scaled shower goodness}, or MSSG. Given the similarity between the CRe and $\gamma$-ray shower development, this goodness of fit MSSG has a rather similar distribution for both primary $\gamma$-rays and CRe. 
For the analysis of $\gamma$-ray sources, the cut value on MSSG is chosen as a compromise between statistics of $\gamma$-ray candidates and background level, given that the remaining background can be reliably determined from off-source regions, and subtracted. For the investigation of the diffuse CRe flux, the MSSG is calculated assuming the electron hypothesis, and a hard cut of MSSG\,$<\,-0.6$ is used, together with additional selections cuts (see \cite{SM}). This drastically reduces the background of cosmic-ray protons and nuclei (CRn) among the CRe candidate events, at the expense of signal statistics, resulting in a proton rejection of better than $10^4$ at a few TeV. 

This technique crucially relies on the ability of the simulations to accurately reproduce the MSSG distribution for electromagnetic air showers. 
This was validated using the MSSG distribution for
$\gamma$-rays from the blazar PKS~2155-304 (using a data set of 755 runs), after subtracting the cosmic-ray background. The distribution is compared
 to $\gamma$-ray simulations obtained within the run-wise scheme. Identical cuts were used for selection of runs, images, and events as for the CRe sample.
As illustrated in Fig.~\ref{fig:mssg_pks2155}, the measured distribution is indeed well reproduced by simulations, see also \cite{SM}.

In Fig.~\ref{mssg} the measured distribution in MSSG is shown,
in three energy ranges (ochre histograms), in comparison to electron run-wise simulations (blue curves). 
The data distributions exhibit a clear peak consistent with the electron simulations,
on top of a much broader distribution that corresponds to CRn.
The data are modelled as a sum of the simulated MSSG distribution of electrons 
and an analytical parameterization 
(dotted line) to account for the CRn background. The resulting distribution -- shown as solid black line --
is in good agreement with the data, although it should be noted that there is some
freedom in the choice of the analytical parameterization representing the hadronic background. 

The analysis presented here deliberately avoids using distributions from 
simulated CRn, and does not attempt to subtract the CRn remaining after the CRe selection cuts. CRn simulations have considerably larger uncertainties than those of electromagnetic cascades, due to the model dependence in the simulation of hadronic interactions. Moreover, the fraction of hadronic showers triggering the instrument and passing the selection cuts is much lower
than for electromagnetic showers, making the production of a CRn Monte-Carlo data set excessive in terms of computing time. 

From Fig.~\ref{mssg}, an estimate of the remaining CRn contamination as a
function of the MSSG cut can be obtained.
Below $\mathrm{MSSG} \simeq -1$, where very little CRn contamination is expected, the measured MSSG distribution is in excellent agreement with the predictions for electrons. For $\mathrm{MSSG} \gtrsim -1$, an increasing excess over the electron simulations is seen, which is attributed to CRn. The applied cut of $\mathrm{MSSG} \leq -0.6$ results in a CRn contamination in the CRe dataset of less than 
25\% for the energy range of $0.1-1$~TeV, and less than $30\%$ for $1-3$~TeV. Beyond 3~TeV a dominant background contribution cannot be excluded \cite{SM}.

\section{\label{sec:Spectrum}Spectrum of Electron + Positron Candidate Events}
%
\begin{figure}[htb!]
\includegraphics[scale=0.45]{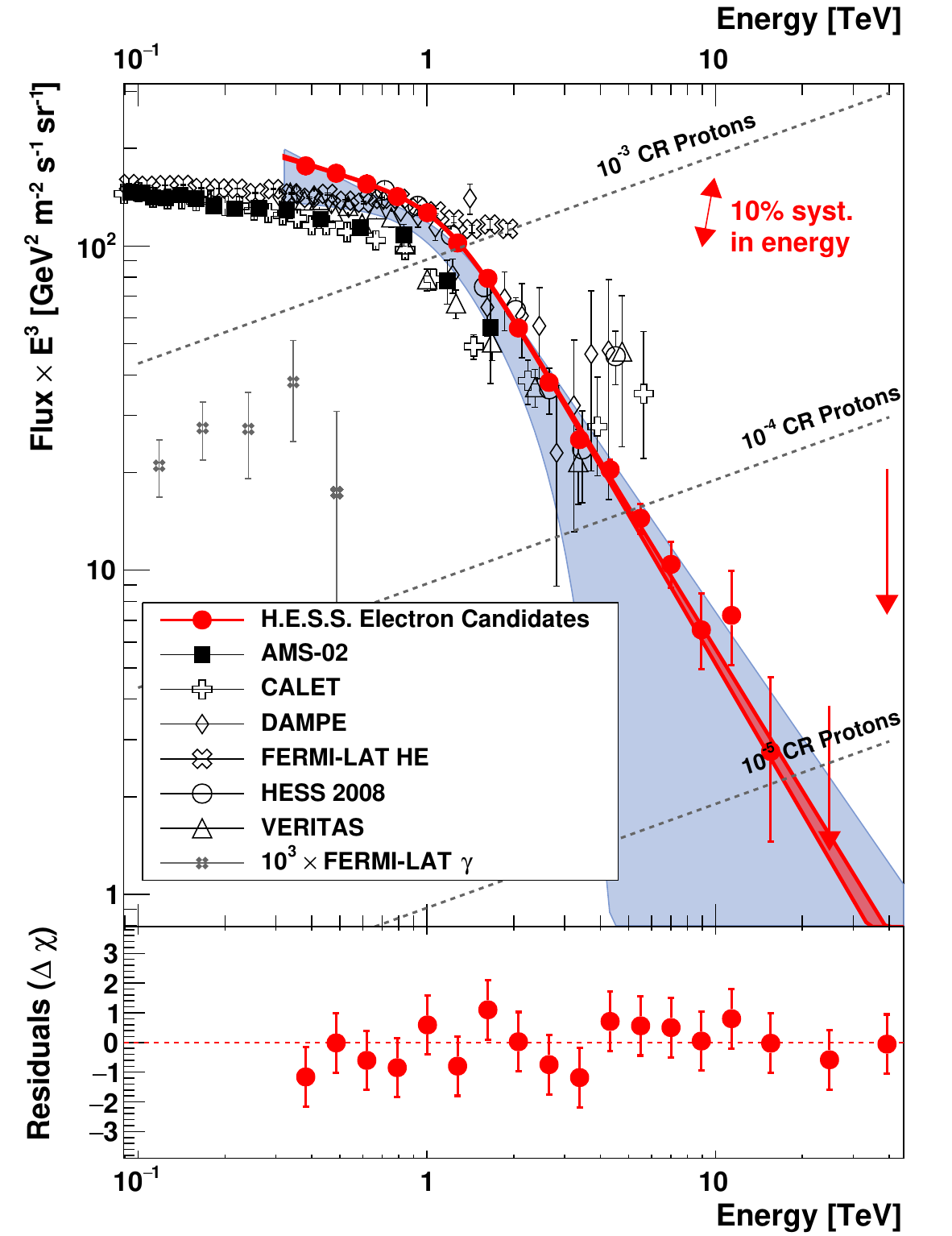} 
\caption{
Filled red circle data points: spectrum of cosmic-ray electron/positron (CRe) candidate events measured by H.E.S.S. The data set still contains a residual background from cosmic-ray nucleons (CRn) and therefore places an upper limit on the true CRe flux. 
The dark-red band indicates the broken-power-law fit to the data (Eqn. 1), with the width of the band corresponding to statistical errors.
The light-blue band denotes the estimated range of the true CRe flux, considering the CRn contamination as well as the statistical errors and systematic errors. Separately shown is the systematic error on the global energy scale, that also impacts the normalisation of $E^3 F(E)$, as visualised by the red arrow. Included are CRe measurements by AMS-02~\cite{AGUILAR20211}, {\it Fermi}-LAT~\cite{2017PhRvD..95h2007A}, CALET~\cite{2023PhRvL.131s1001A}, DAMPE~\cite{2017Natur.552...63D}, VERITAS~\cite{2018PhRvD..98f2004A}, and previous H.E.S.S. measurements~\cite{2008PhRvL.101z1104A,2009A&A...508..561A}.
Also shown is the CR proton flux (scaled down by $10^{-3}-10^{-5}$) and the {\it Fermi}-LAT diffuse extragalactic $\gamma$-ray flux \cite{2015ApJ...799...86A} (scaled up by $10^3$).
The bottom panel shows the residuals expressed in $\Delta \chi = (\phi_i - \phi(E))/\sigma_i$ where $\phi_i$ and $\sigma_i$ are respectively the
flux measured in bin $i$ and its statistical uncertainty and $\phi(E)$ is the expected flux from Eqn. \ref{eq:fit}.
\label{fig:spectrum}}
\end{figure}

\begin{figure}[htb!]
  \includegraphics[scale=0.45]{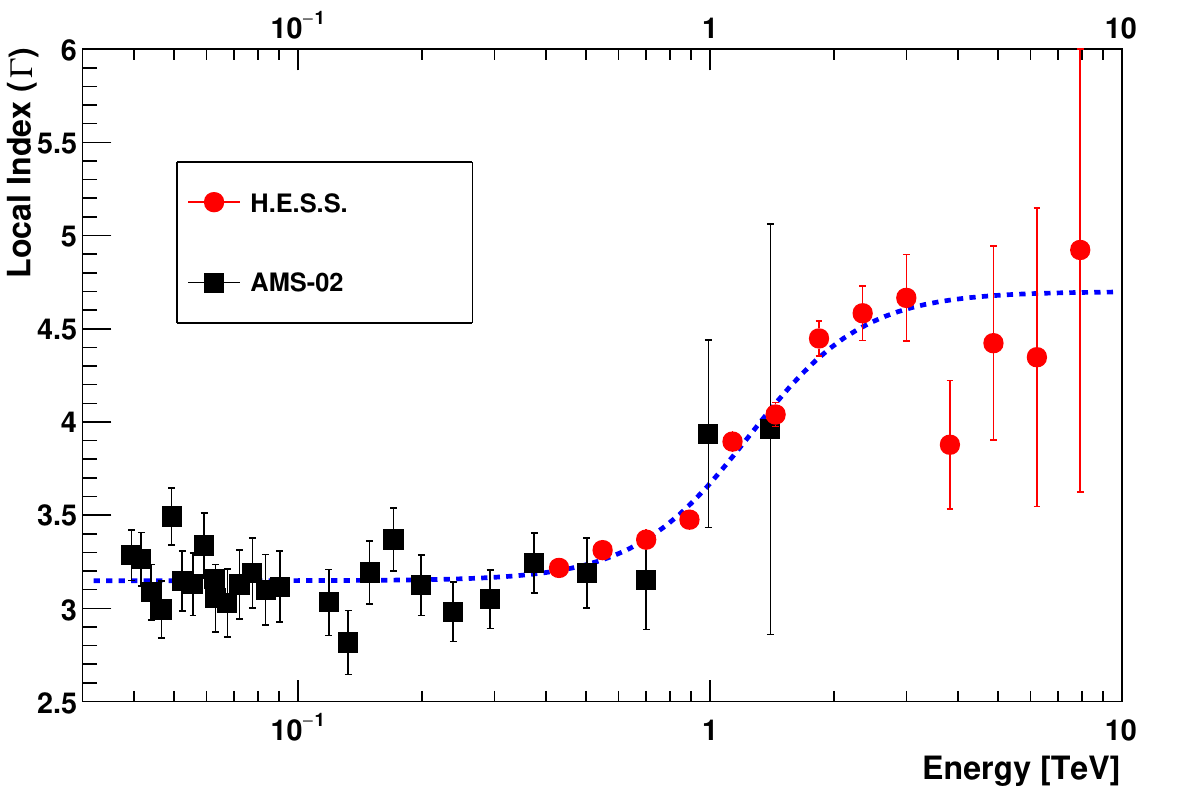
  } \caption{Local spectral index, calculated from adjacent data points as $\Gamma(E) = \Delta \log \phi / \Delta \log E$, for the high-statistics H.E.S.S. and AMS-02 \cite{AGUILAR20211} spectra. H.E.S.S. data are described by a sigmoid function in log E, shown as blue dashed line.}  
 \label{fig:local_index}
\end{figure}

This analysis resulted in the selection of 265\,574 
electron-like events from 0.3 TeV to 
40 TeV. The energy spectrum for these events is derived using a 
forward-folding procedure assuming that all selected events are electrons and using instrument response functions computed from electron simulations described 
above.
 The red data points in Fig.~\ref{fig:spectrum} show the resulting spectrum of these CRe candidate events along with previous CRe measurements. Numerical values are provided in \cite{SM}.

The spectrum of CRe candidate events is consistent with an otherwise featureless broken power law with a break around 1~TeV:
\begin{equation}
F(E)~=~F_0 
\left(\frac{E}{1~\text{TeV}}\right)^{ - \Gamma_1} \left(1 + \left(\frac{E}{E_{b}}\right)^{\frac{1}{\alpha}}\right)^{-(\Gamma_2 - \Gamma_1)\alpha}
\label{eq:fit}
\end{equation}
with:

\begin{itemize}
	\item $F_0 = (126.1 \pm 0.5 \,_{\mathrm{stat}}\, \pm 13\,_\mathrm{sys})\,
 \mathrm{GeV}^2\, \mathrm{m}^{-2} \, \mathrm{sr}^{-1} \, \mathrm{s}^{-1}$ as flux normalisation,
	\item $\Gamma_1 = 3.25 \pm 0.02\,_\mathrm{stat}\, \pm 0.2\,_\mathrm{sys}
 $ as the low-energy index of the power law,
	\item $\Gamma_2 = 4.49 \pm 0.04\,_\mathrm{stat}\,\pm 0.2\,_\mathrm{sys}
 $ as the high-energy index of the power law,
	\item $E_b = (1.17 \pm 0.04\,_\mathrm{stat}\,
 \pm 0.12\,_\mathrm{sys})
 $~TeV as the break energy,
	\item $\alpha = 0.21 \pm 0.02\,_\mathrm{stat}\,
 ^{+0.10_\mathrm{sys}}_{-0.06_\mathrm{sys}}\,
 $ as the sharpness of the break.
\end{itemize}
The dark-red band in Fig.~\ref{fig:spectrum} indicates the best fit to the data, with the width of the band corresponding to the statistical errors of the data.

Systematic errors of the parameters described in Eqn.~\ref{eq:fit} are derived by varying analysis cuts as well as accounting for the variation of the observed flux with zenith angle, with atmospheric conditions and with time, i.e. aging of the instrument, see \cite{SM}. An additional error on $F_0$ and $E_b$ arises from a 10\% uncertainty on the global energy scale, which translates into a 21\% uncertainty in the energy-weighted $F_0$, as indicated by the double-headed red arrows in Fig.~\ref{fig:spectrum}.

In the energy range of overlap, the measured flux $E^3 F(E)$ of CRe candidate events is overall about 30\% higher compared to other measurements. 
Taking into account systematic uncertainties 
and the 
estimates of a typical 15\% hadronic contamination at energies below 3~TeV \cite{SM}, the estimated range of the true CRe spectrum (visualised as blue-shaded area in Fig.~\ref{fig:spectrum}) is compatible to other measurements.

The high statistics of the H.E.S.S. data allow us to determine a local spectral index of CRe candidate events, calculated as $\Gamma(E) = \Delta \log \phi / \Delta \log E$ for data points adjacent in $E$ (Fig. \ref{fig:local_index}). The calculation uses the resolution-unfolded spectra; given the energy resolution of better than 10\% (see Fig.~5 of \cite{SM}), resolution corrections are modest. Also included in Fig. \ref{fig:local_index} is the local spectral index calculated for AMS-02 data \cite{AGUILAR20211}. The variation of the spectral index across the break is well described by a sigmoid function in $\log E$. 

\section{Discussion and conclusion} 

The analysis of an extended data set from H.E.S.S. led to the identification of a vastly increased number of 
CRe-like events compared to the previous H.E.S.S. measurement \cite{2008PhRvL.101z1104A}, extending to energies of 
40 TeV.
The spectrum of CRe candidate events is well described by a broken power law. The break at 1.17 TeV is relatively sharp; the index change -- consistent with a $\Delta \Gamma = 1$ cooling break -- occurs over a factor 
3 in energy.

This sample of CRe candidate events contains a contamination of 
CRn. For energies up to $\sim 3$ TeV -- well above the break -- the contamination can be estimated from the 
shape of the MSSG distribution (Fig. \ref{mssg}) and is 
smaller than $ 30\%$; in this domain the measured spectral shape can be taken as representative for the 
true CRe. At higher energies 
the contamination increases 
in accordance with the much steeper CRe spectrum compared to the CRn spectrum with its spectral index of about 2.7. With increasing energy it is therefore increasingly difficult to limit the contamination (see Fig. \ref{fig:spectrum}). The measured spectrum at the highest energies should be considered an upper limit for the true CRe flux; the range of the true CRe flux is indicated by the light-blue area in Fig.~\ref{fig:spectrum}. 

Beyond CRe, $\gamma$-rays from the diffuse $\gamma$-ray background are also likely to pass the event selection cuts. However, given the measured isotropic diffuse flux \cite{2015ApJ...799...86A}, this contamination is plausibly negligible across the energy range covered (see Fig. \ref{fig:spectrum}). Taking the well-measured {\it Fermi}-LAT results as indicative for the TeV range \cite{2012ApJ...750....3A}, also the flux of high-latitude Galactic diffuse emission is still significantly lower than the measured CRe spectrum. This is supported by the lack of variation of the spectrum normalisation as function of Galactic latitude (see \cite{SM}, Fig. \ref{fig:latitude}).

For the sub-TeV CRe flux, there is a $\sim$ 30\% significant difference between AMS-02/CALET and {\it Fermi}-LAT/DAMPE data. The systematic errors of the H.E.S.S. flux normalisation do not give preference to one or the other group of data.
The measured low-energy spectral index of $\Gamma_1 = 3.25$ is slightly larger than the indices in the range of $\sim 3.1$ ({\it Fermi}-LAT, DAMPE) to $\sim 3.2$ (AMS-02) measured by other experiments, but fully compatible within statistical and systematic errors.
The measured break energy of $E_b = 1.17$ TeV is, within systematic errors, marginally compatible with the DAMPE value of 0.91 TeV, but is significantly larger than the break of 0.71 TeV quoted by VERITAS. Also the {\it Fermi}-LAT lower limit (95\% C.L.) for the break energy of 1.8 TeV is not compatible with the clear and highly significant spectral break observed with H.E.S.S. and other experiments.

At high energy, the H.E.S.S. spectrum of CRe candidate events extends well beyond the direct measurements presented recently \cite{2023PhRvL.131s1001A, 2017Natur.552...63D}, and other indirect measurements. The high-energy index of $\Gamma_2 = 4.49$ obtained with H.E.S.S. for the spectrum of CRe candidates is larger than the values of 3.92 of DAMPE and 4.1 of VERITAS. Given that the CRe candidates include a CRn background whose relative importance is increasing with energy, the measured value should be interpreted as a lower limit for the true CRe high-energy index, and a high-energy cutoff is not excluded.

The H.E.S.S. data do not confirm the existence of a 1.4~TeV peak in the spectrum, which has been associated with a dark matter signal \cite{Yuan:2017ysv}, nor the rise observed by various experiments in their last energy points around 5~TeV. 

The detection of a multi-TeV CRe flux and the associated spectral break probe local CRe accelerators. The rapid cooling at these energies imposes limitations on both the electrons' energy-loss time ($\sim 100$~kyr) and their propagation distance (a few hundred parsecs).  
The H.E.S.S. results dismiss the presence of a strong close-by source causing a rise in $E^3 F(E)$ at multi-TeV energies (e.g. \cite{2004ApJ...601..340K}) and impose limits on local sources: In a burst-like scenario, a Vela-type source with a distance of 300~pc and an age of 11~kyears is limited to a maximum energy of $\sim 2\cdot 10^{46}$~erg released in electrons (see \cite{SM}). 

The high statistics of the H.E.S.S. data allow us to characterize the shape of the observed spectral break. A sharp break ($\alpha=0$ in Eqn. 1) is excluded with high confidence; the break can be characterized by an 
$\alpha$ of 0.21, implying that the index changes by about unity over a factor 
3 in energy, as visible in Fig. \ref{fig:local_index}. Uncertainties in the CRn background at higher energies prevent us from ruling out an exponential cut-off above a few TeV. Nevertheless, the identification of a break at about $1$~TeV remains a robust finding, and carries important information regarding CRe acceleration and propagation within the local Galaxy. For synthetic models with distributed ensembles of sources
(e.g. \cite{2018JCAP...11..045M,2021PhRvD.103h3010E,2022ApJ...926....5A}), and a corresponding spread of propagation times, one tends to obtain significantly smoother spectra than observed, with a smeared-out break.
The observed, still relatively sharp break may therefore favor a scenario in which -- at energies around one TeV -- a single nearby source, with a burst-like release of electrons (e.g. \cite{Mauro_2014, Recchia_2019}), takes over 
a population of CRe escaping from distributed sources 
(e.g. \cite{2011MNRAS.415.1807D,2013MNRAS.431..415B}), possibly resulting also in an observable anisotropy at these energies. 
The steep fall-off of the CRe candidate spectrum limits the capabilities of space-borne instruments to measure the multi-TeV CRe spectrum. The H.E.S.S. measurement can potentially be further improved using machine learning to select CRe; this however implies simulation of huge CRn background samples, that is computationally extremely expensive.
  
\begin{acknowledgments}
The support of the Namibian authorities and of the University of
Namibia in facilitating the construction and operation of H.E.S.S.
is gratefully acknowledged, as is the support by the German
Ministry for Education and Research (BMBF), the Max Planck Society,
the Helmholtz Association, the French Ministry of
Higher Education, Research and Innovation, the Centre National de
la Recherche Scientifique (CNRS/IN2P3 and CNRS/INSU), the
Commissariat à l’énergie atomique et aux énergies alternatives
(CEA), the U.K. Science and Technology Facilities Council (STFC),
the Irish Research Council (IRC) and the Science Foundation Ireland
(SFI), the Polish Ministry of Education and Science, agreement no.
2021/WK/06, the South African Department of Science and Innovation and
National Research Foundation, the University of Namibia, the National
Commission on Research, Science \& Technology of Namibia (NCRST),
the Austrian Federal Ministry of Education, Science and Research
and the Austrian Science Fund (FWF), the Australian Research
Council (ARC), the Japan Society for the Promotion of Science, the
University of Amsterdam and the Science Committee of Armenia grant
21AG-1C085. We appreciate the excellent work of the technical
support staff in Berlin, Zeuthen, Heidelberg, Palaiseau, Paris,
Saclay, Tübingen and in Namibia in the construction and operation
of the equipment. This work benefited from services provided by the
H.E.S.S. Virtual Organisation, supported by the national resource
providers of the EGI Federation.

\end{acknowledgments}

\begin{titlepage}
{\vskip 60pt
\centering
\large{\bf High-Statistics Measurement of the Cosmic-Ray Electron Spectrum with H.E.S.S.}
\vskip 20pt
\large{\bf Supplemental Material\\}
\vskip 30pt}
\end{titlepage}

\section{Overview}

The supplemental material provides additional information on the following topics:
\begin{itemize}
\item Numerical values for the data points of the CRe candidate flux;
\item Additional information on the data set used;
\item Instrument response functions; 
\item Validation of the instrument simulation using background-subtracted $\gamma$-ray data;
\item Contamination by diffuse $\gamma$-rays and hadronic cosmic rays;
\item Match between CRe data and simulations;
\item Stability of the electron spectrum and its parameters if data are divided into subsets in time, zenith angle and Galactic latitude, and the validation of the systematic errors quoted;
\item { Implications of the measurement for local CRe accelerators.}
\end{itemize}

\section{CRe flux data points}

Numerical values for the CRe flux data shown in Fig. \ref{fig:spectrum} of the main paper are given in Table \ref{table:spectrum}.
\begin{table}
\caption{CRe candidate flux values, without subtraction of a hadronic contamination.  Upper and lower flux values $\phi_{min}$ and $\phi_{max}$  correspond to  $1\sigma$ statistical errors; if only upper values are given, they represent the 99\% C.L. upper limit.} 
\label{table:spectrum}
\begin{ruledtabular}
\begin{tabular}{cccc}

$E$ & $\phi$  & $\phi_{min}$ & $\phi_{max}$ \\ 
\hline

TeV & \multicolumn{3}{c}{m$^{-2}$ s$^{-1}$ sr$^{-1}$ TeV$^{-1}$} \\ 
\hline

0.38	& 3.2e-03	& 3.2e-03	& 3.2e-03 \\
0.49	& 1.5e-03	& 1.5e-03	& 1.5e-03 \\
0.62	& 6.5e-04	& 6.5e-04	& 6.7e-04 \\
0.79	& 2.9e-04	& 2.9e-04	& 2.9e-04 \\
1.0	& 1.2e-04	& 1.2-04	& 1.3e-04 \\
1.3	& 4.9e-05	& 4.8e-05	& 4.9e-05 \\
1.6	& 1.8e-05	& 1.8e-05	& 1.9e-05 \\
2.1	& 6.3e-06	& 6.1e-06	& 6.4e-06 \\
2.7	& 2.0e-06	& 2.0e-06	& 2.1e-06 \\
3.4	& 6.6e-07	& 6.3e-07	& 6.9e-07 \\
4.3	& 2.5e-07	& 2.4e-07	& 2.7e-07 \\
5.5	& 8.8e-08	& 7.9e-08	& 9.7e-08 \\
7.0	& 3.2e-08	& 2.7e-08	& 3.7e-08 \\
8.9	& 8.7e-09	& 6.6e-09	& 1.1e-08 \\
11.3	& 5.7e-09	& 4.2e-09	& 7.6e-09 \\
15.5	& 7.0e-10	& 3.7e-10	& 1.2e-09 \\
24.8	& U.L.				& & 2.7e-10 \\
39.2	& U.L.				& & 3.7e-10 \\

\end{tabular}
\end{ruledtabular}
\end{table}

\section{Data set}

The analysis presented here uses all data taken with H.E.S.S. CT1-4 before the camera upgrade, from December 2003 to June 2015. The larger CT5 telescope was available from 2013 on but has a smaller field of view than CT1-4 and is therefore not used for the CRe analysis. Data taken with the upgraded CT1-4 cameras could increase the data set, but require additional simulations and a dedicated study of instrument systematics for the upgraded cameras. These data may prove useful if novel algorithms allow improved rejection of CRn backgrounds at multi-TeV energies; currently, background systematics rather than CRe statistics are the limitation. We also note that given the steep high-energy spectrum, a doubling of the data set only extends the enerygy range by about 20\%.

The general selection criteria for{ observation runs used in the analysis} are described in the main paper. Table \ref{table_cuts} summarises the run selection,{ image selection and} event-selection cuts, both the usual cuts applied in $\gamma$-ray analyses and the more stringent cuts for the CRe analysis. 
For maximum data quality, only runs with all four telescopes CT1-4 operational are used.
The atmospheric transparency to Cherenkov radiation, relative to a nominal atmosphere, is required to be $> 0.6$. Atmospheric transparency is inferred from the trigger rate corrected for the zenith angle, average pixel gain and optical throughput \cite{2014APh....54...25H}. This cut rejects less than 10\% of all runs; while atmospheric transparency is corrected on a run-by-run basis in the data analysis, studies have shown that the correction becomes less reliable once the transparency is below 0.6, i.e. when the correction factor becomes large.

Image selection cuts include the usual requirements of at least 5 pixels above 5 photoelectrons, and at least 60 photoelectrons in the image, but in addition require that images are well contained in the camera. Image containment is the fraction of the best-fit
shower image that is contained in the camera. High-energy showers tend to produce longer images that are less contained in the camera.

Event selection cuts require at least 3 good images, and exclude showers with core far outside the telescope array, where reconstruction quality degrades. Further selection cuts are applied to NSB goodness, the MSSG range, the reconstructed depth of the first interaction, and quality of reconstruction of the shower direction.
{NSB goodness is a measure of how compatible with pure night sky background (NSB) the image is. Low NSB goodness indicates noisy images compatible with pure noise. 
 The Mean Scaled Shower Goodness (MSSG) describes how well the shower images match the templates (see Fig. \ref{mssg} of main paper). 
 The depth of $1^{\mathrm{st}}$ interaction is a fit parameter determined from the image shape, characterizing the starting depth of the shower, see \cite{2009APh....32..231D}.
 Direction error is the estimated precision with which the shower axis is reconstructed; cosmic-ray induced events tend to be more poorly reconstructed than CRe or $\gamma$-ray induced events.}

\begin{table*}[htp]
\caption{{ Summary of the cuts for run selection, image selection and event selection. See text for explanation of the selection parameters.}
  }
\begin{center}
\begin{ruledtabular}
\begin{tabular}{lcc}
Cut parameter & $\gamma$-ray cut values & electron cut values \\
\hline
\multicolumn{3}{l}{Run selection cuts (in addition to general run quality cuts)} \\
\hline
Minimum number of operational telescopes & 3 & 4 \\
Relative atmospheric transparency & n.a. & $>$0.6 \\
\hline
\multicolumn{3}{l}{Image selection cuts} \\
\hline
Minimum number of pixels above 5 { photoelectrons} per image & 5 & 5 \\
Minimum number of { photoelectrons} per image & 60 & 60 \\
Image containment (\%) & $>$50 & $>$20 \\
\hline
\multicolumn{3}{l}{Event selection cuts} \\
\hline
Minimum number of telescope images & 2 & 3 \\
Core distance (m) & n.a. & $<$ 200\\
NSB goodness & $>$30 & $>$40 \\
MSSG range & -4 to 0.9 & -4 to -0.6 \\
Depth of $1^{\mathrm{st}}$ interaction ({ in radiation lengths}) & -1.1 to 3.4 & -1.5 to 0.5 \\
Direction error (deg.) & $< $ 0.1 & $<0.03$ \\
\end{tabular}
\end{ruledtabular}
\end{center}
\label{table_cuts}
\end{table*}%

\begin{figure}[htbp] 
   \centering
   \includegraphics[width=0.46\textwidth]
   {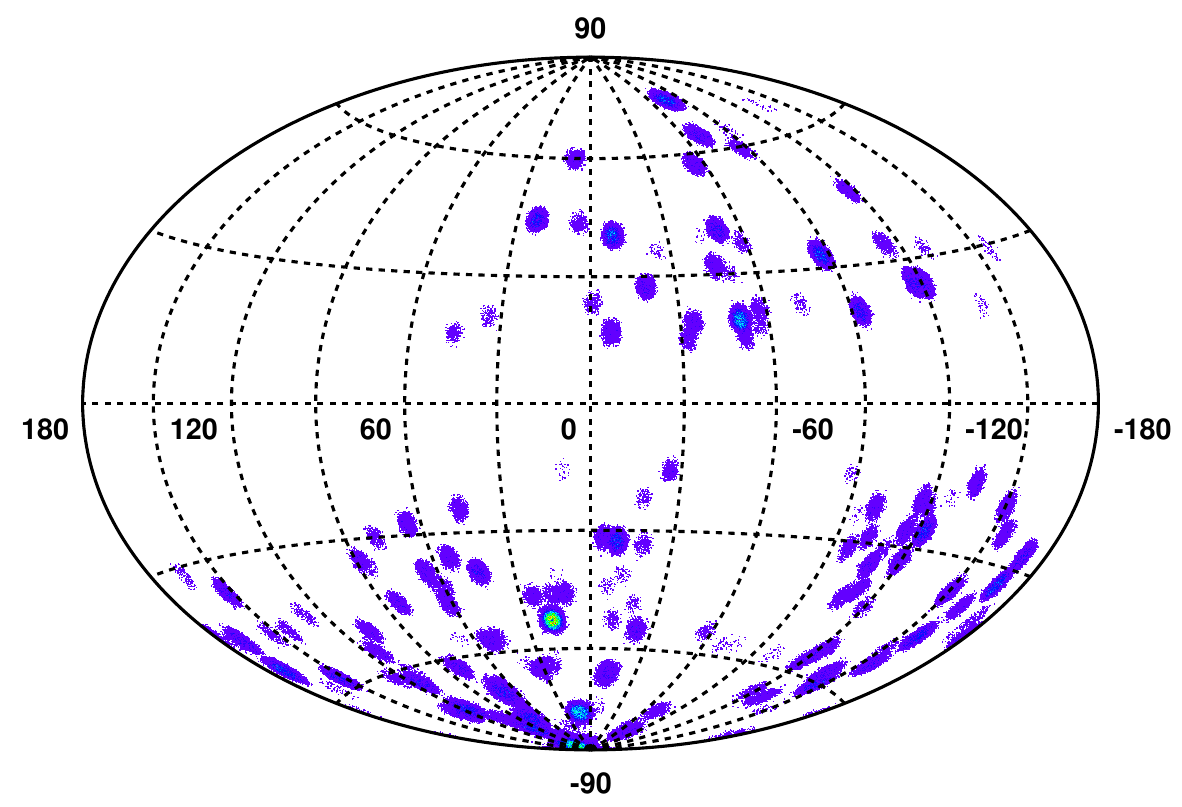} 
   \caption{Distribution in Galactic coordinates of the selected electron candidate events.}
   \label{fig:sky}
\end{figure}

The sky distribution of selected CRe events is shown in Fig. \ref{fig:sky}, in Galactic coordinates. The distribution represents the sky coverage of H.E.S.S. targets, { apart from the data removed to avoid contamination by $\gamma$-rays (data taken within $\pm 15^\circ$ from the Galactic Plane and runs within $5^\circ$ from the LMC)}, and {runs} with zenith angles $> 45^\circ$; part of the sky is not visible from Namibia.

\begin{figure}[htbp] 
   \centering
   \includegraphics[width=0.46\textwidth]{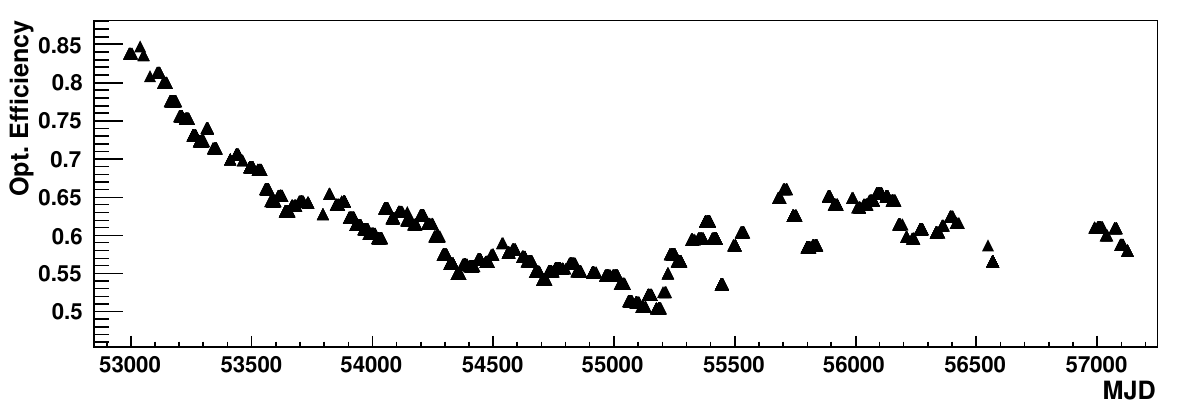} 
   \caption{Optical efficiency of the telescopes as a function of time (averaged over telescopes). The optical efficiency is determined from the intensity of muon rings.}
   \label{fig:opt_eff}
\end{figure}

The data used span more than a decade; during this time, the telescopes aged. This is visible in the optical efficiency shown in Fig. \ref{fig:opt_eff}. The optical efficiency reflects the state of mirrors and Winston cones, and is degrading steadily over time, partly recovering when optical elements are cleaned or re-coated. 

\begin{figure}[htbp] 
   \centering
   \includegraphics[width=0.46\textwidth]{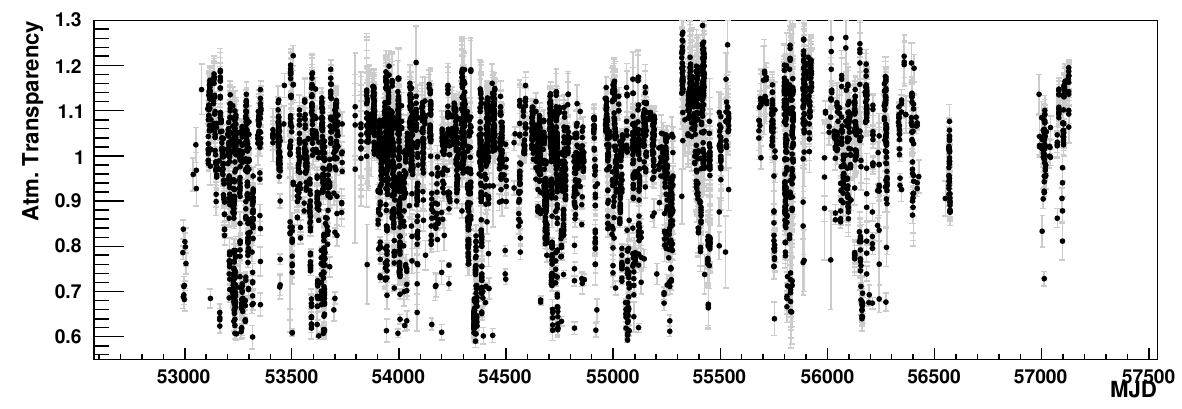} 
   \caption{Relative transparency of the atmosphere for the selected runs (runs with transparency below 0.6 are removed). The transparency is determined from the telescope trigger rates \cite{2014APh....54...25H}.}
   \label{fig:transp_coeff}
\end{figure}

Also varying with time is the transparency of the atmosphere (Fig. \ref{fig:transp_coeff}). A seasonal pattern is visible, with nights of low transparency due to dust from large-scale biomass burning in { the third quarter} of most years.

The optical efficiency of the individual telescopes, the transparency of the atmosphere and many other parameters such as the level of night sky background or the number and distribution of inactive pixels in each telescope are taken into account in the determination of instrument response functions (IRFs), on a run-wise basis.

\section{Instrument response functions}

\begin{figure}[htbp] 
   \centering
   \includegraphics[width=0.46\textwidth]{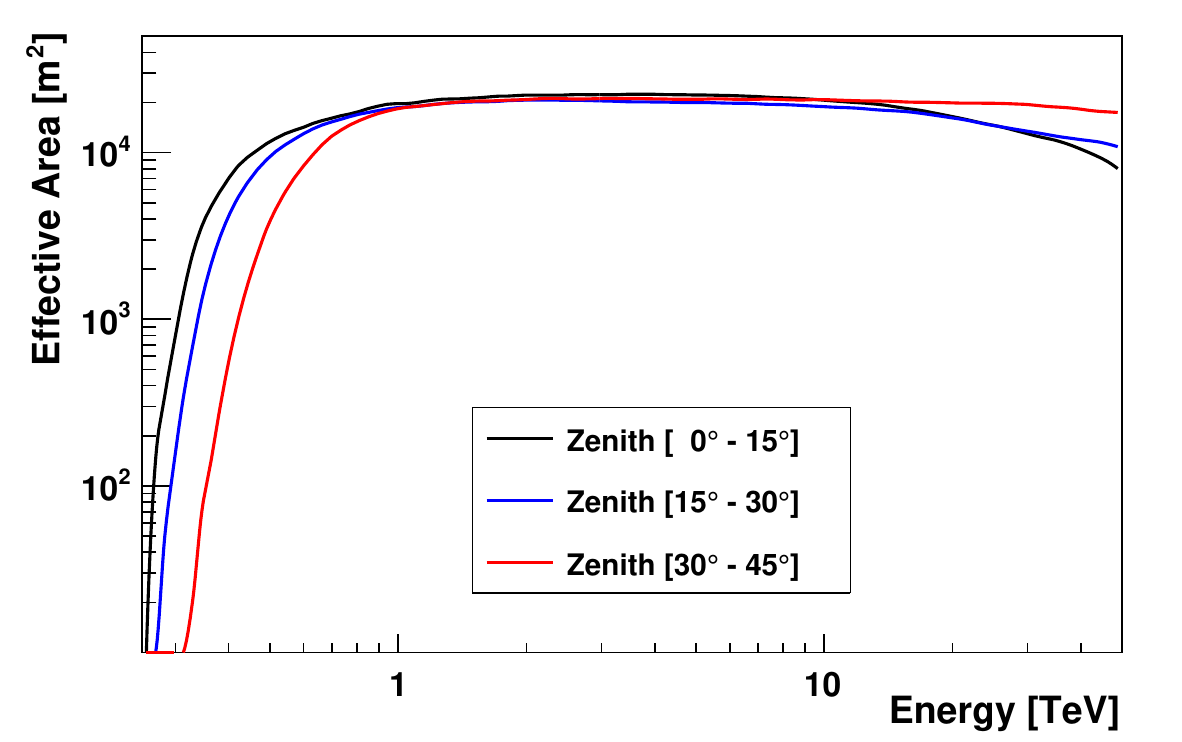} 
   \caption{Effective detection area for CRe, after selection cuts, as a function of energy, for three different ranges in zenith angle. Effective detection areas are evaluated on a run-wise basis; the averaged areas are shown. }
   \label{fig:effarea}
\end{figure}

\begin{figure}[htbp] 
   \centering
   \includegraphics[width=0.46\textwidth]{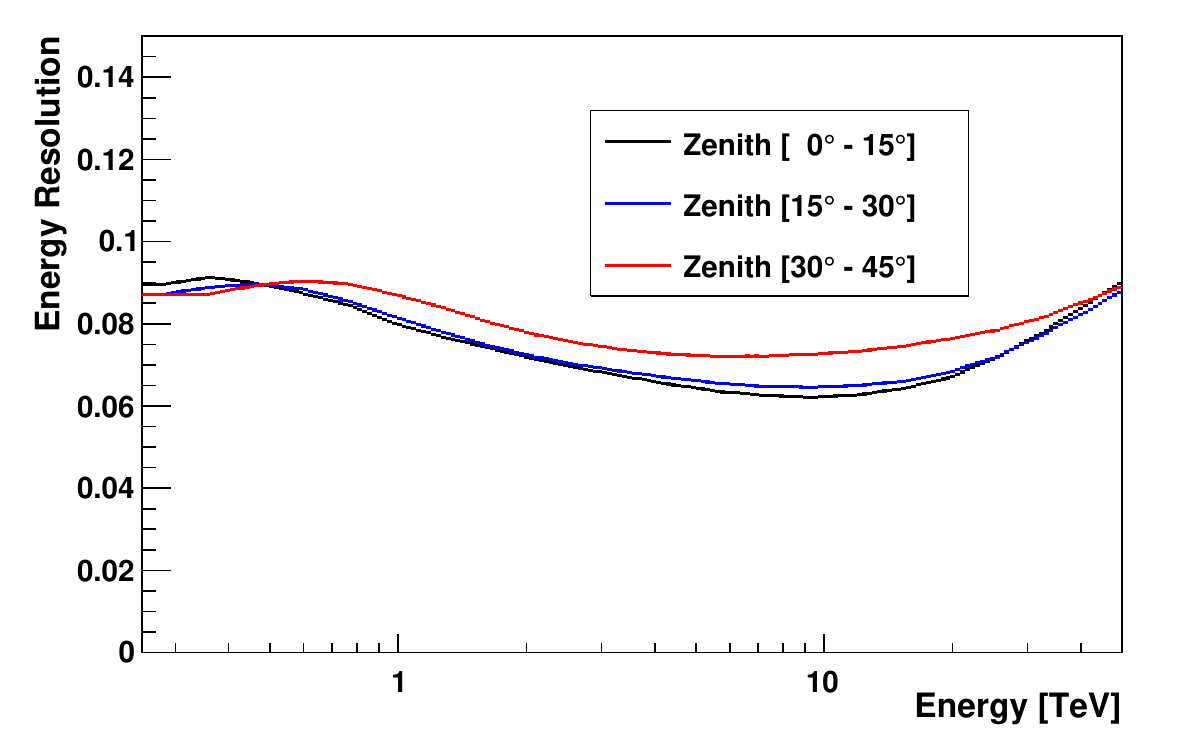} 
   \caption{Energy resolution for CRe, as a function of energy, for three different zenith angle ranges. {Energy resolution is defined as $\sigma(\ln{E}) \approx \sigma(E)/E$} and is evaluated on a run-wise basis; the run-averaged resolution is shown.}
   \label{fig:energyres}
\end{figure}

The detection capabilities of the H.E.S.S. instrument are characterized by instrument response functions. For the measurement of the CRe flux, the relevant response functions include the effective detection area after electron selection cuts (Fig. \ref{fig:effarea}) and the energy resolution (Fig. \ref{fig:energyres}). 

For the zenith angle range used -- up to $45^\circ$ -- the effective detection area is essentially constant in the range between 1 TeV and 20 TeV. Due to the tight selection cuts, the saturation value of the effective area is almost an order of magnitude below the geometrical area of $1.3 \cdot 10^5$\,m$^2$ corresponding to the 200 m core distance cut. At lower energies, the effective area rises steeply, reflecting trigger efficiencies and -- more importantly -- the selection cuts on the number of images and the image sizes (see Table \ref{table_cuts}). The drop of effective areas at energies above 20 TeV is caused by very long images that are less well contained in the camera. 

The energy resolution is shown in Fig. \ref{fig:energyres}; at all energies and zenith angles, the (run-averaged) CRe energy resolution is better than 10\%. As one might expect, the energy resolution improves with energy, except for the highest energies where events from large impact distances contribute the most.

\section{Validation of instrument simulation using $\gamma$-rays} 

\begin{figure*}
     \centering

     \begin{subfigure}[b]{\textwidth}
         \centering
         \includegraphics[width=0.38\textwidth]{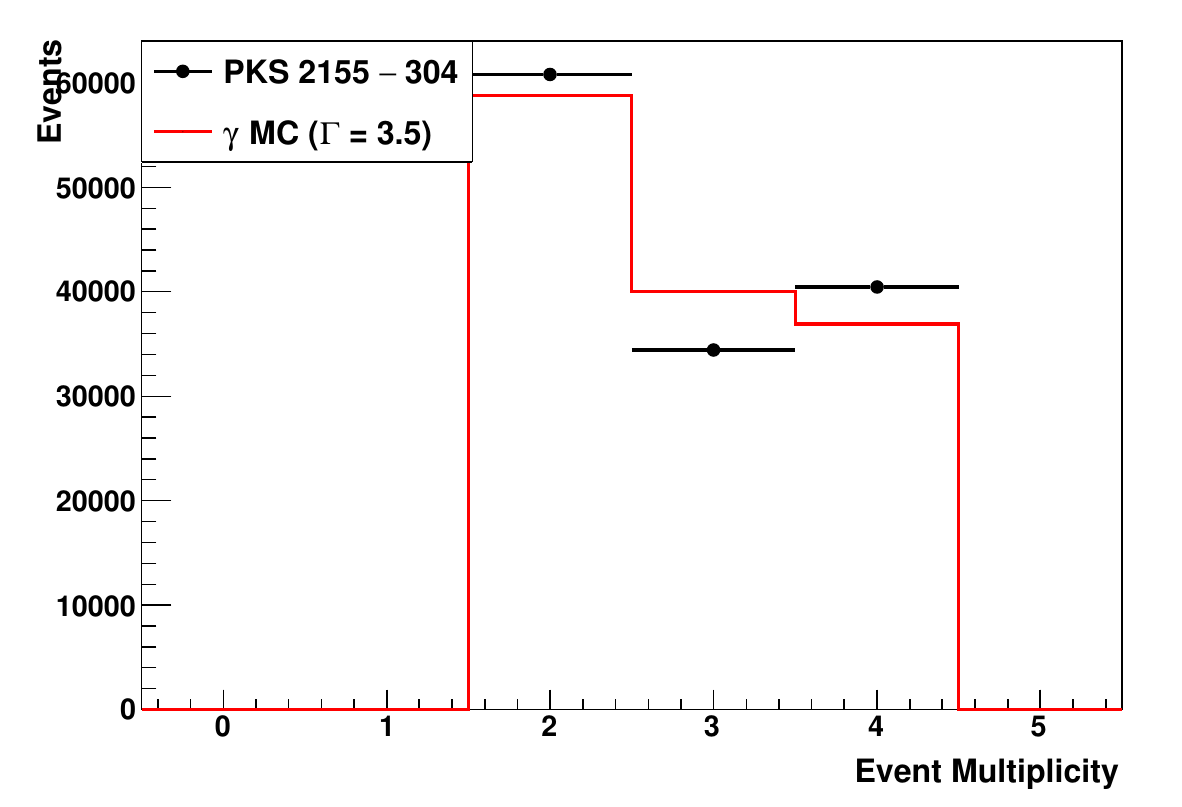} 
          \includegraphics[width=0.38\textwidth]{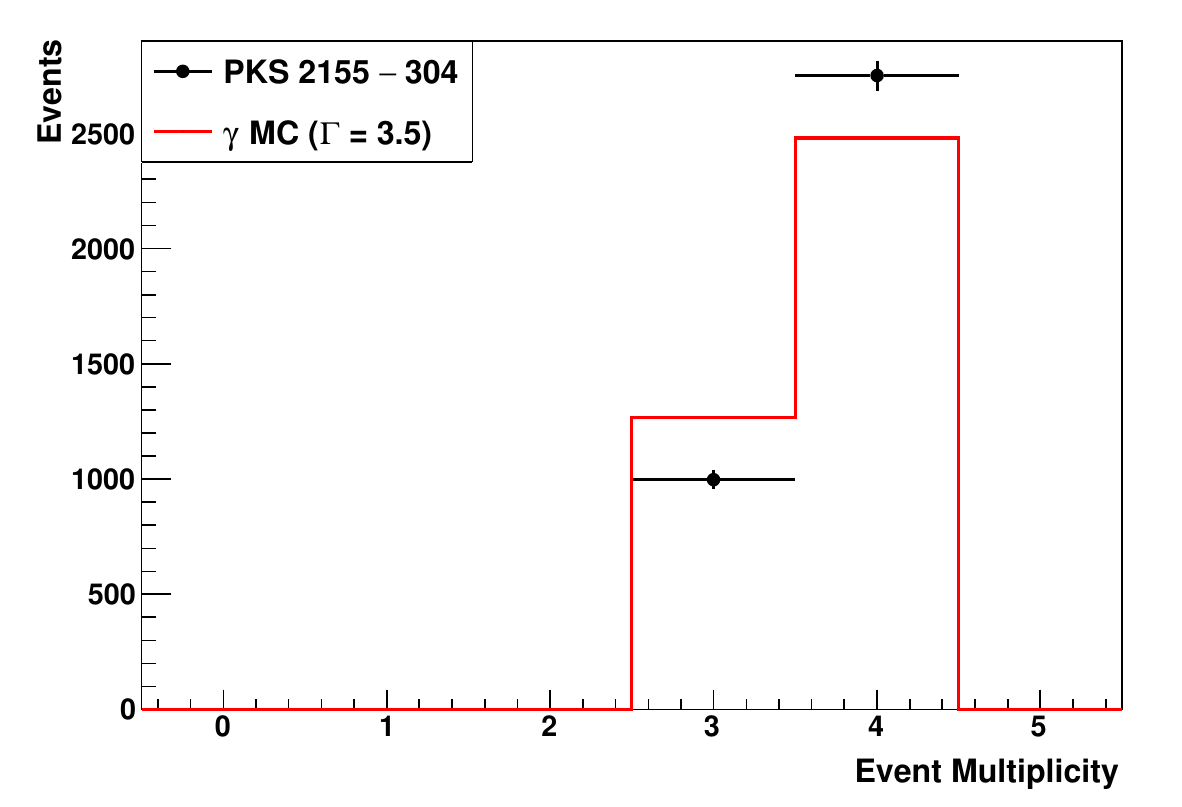}
         \caption{Distribution in telescope multiplicity}
     \end{subfigure}

     \begin{subfigure}[b]{\textwidth}
         \centering
         \includegraphics[width=0.38\textwidth]{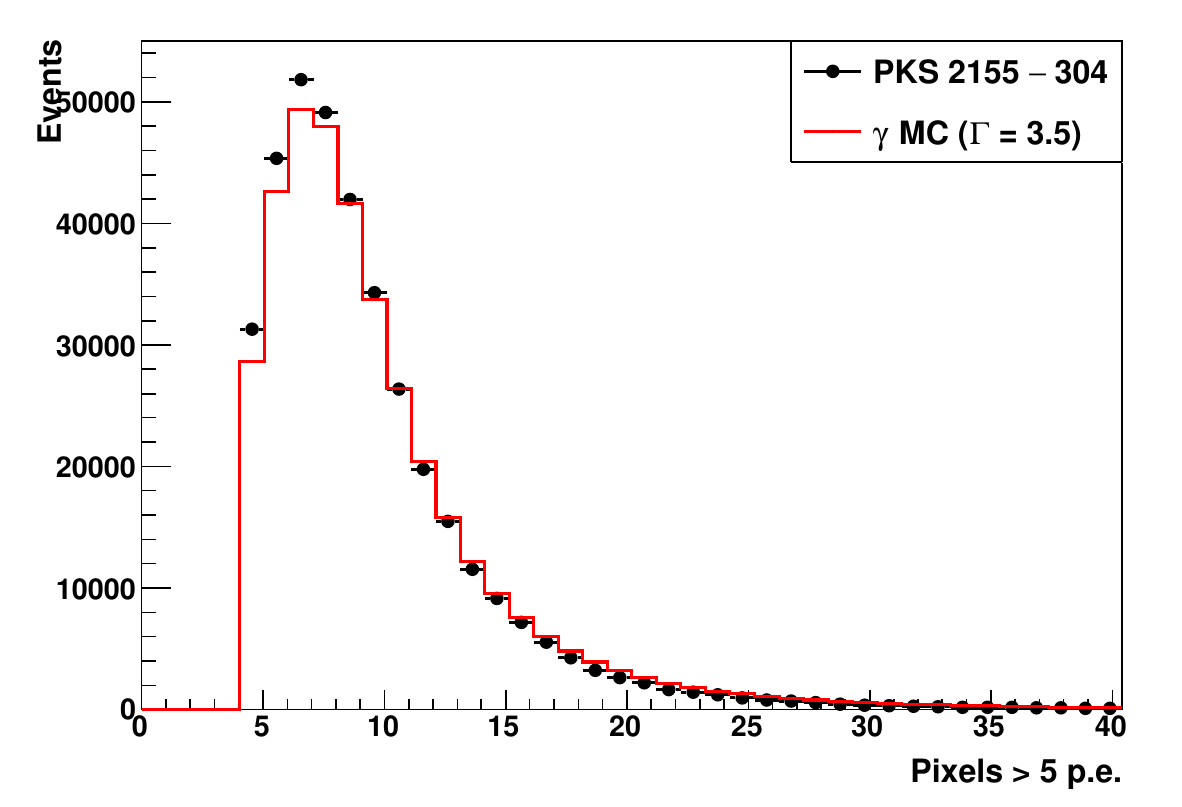} 
          \includegraphics[width=0.38\textwidth]{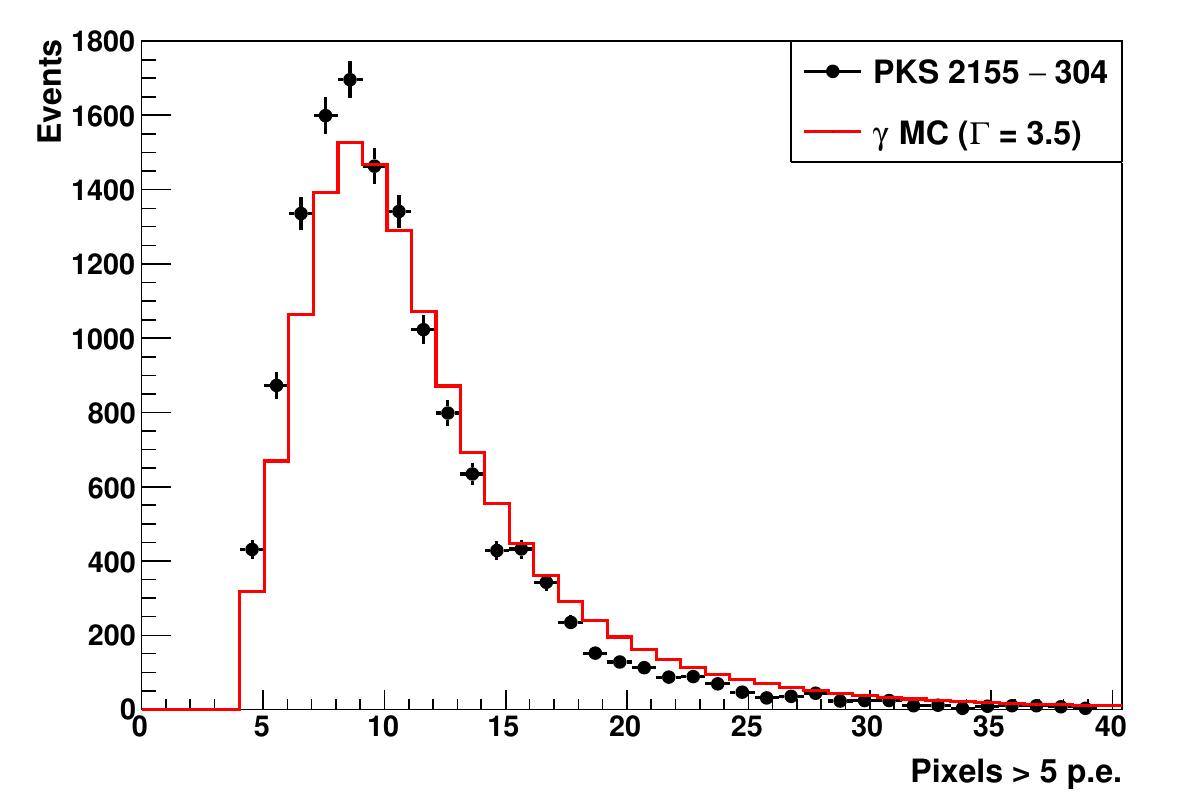}
         \caption{Distribution in number of image pixels}
     \end{subfigure}

     \begin{subfigure}[b]{\textwidth}
         \centering
         \includegraphics[width=0.38\textwidth]{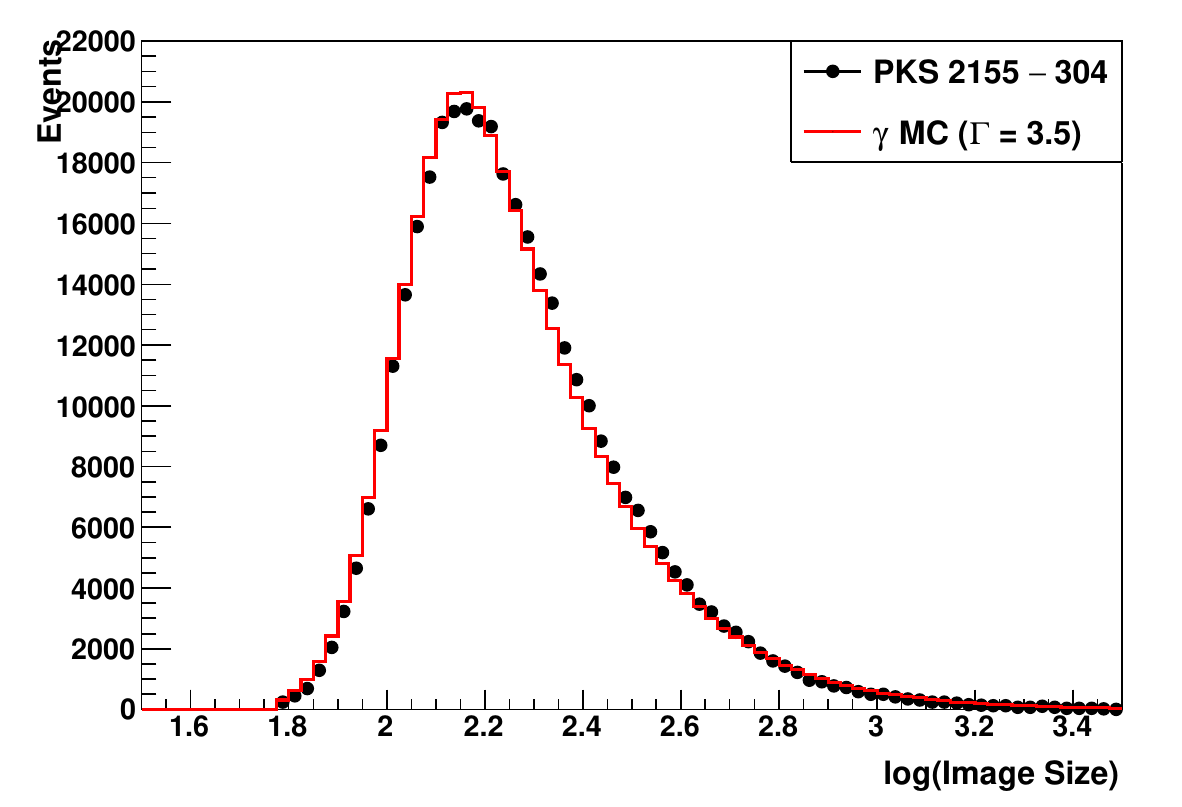} 
          \includegraphics[width=0.38\textwidth]{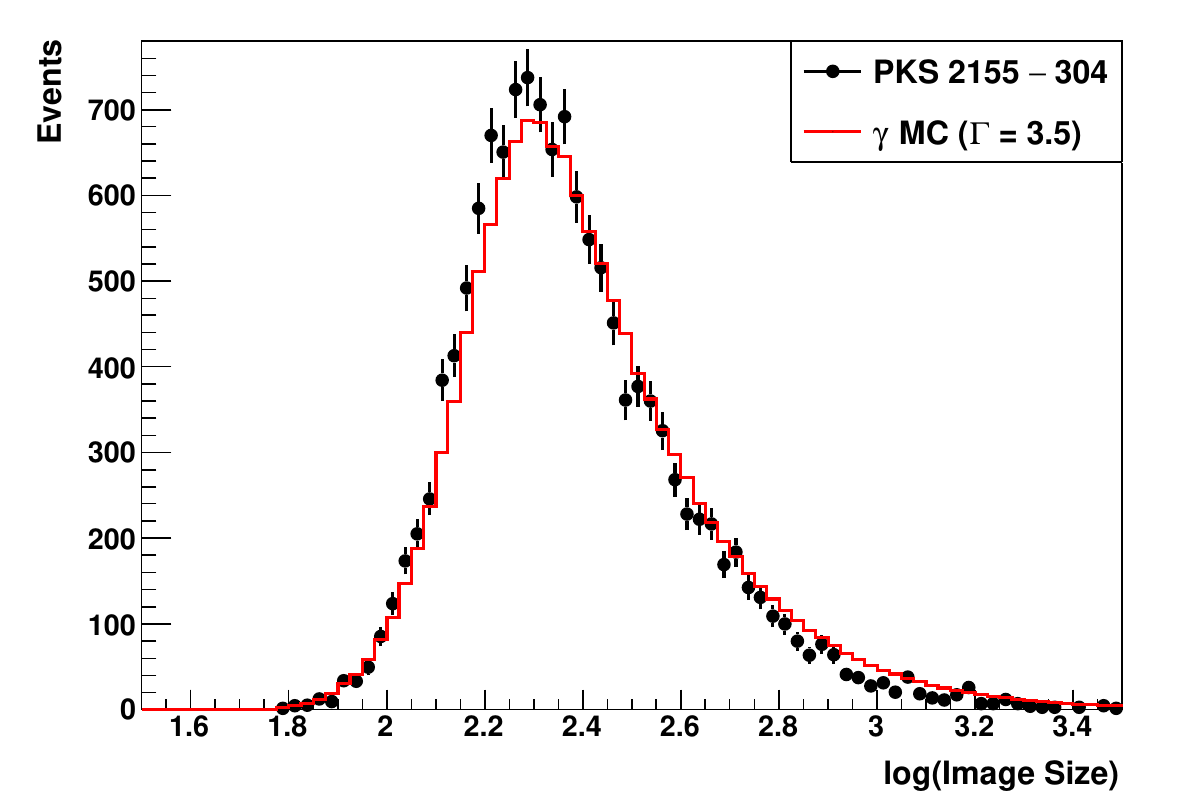}
         \caption{Distribution in log(image size)}
     \end{subfigure}

     \begin{subfigure}[b]{\textwidth}
         \centering
         \includegraphics[width=0.38\textwidth]{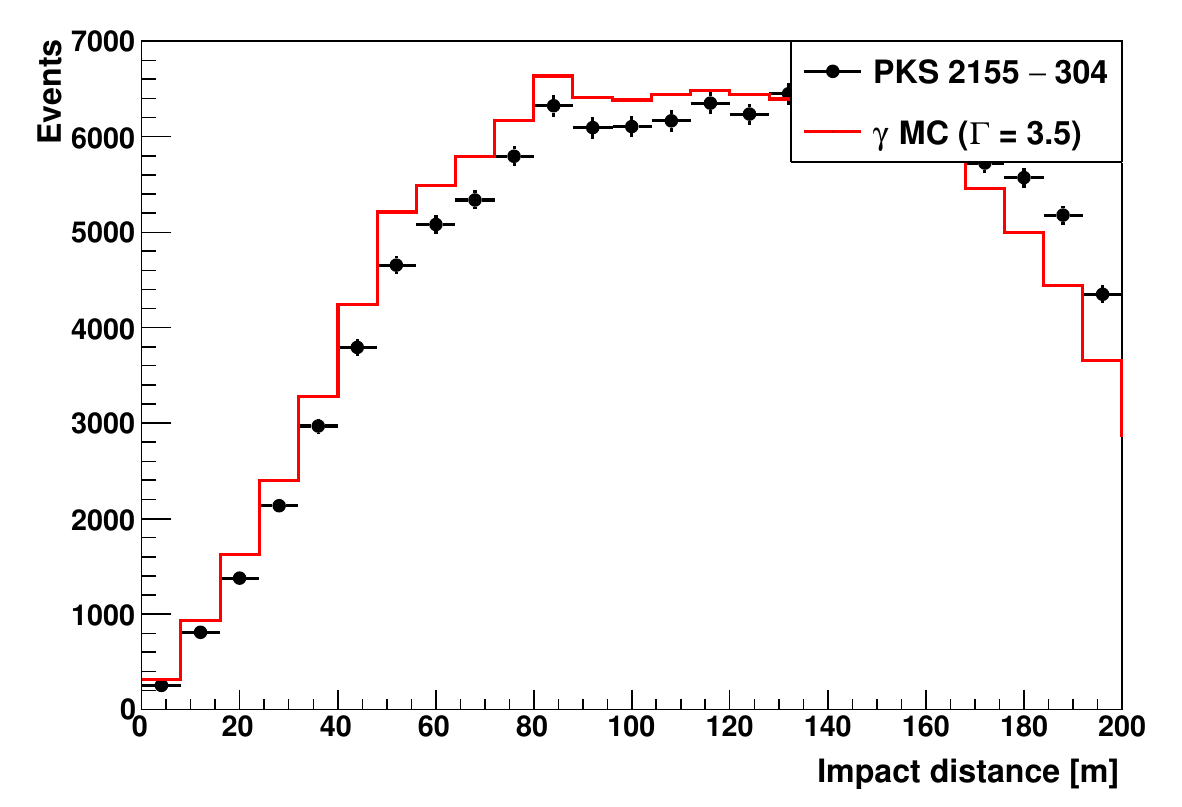} 
          \includegraphics[width=0.38\textwidth]{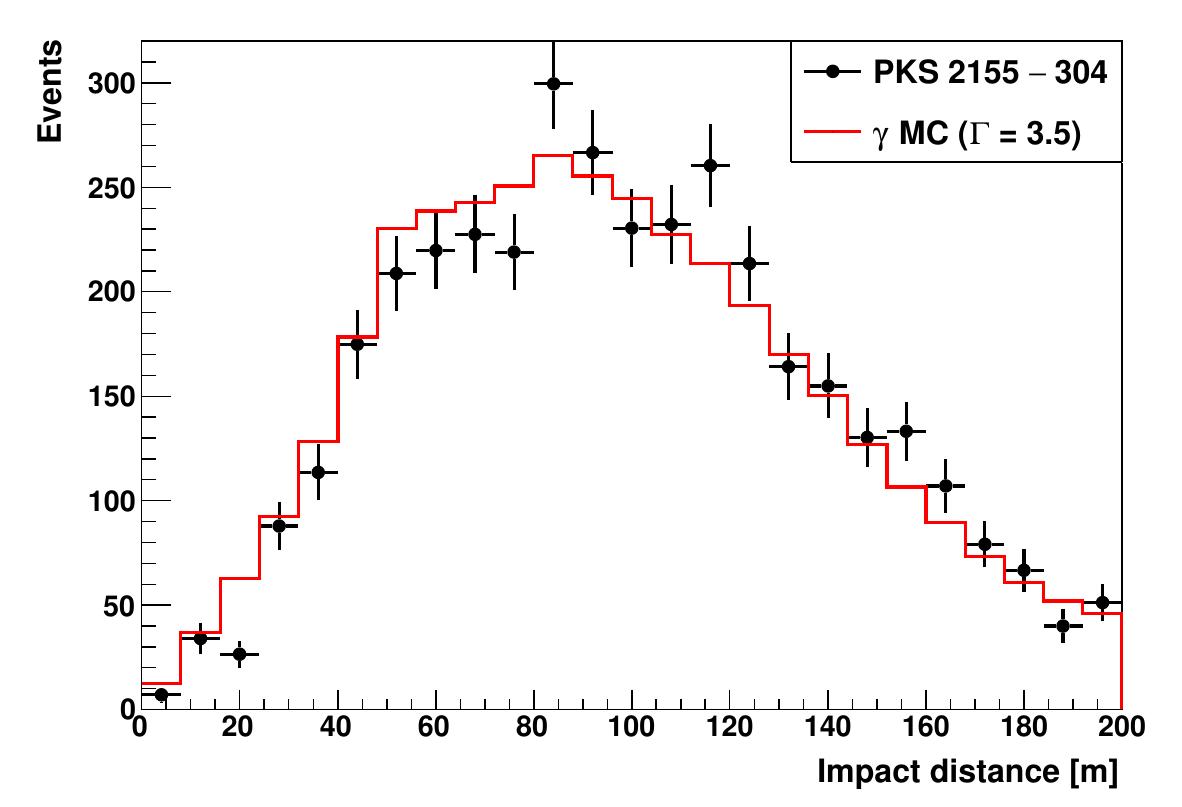}
         \caption{Distribution in core distance}
     \end{subfigure}

        \caption{Properties of $\gamma$-ray events from PKS 2155-304, after background subtraction (data points), compared to simulations (red histogram). Left column: with selection cuts as usually applied for $\gamma$-rays; right column: with the harder selection cuts applied for the CRe sample. Shown are (a): Number of triggered telescopes (telescope multiplicity), (b) Number of image pixels in each telescope, (d) log$_{10}$ of image intensity in photoelectrons, (d): Core distance to center of the array, calculated orthogonal to the shower axis. }
        \label{fig:sim_1}
\end{figure*}

\begin{figure*}
     \centering
     \begin{subfigure}[b]{\textwidth}
         \centering
         \includegraphics[width=0.38\textwidth]{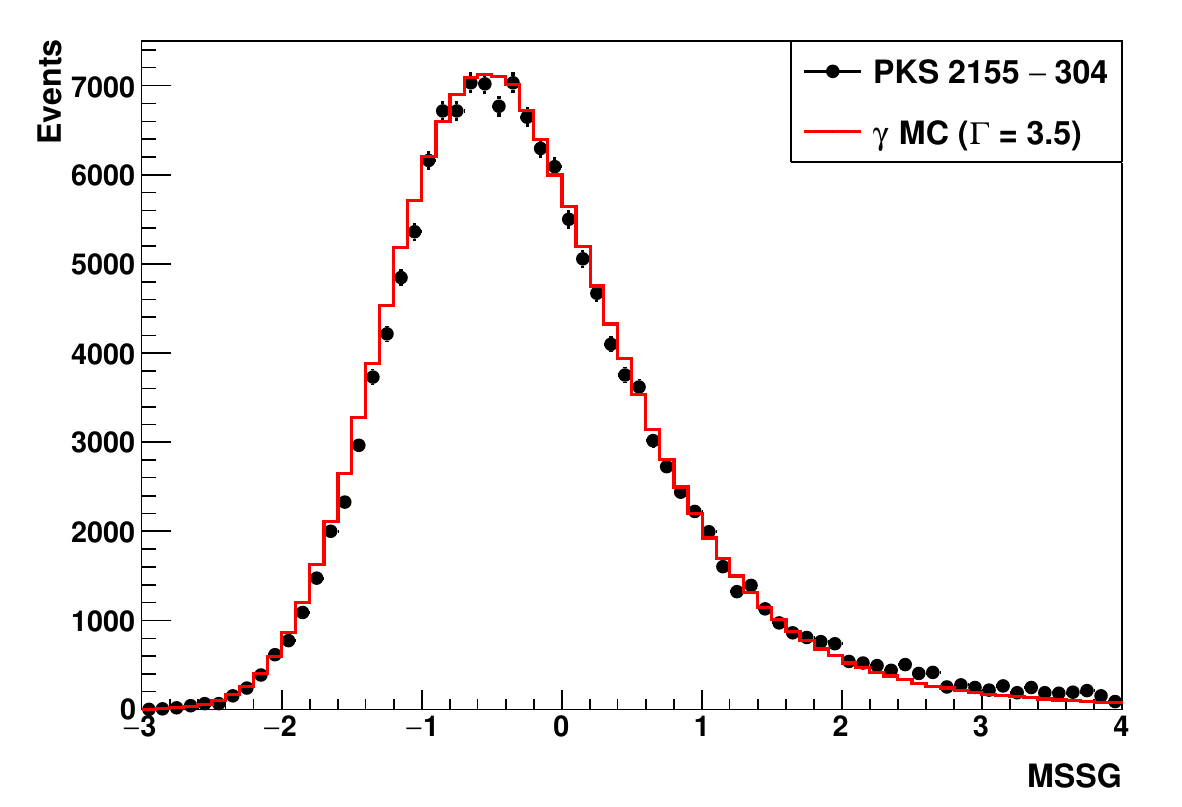} 
          \includegraphics[width=0.38\textwidth]{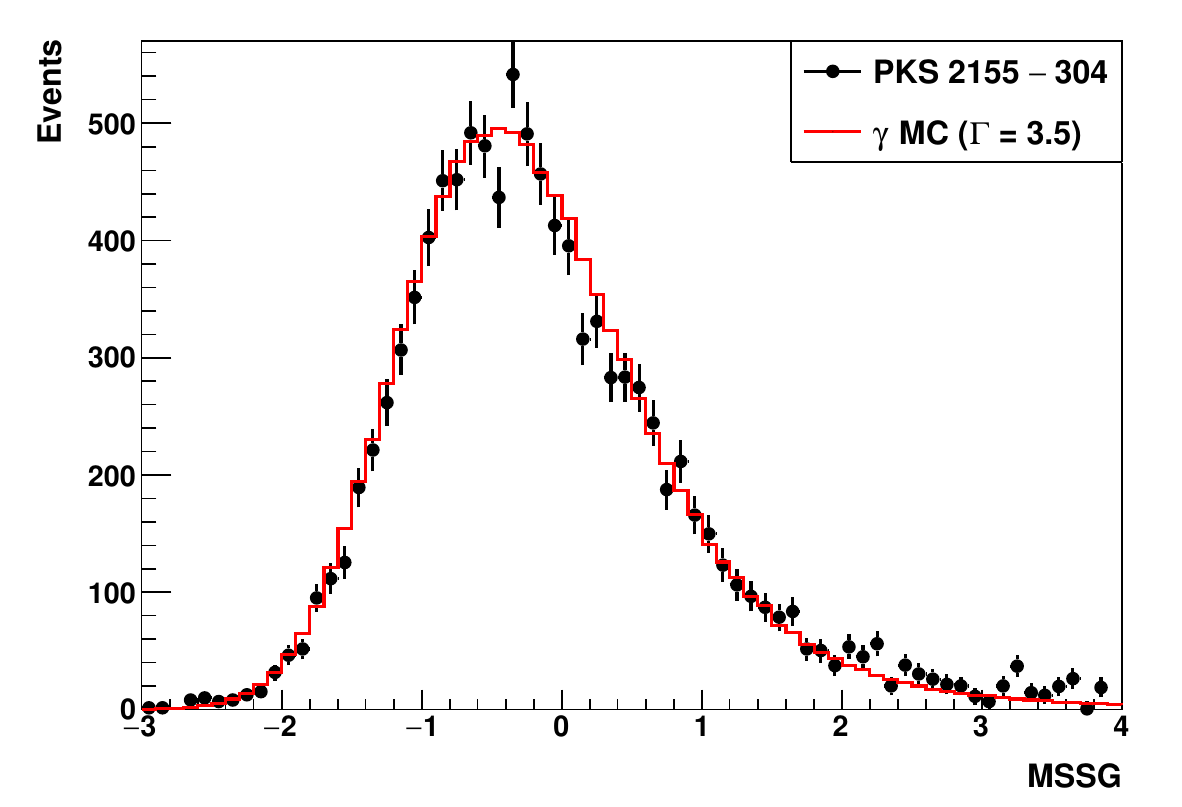}
         \caption{Distribution in Mean Scaled Shower Goodness (MSSG)}
     \end{subfigure}

     \begin{subfigure}[b]{\textwidth}
         \centering
         \includegraphics[width=0.38\textwidth]{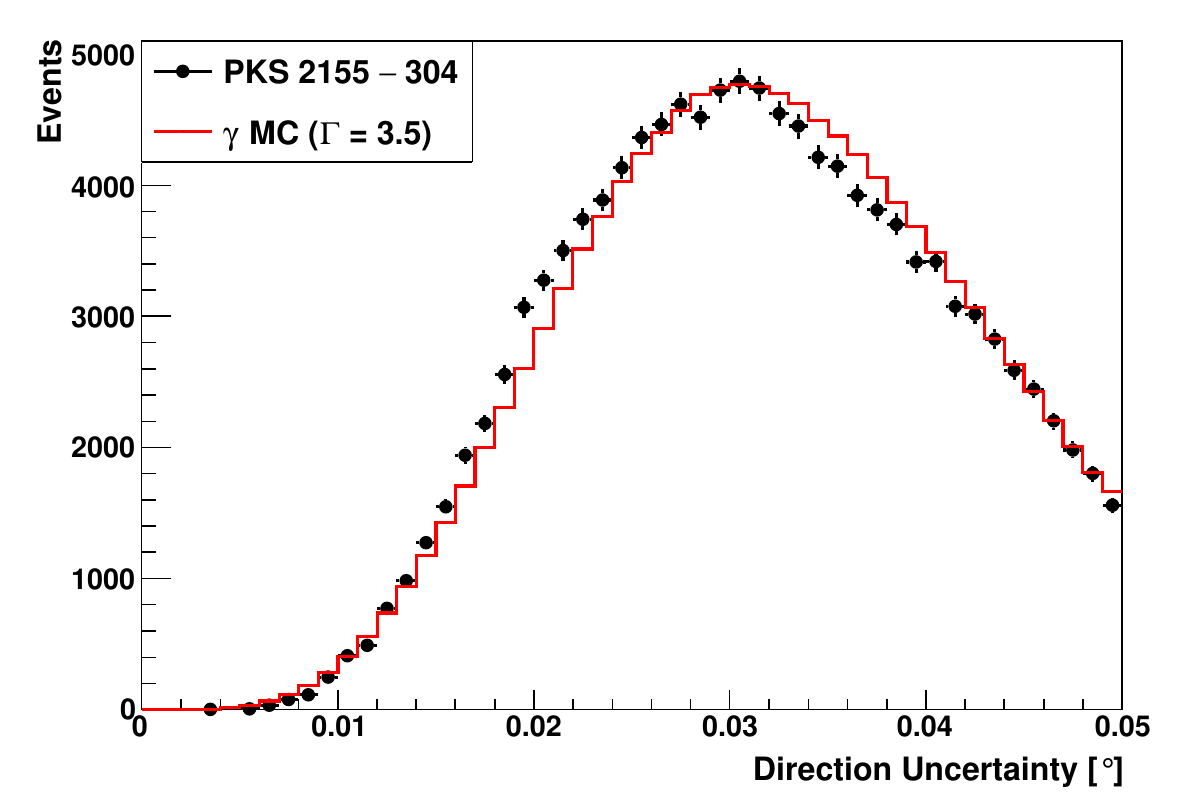} 
          \includegraphics[width=0.38\textwidth]{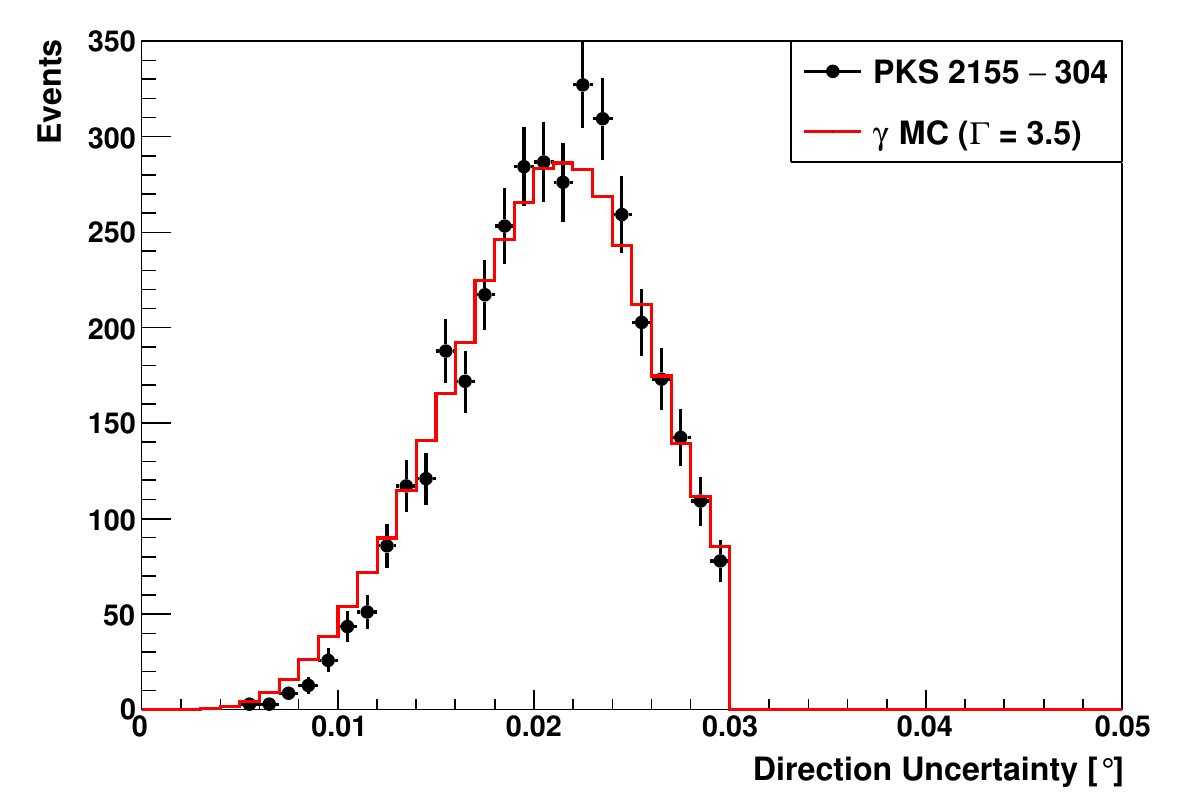}
         \caption{Distribution in direction error}
     \end{subfigure}

     \begin{subfigure}[b]{\textwidth}
         \centering
         \includegraphics[width=0.38\textwidth]{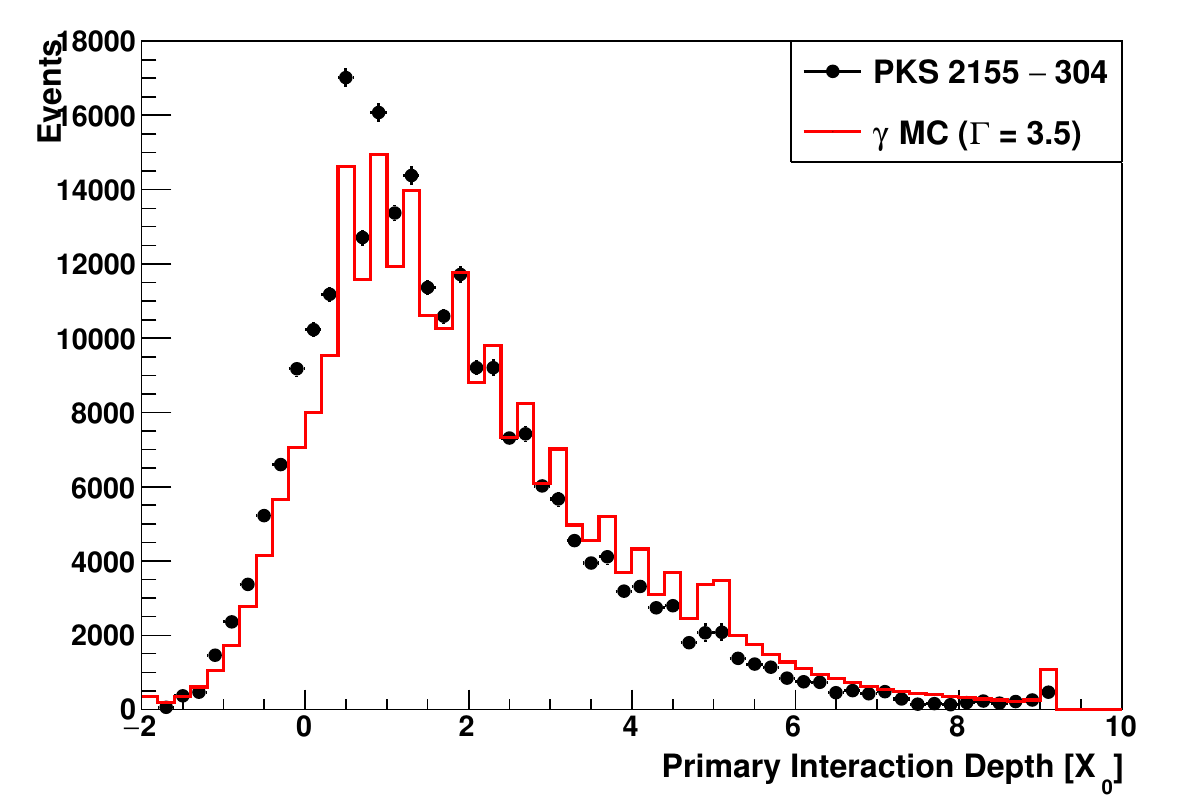} 
          \includegraphics[width=0.38\textwidth]{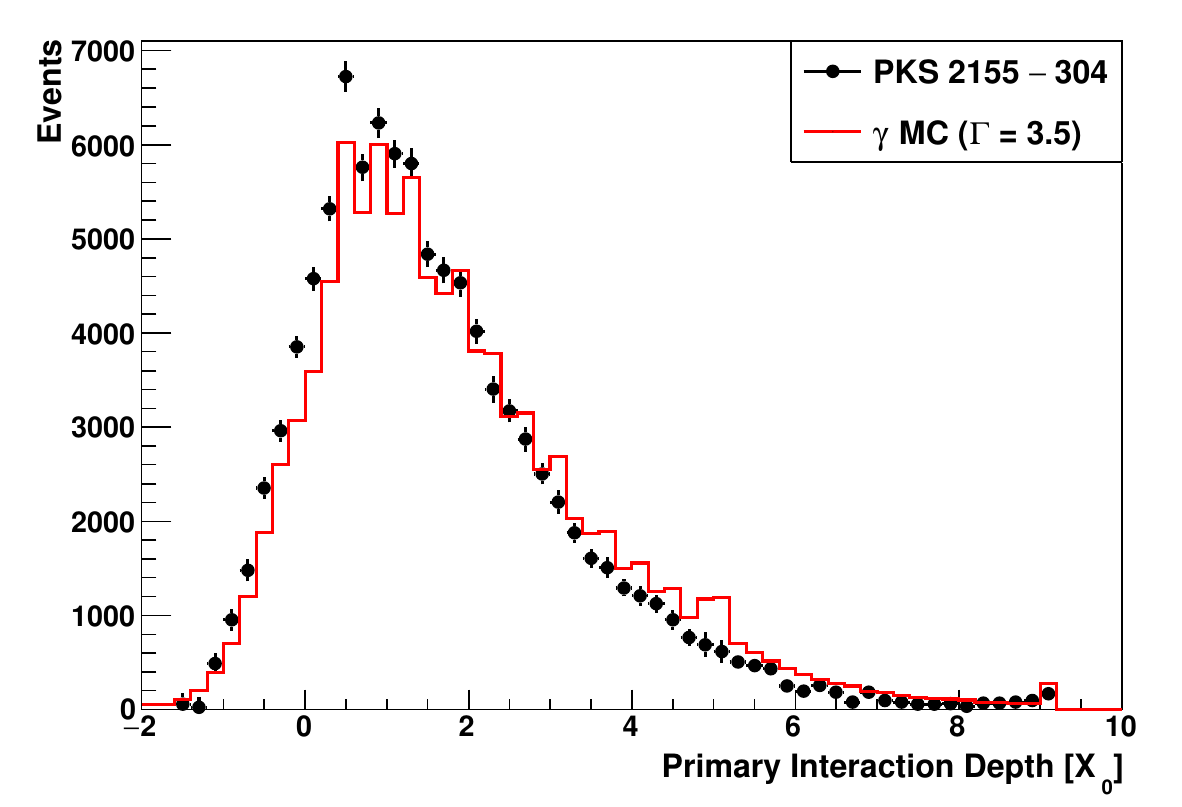}
         \caption{Distribution in reconstructed depth of $1^{\mathrm{st}}$ interaction, for pre-selected events}
     \end{subfigure}

     \begin{subfigure}[b]{\textwidth}
         \centering
         \includegraphics[width=0.38\textwidth]{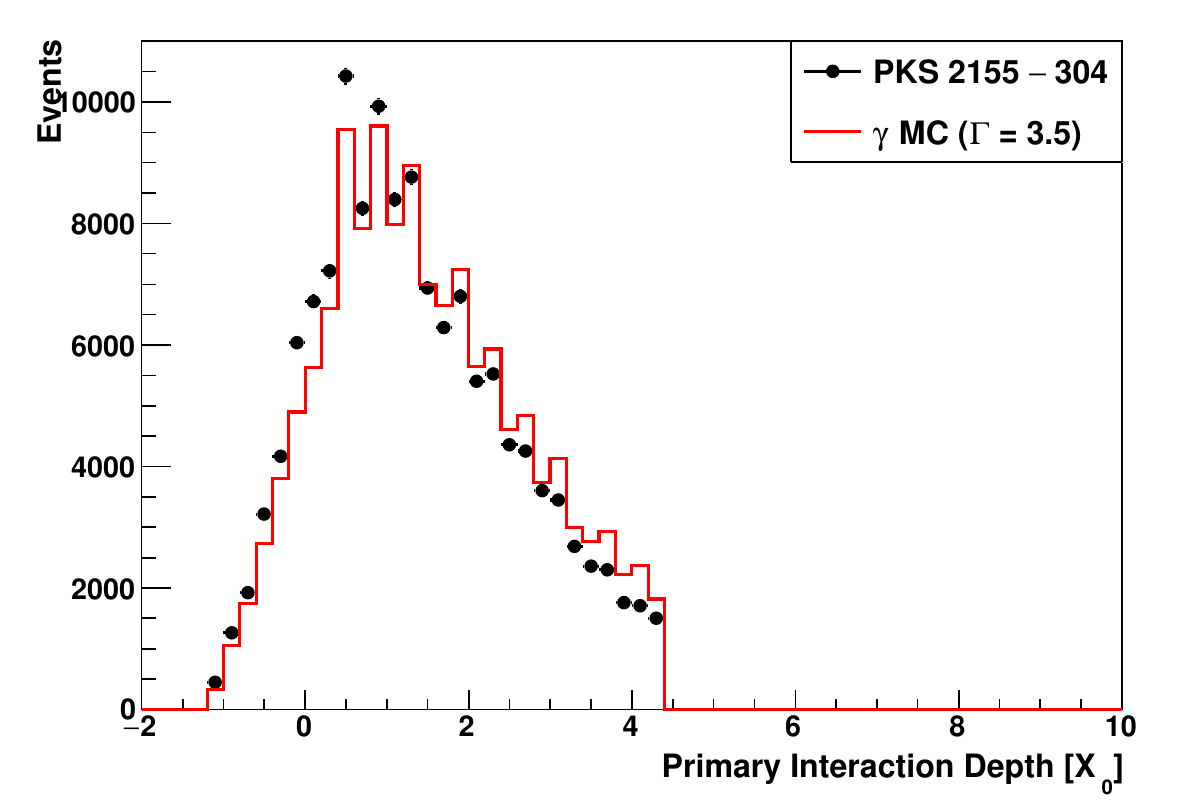} 
          \includegraphics[width=0.38\textwidth]{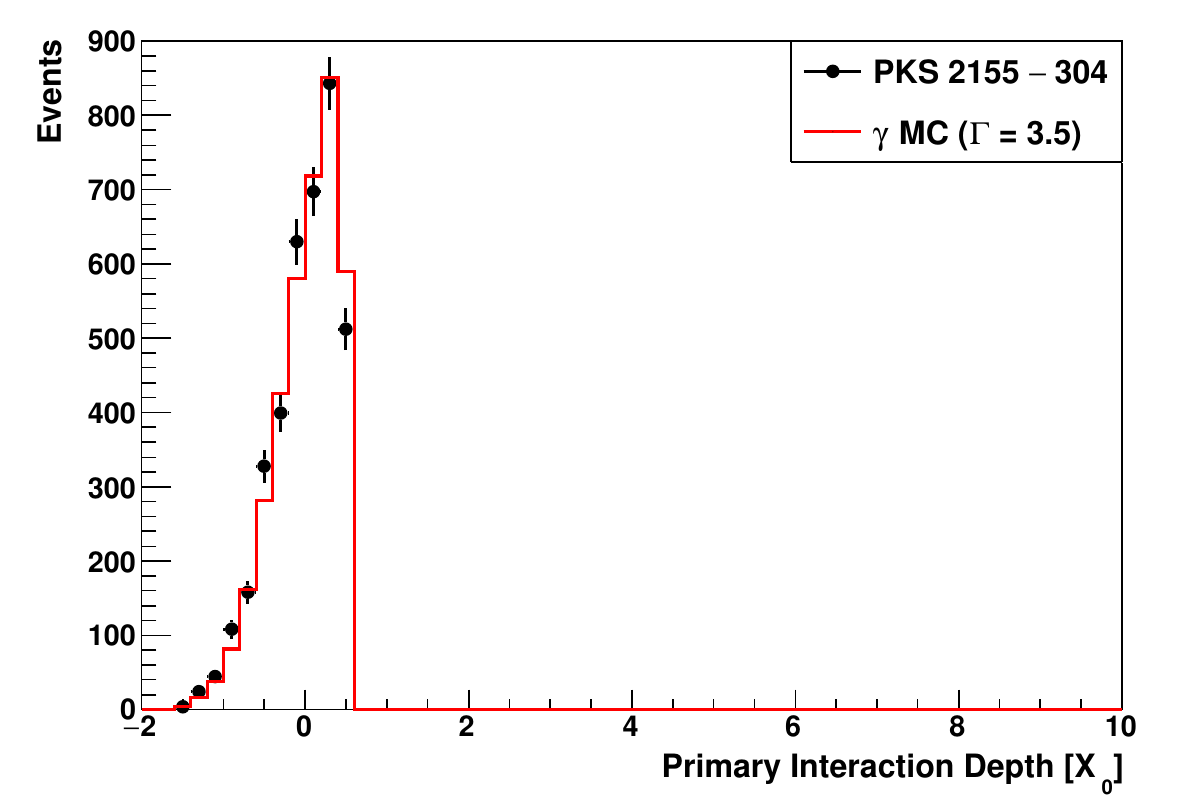}
         \caption{Distribution in reconstructed depth of $1^{\mathrm{st}}$ interaction, for selected events}
     \end{subfigure}

        \caption{Properties of $\gamma$-ray events from PKS 2155-304, after background subtraction (data points), compared to simulations (red histogram). Left column: with selection cuts as usually applied for $\gamma$-rays; right column: with the more stringent selection cuts applied for the electron sample. Shown are (a): Distribution in Mean Scaled Shower Goodness (MSSG), (b) Distribution in direction error, (c) Distribution in reconstructed depth of $1^{\mathrm{st}}$ interaction, for pre-selected events prior to final cuts (d): Same for selected events. }
        \label{fig:sim_2}
\end{figure*}

The selection of electron-like events and the determination of the effective detection areas after the stringent selection cuts relies on an accurate simulation of the detector. Since $\gamma$-ray induced showers are rather similar to electron-induced showers, a good way to demonstrate the quality of the simulation is to use data from a $\gamma$-ray source, where -- after statistical subtraction of the background under the $\gamma$-ray signal -- a pure $\gamma$-ray sample can be obtained. Data from observations of the active galactic nucleus PKS 2155-304 are used, a $\gamma$-ray point source with a spectral index $\Gamma \approx 3.5$, in the same range as the electron spectral index. Using the identical scheme as for the electron sample, 755 28-min runs on PKS~2155-304 were processed, with simulation parameters for the instrument and the atmosphere adapted to the specific conditions of each run \cite{RunWise2020}. The distribution in zenith angle of the PKS 2155-304 data is similar to that of the CRe data.
In Fig.~\ref{fig:sim_1} and \ref{fig:sim_2}, the $\gamma$-ray data distributions are compared to $\gamma$-ray simulations, for various crucial event parameters and for two sets of selection cuts: the normal cuts used for $\gamma$-ray analysis (left column in Figs. \ref{fig:sim_1} and \ref{fig:sim_2}), and the more stringent cuts applied to derive a sample of CRe candidates (right column). The cuts are summarized in Table \ref{table_cuts}.

Fig. \ref{fig:sim_1} compares $\gamma$-ray data and simulation for general shower properties, such as the number of telescopes triggered in a given event, the number of image pixels containing at least 5 photoelectrons, the distribution in the total image intensity in units of photoelectrons, and the distance between the reconstructed shower core and the center of the H.E.S.S. array. These quantities are sensitive to the modelling of the telescope and array energy thresholds; the fact that the distributions are well reproduced by the simulation implies a reliable modelling of effective detection areas. The position of the peaks in the distributions of pixel numbers and image intensity -- that characterise the threshold behaviour -- are very well reproduced by the simulation.

Fig. \ref{fig:sim_2} shows distributions in variables that are sensitive to distinguish electron- (and $\gamma$-)like showers from the hadronic background. Variables include the Mean Scaled Shower Goodness (MSSG) describing how well the shower images match the templates, the error on the reconstructed shower direction (which is another measure for the quality and consistency of the shower reconstruction), and the reconstructed depth of the first interaction. For these critical variables, the simulations provide a good description of the data, demonstrating that one can rely on the simulations in the selection of events, and in the determination of selection efficiencies and effective detection areas.

\section{Contamination of the CRe sample}

Potential contaminations in the CRe sample include:
\begin{itemize}
\item Diffuse extragalactic $\gamma$-rays
\item Galactic diffuse $\gamma$-rays
\item $\gamma$-rays from unresolved sources, or remaining contamination from known sources
\item Misidentified cosmic-ray nucleons (CRn)
\end{itemize}

\subsection{Contamination by $\gamma$-rays}
{ Fig. \ref{fig:spectrum} of the main paper includes - beyond the measured CRe spectrum  – also the spectrum of diffuse extragalactic $\gamma$-rays as measured by {\it Fermi}-LAT \cite{2015ApJ...799...86A}.} The CRe selection cuts accept the bulk of the extragalactic $\gamma$-rays, but their flux is too low to present a significant contamination, and cuts off beyond a TeV. A significant contamination by Galactic diffuse $\gamma$-rays can be excluded since CRe spectral parameters are consistent across different latitude ranges (see below, Fig. \ref{fig:latitude}). Contamination from known $\gamma$-ray sources was tested by varying the (generous) angular cut used to exclude source regions. Regarding contamination by unresolved sources, it should be noted that the CRe flux, integrated over the $4^\circ$ diameter used field of view, is of the same scale as the flux of the Crab nebula, well above any conceivable flux of unresolved $\gamma$-ray sources.

\subsection{Contamination by hadronic cosmic rays}
{ Fig. \ref{fig:spectrum} of the main paper also includes the cosmic-ray proton spectrum as measured by AMS-02 \cite{AMSProtons} and DAMPE \cite{2019SciA....5.3793A}.} The cosmic-ray proton flux is  up to $10^3$ times the CRe flux for energies below the CRe spectral break; the ratio increases to a few $10^4$ at 10 TeV. Therefore, a proton contamination that increases with energy seems unavoidable; the rejection power of the selection algorithm against protons is of similar scale, but cannot be modeled with sufficient reliability and event statistics, nor determined from the data. At energies around  3~TeV and above, a strong or even dominant proton contamination cannot be excluded.

\begin{figure}[htbp]
   \centering
   \includegraphics[width=0.5\textwidth]{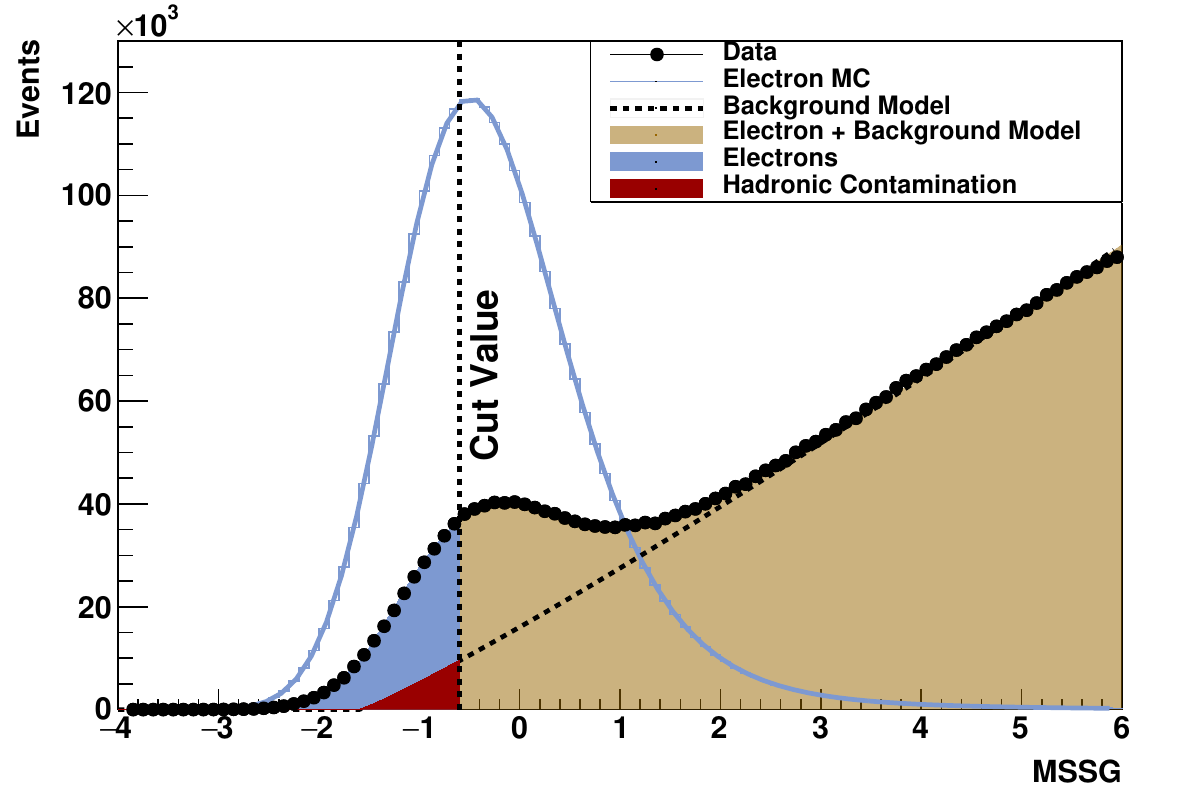}
  
   \caption{Measured MSSG distribution (for the full energy range) fitted by the simulated shape of the distribution for electrons (blue curve) plus an ad-hoc parameterization of the hadronic background (dotted line). The number of CRe candidate events is determined from the number of events with MSSG values below the MSSG cut position. These events are composed of true CRe (blue area) and a tail of the hadronic background (red area).} 
   \label{fig:illust_cont}
\end{figure}

Fig. \ref{fig:illust_cont} illustrates how the hadronic background in the CRe candidate sample is estimated. Since the exact shape of the MSSG distribution of hadronic events in the CRe range cannot be simulated with sufficient reliability and statistics, the estimated contamination depends on the parameterization chosen for the background. 
Fitting the MSSG distribution as a sum of a parametrized background and the simulated electron signal, we have compared a wide range of different functional forms to parameterize the shape of the hadronic background. To account for simulation imperfections also in the electrons, we in addition allowed for a small shift in the electron peak, and a scale factor for the width of the distribution, as free parameters. All background shapes providing statistically acceptable fits were retained, and the contamination in the CRe sample determined. These contamination values were used to estimate the contamination and its range of variation, independently for each energy interval. The resulting limits on the hadronic contamination are $<25\%$ for the energy range of $0.3-1$~TeV and $<30\%$ for $1-3$~TeV. Beyond 3~TeV a dominating background cannot be excluded. Given the background shapes explored, our best estimate for the contamination is $12\% \pm 8\%$ for the energy range of $0.3-1$~TeV and $17\% \pm 10\%$ for $1-3$~TeV.

\section{Match between CRe data and simulations}

\begin{figure*}
     \centering

     \begin{subfigure}[b]{\textwidth}
         \centering
         \includegraphics[width=0.30\textwidth]{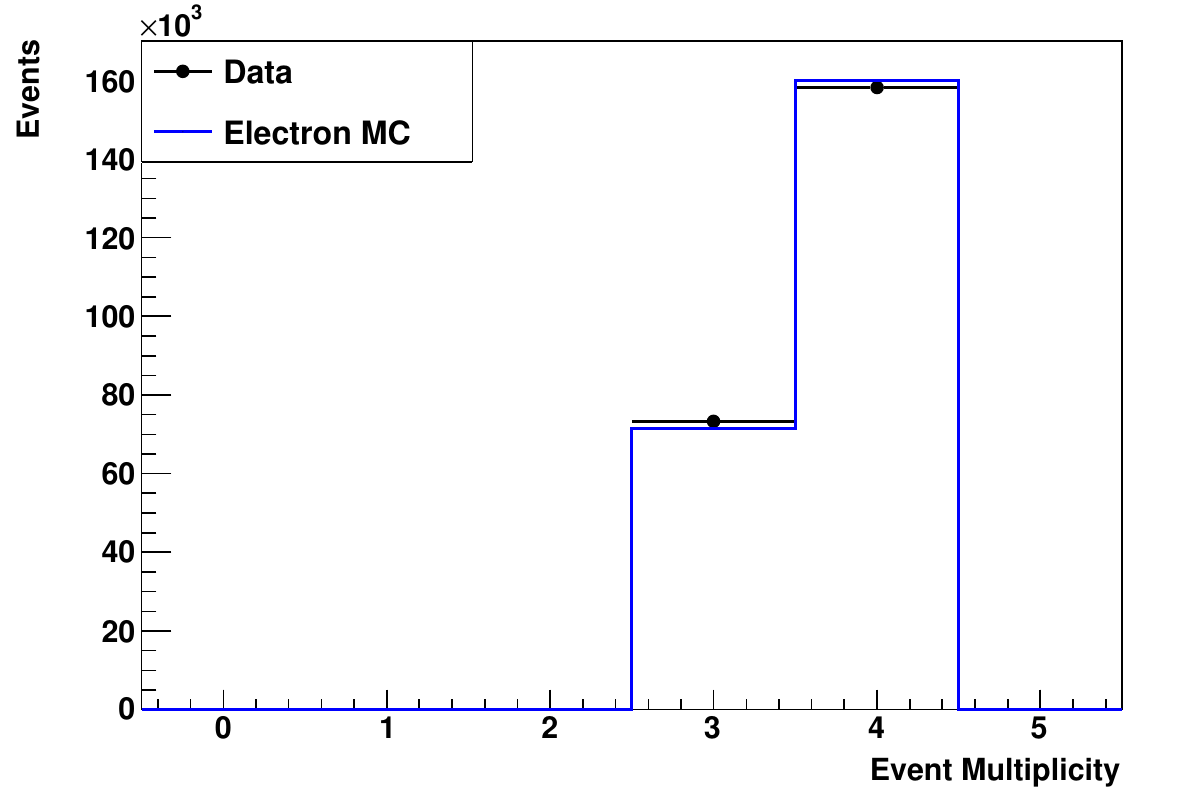}
          \includegraphics[width=0.30\textwidth]{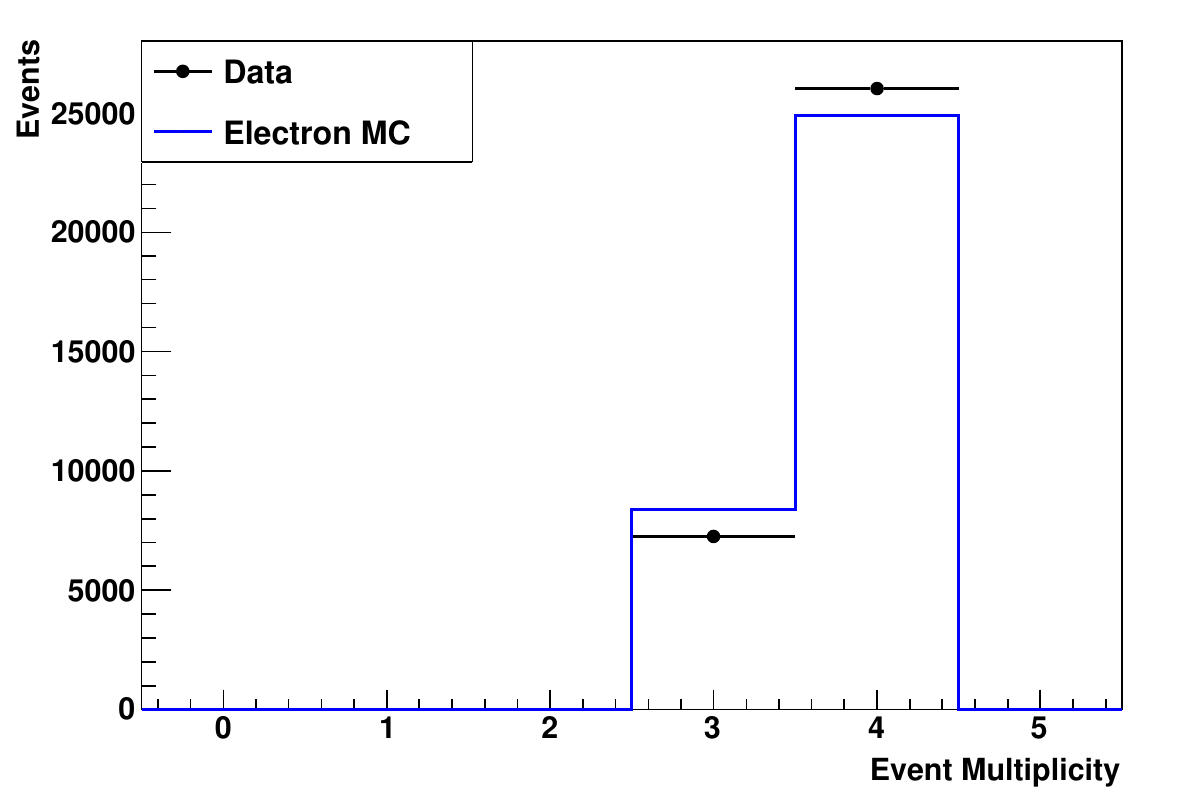}
          \includegraphics[width=0.30\textwidth]{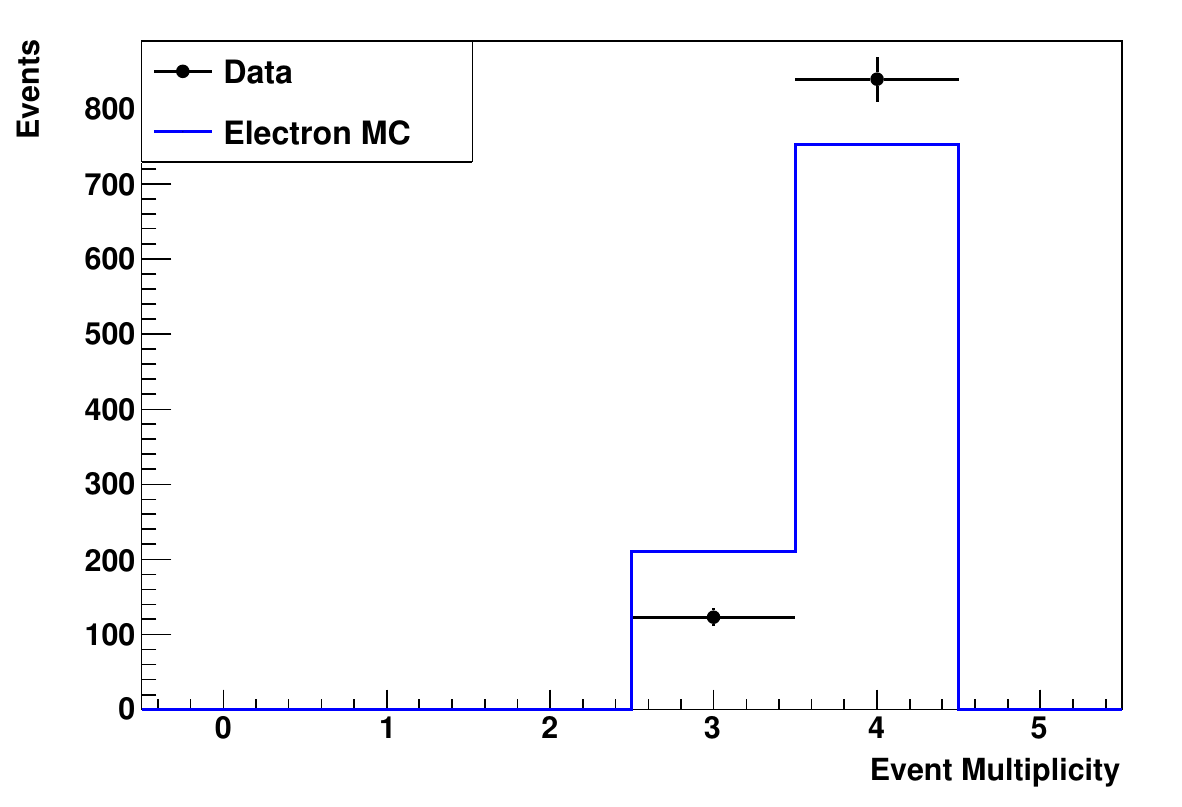}
         \caption{Distribution in telescope multiplicity}
     \end{subfigure}

     \begin{subfigure}[b]{\textwidth}
         \centering
         \includegraphics[width=0.30\textwidth]{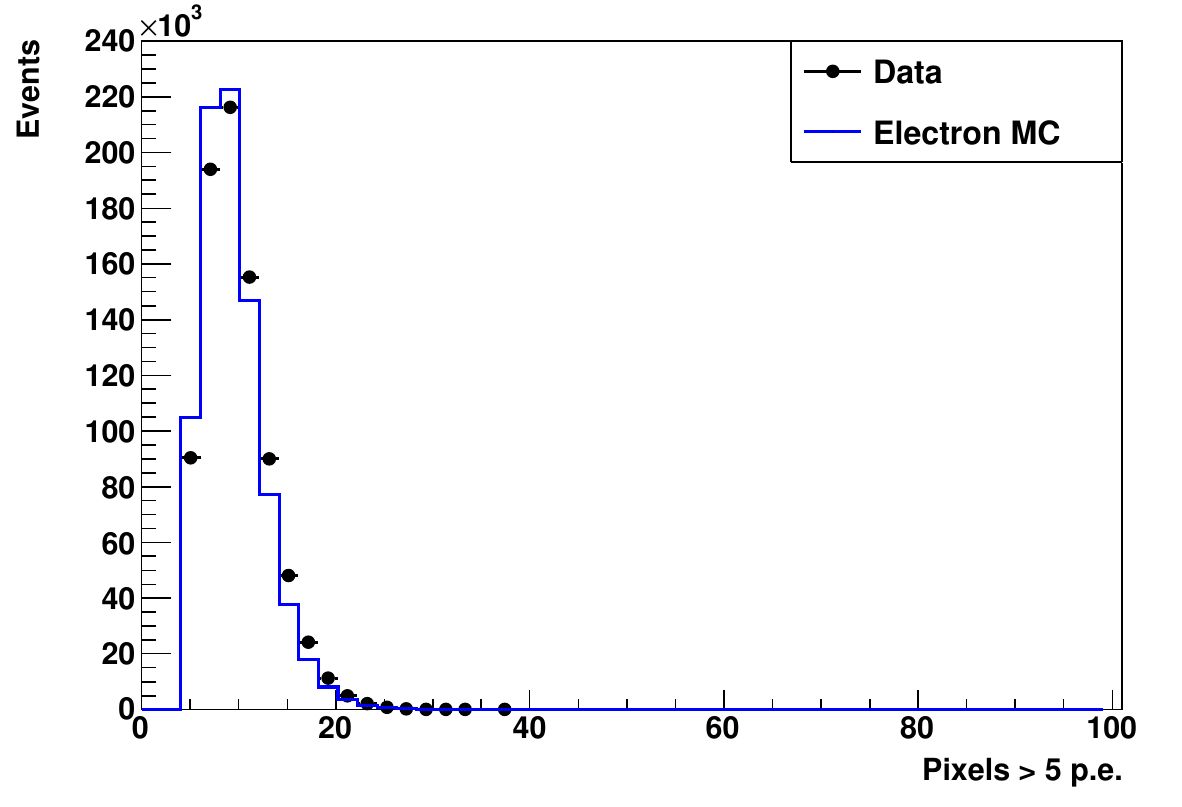}
          \includegraphics[width=0.30\textwidth]{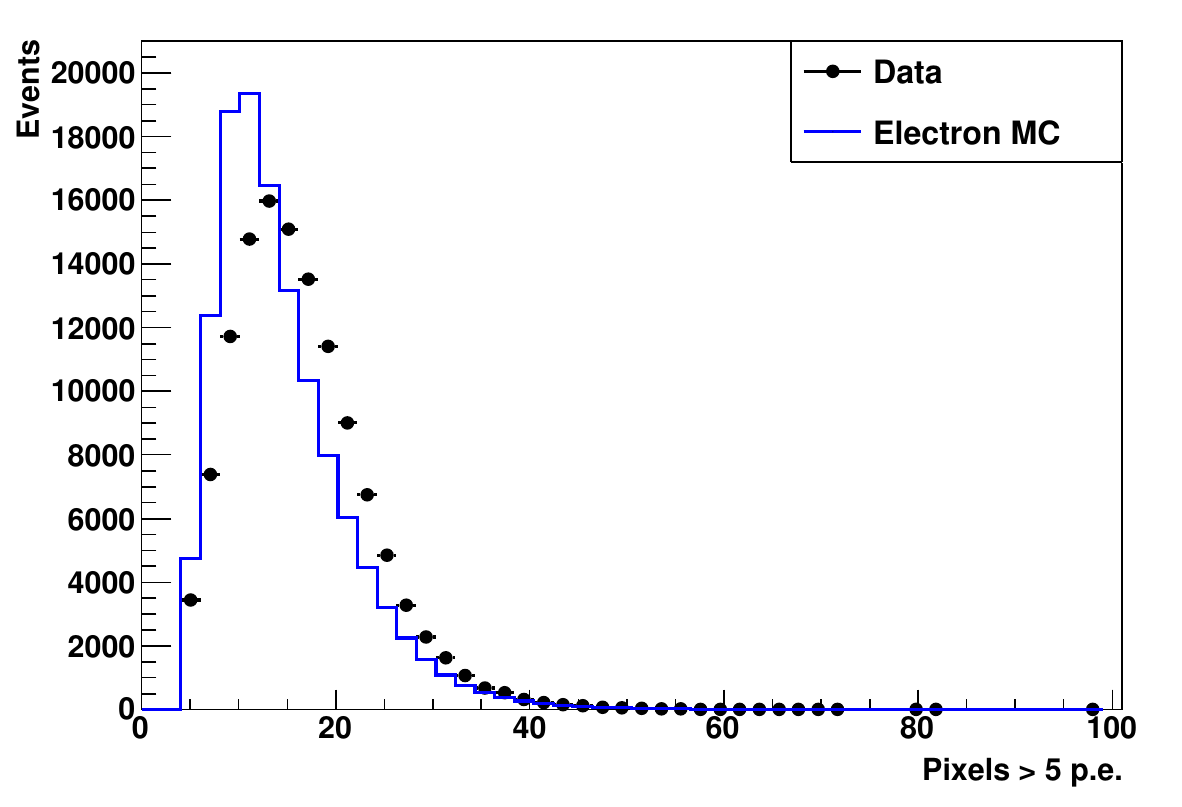}
          \includegraphics[width=0.30\textwidth]{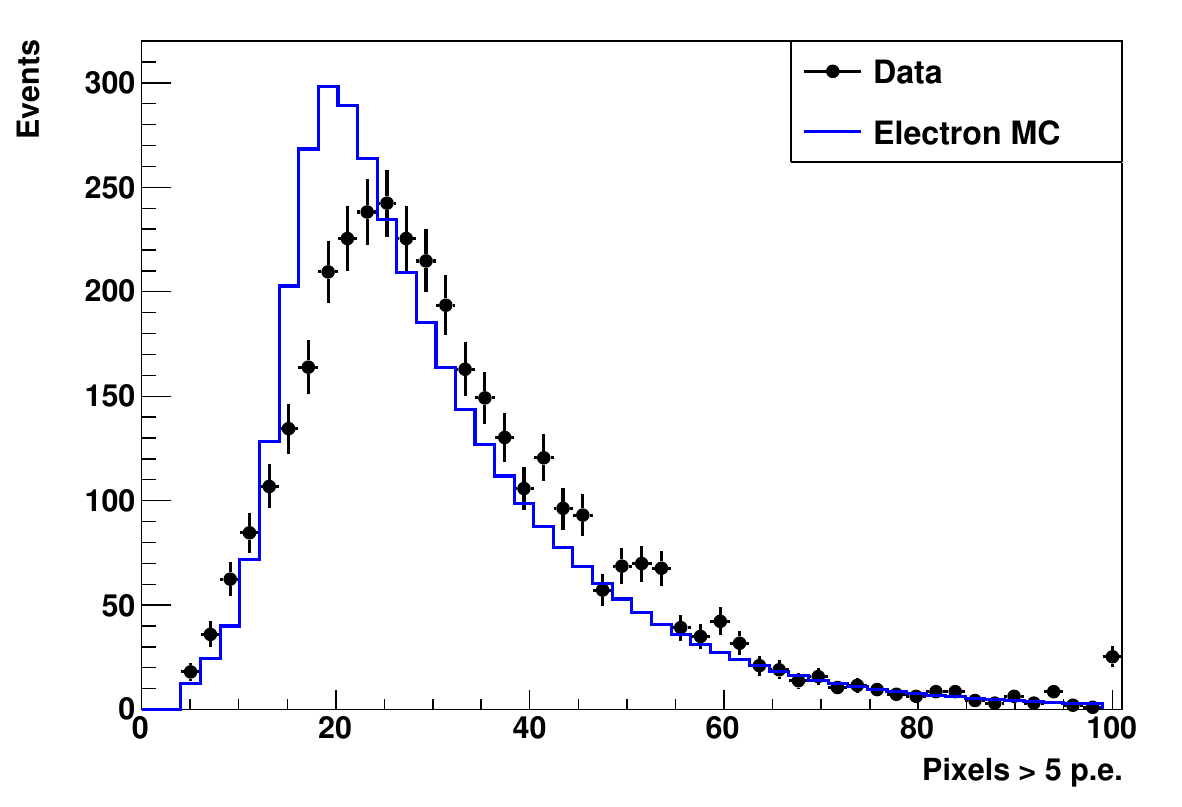}
         \caption{Distribution in number of image pixels}
     \end{subfigure}

     \begin{subfigure}[b]{\textwidth}
         \centering
         \includegraphics[width=0.30\textwidth]{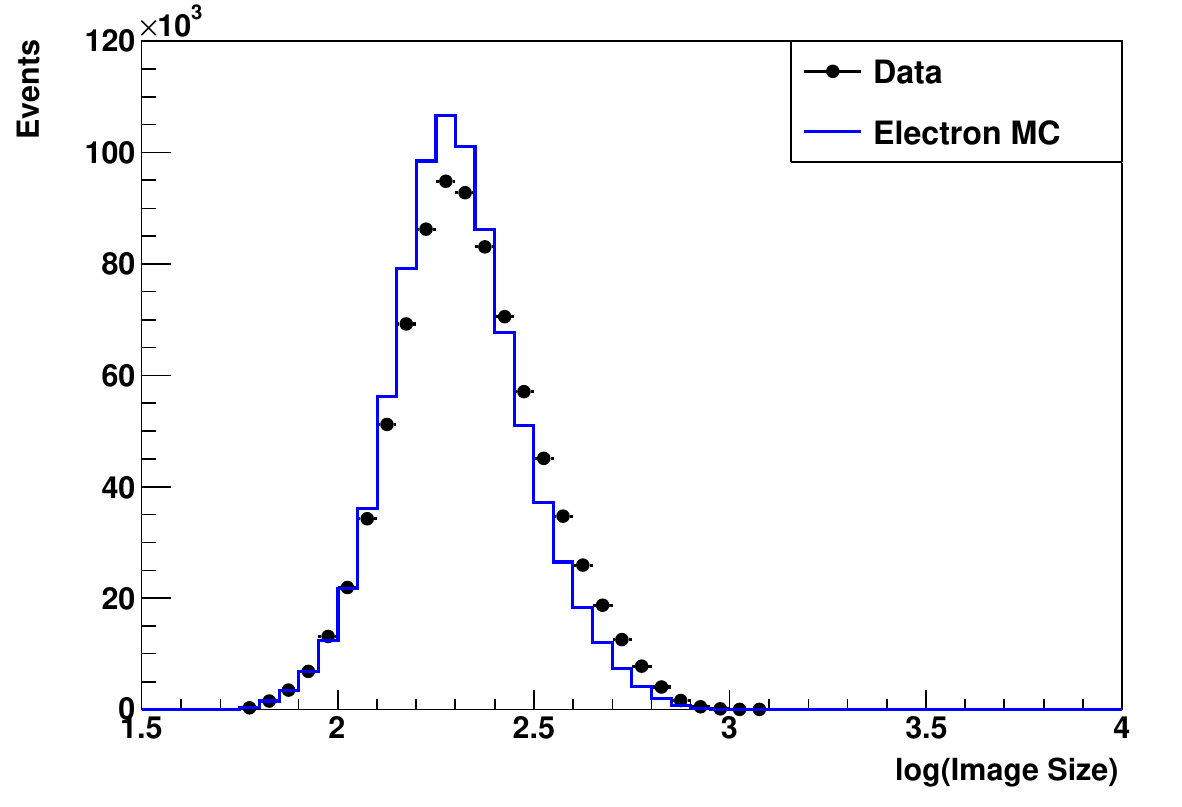}
          \includegraphics[width=0.30\textwidth]{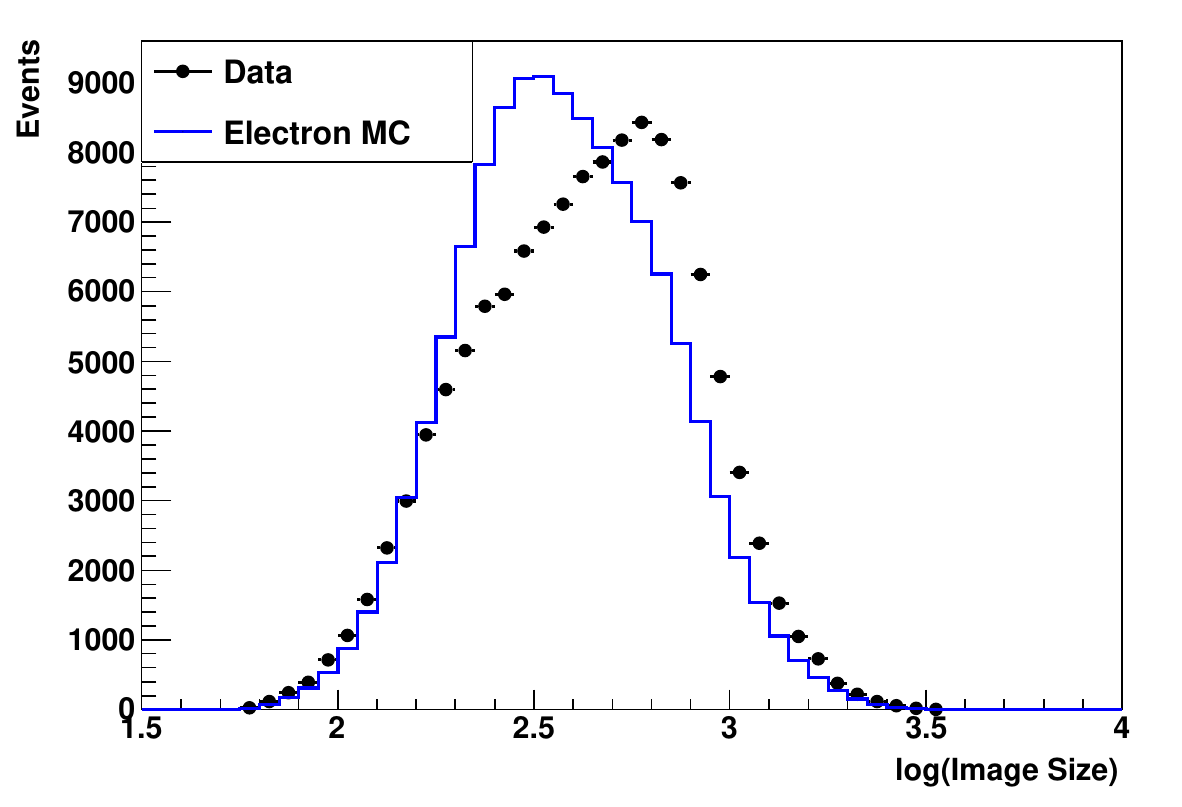}
          \includegraphics[width=0.30\textwidth]{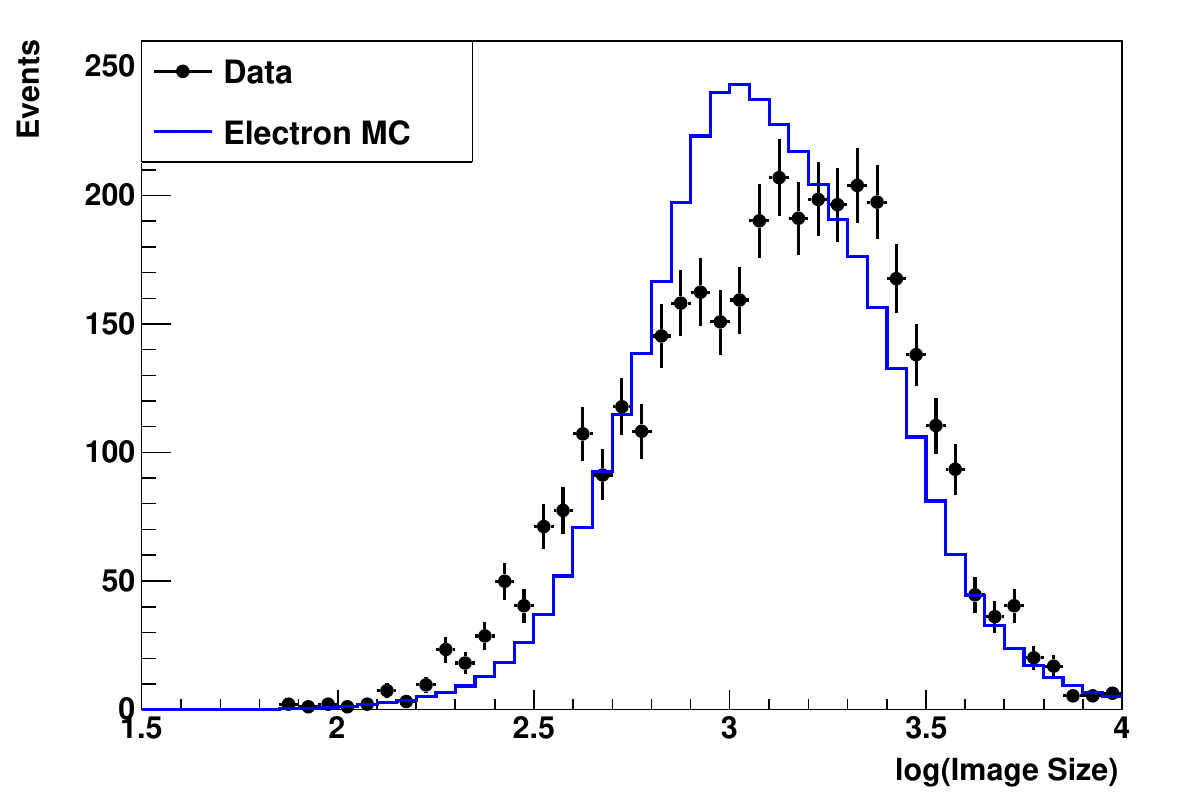}
         \caption{Distribution in log(image size)}
     \end{subfigure}
     
     \caption{Distributions in event parameters after CRe selection (data points), compared to electron simulations (blue histogram). Left column: energy range $0.3 - 1$ TeV; middle column: $1 -3$ TeV; right column: $>$ 3 TeV. Shown are (top to bottom): (a) telescope multiplicity, (b) number of image pixels, (c) image intensity. { Deviations towards higher energies increase due to the increased hadronic contamination.}}
     \label{fig:elecdis_1}
 
 \end{figure*}

\begin{figure*}
     \centering

     \begin{subfigure}[b]{\textwidth}
         \centering
         \includegraphics[width=0.30\textwidth]{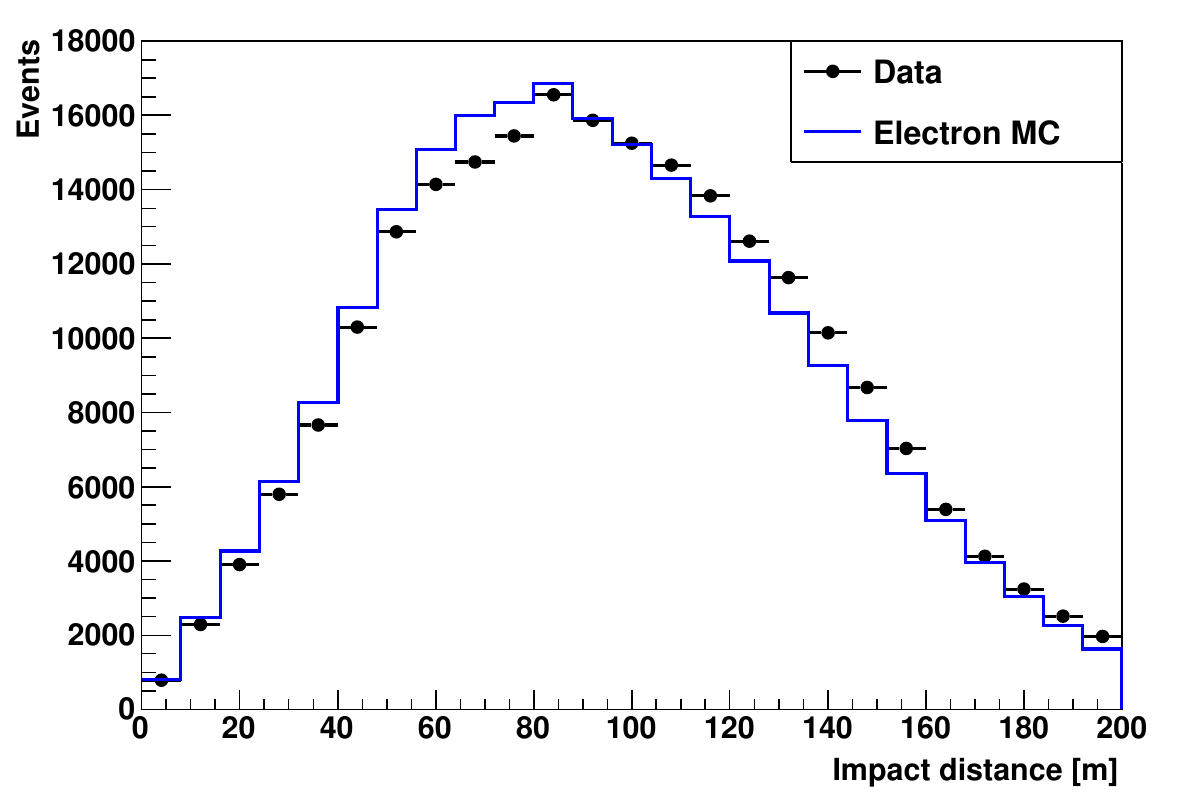}
          \includegraphics[width=0.30\textwidth]{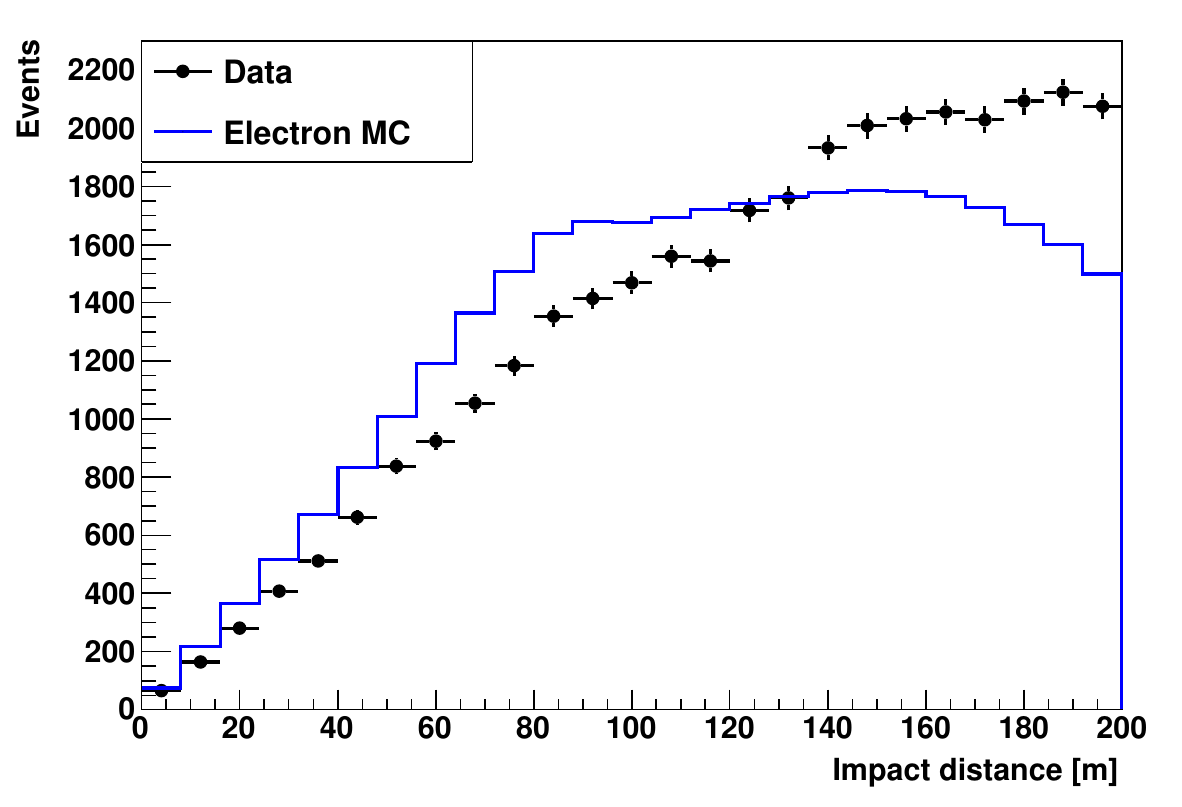}
          \includegraphics[width=0.30\textwidth]{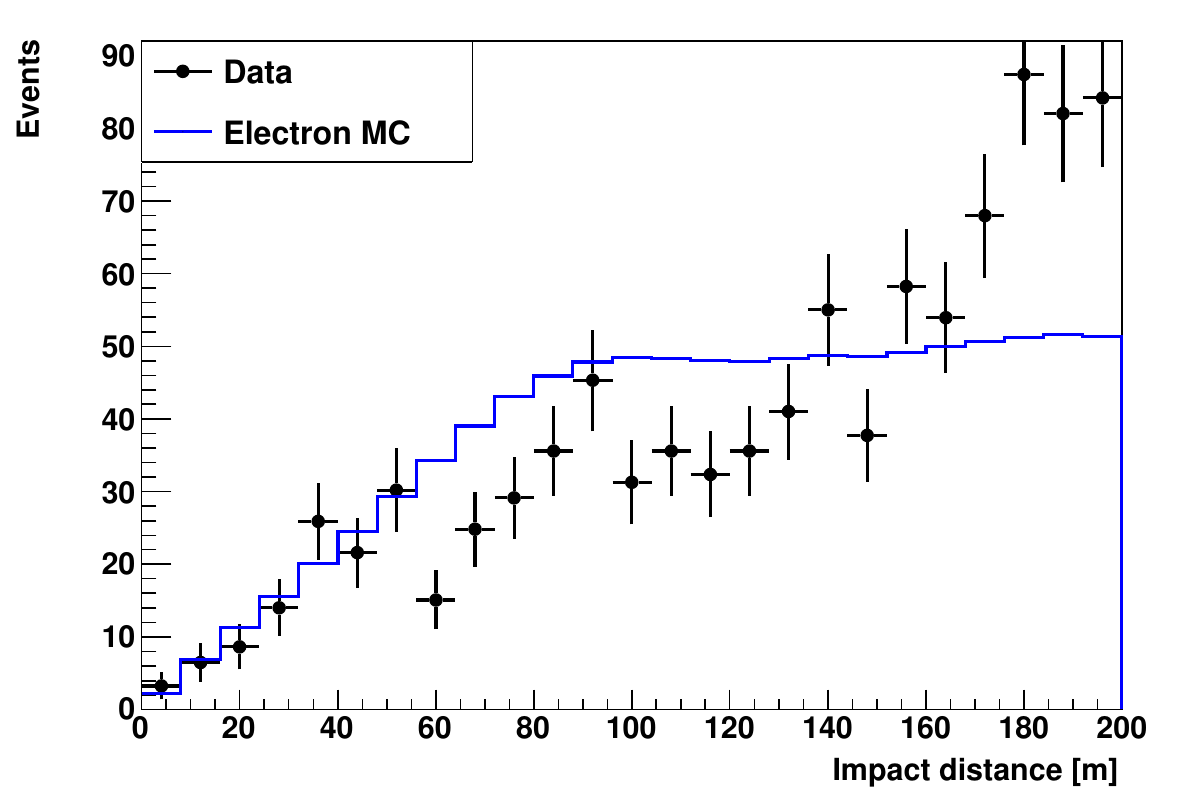}
         \caption{Distribution in core distance}
     \end{subfigure}
     
     \begin{subfigure}[b]{\textwidth}
         \centering
         \includegraphics[width=0.30\textwidth]{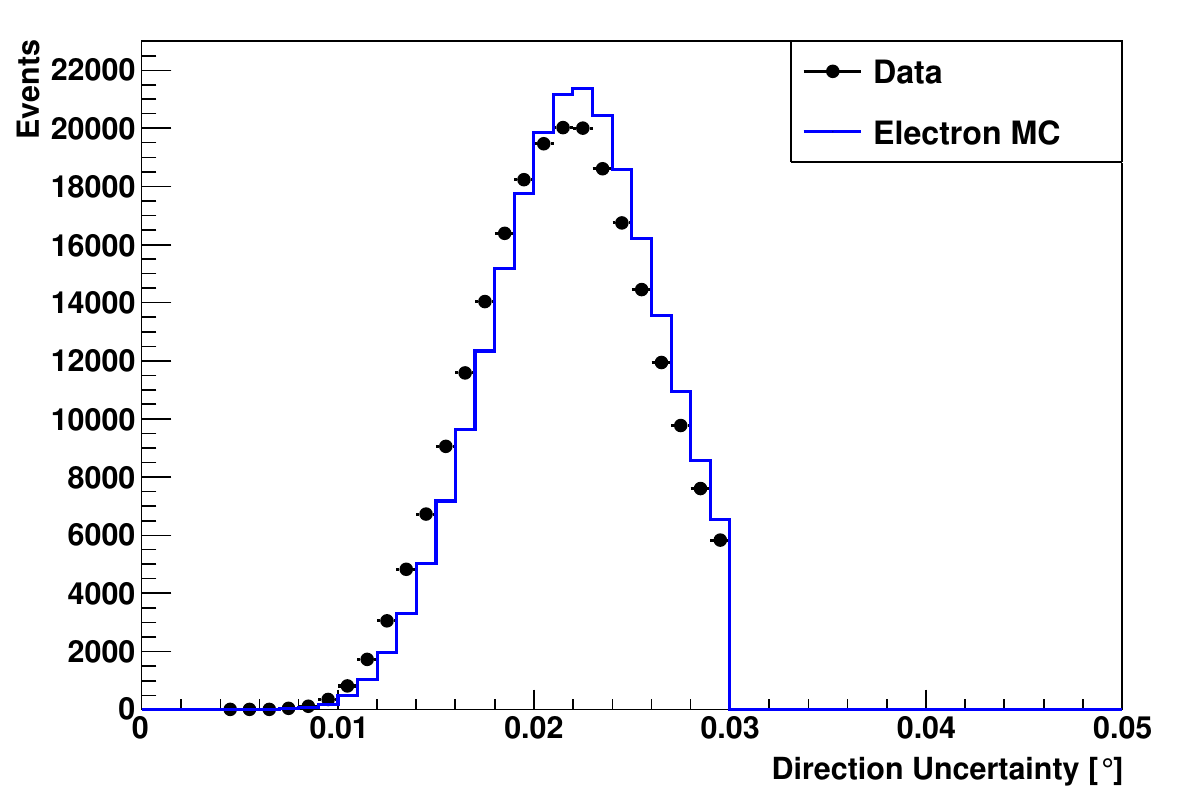}
          \includegraphics[width=0.30\textwidth]{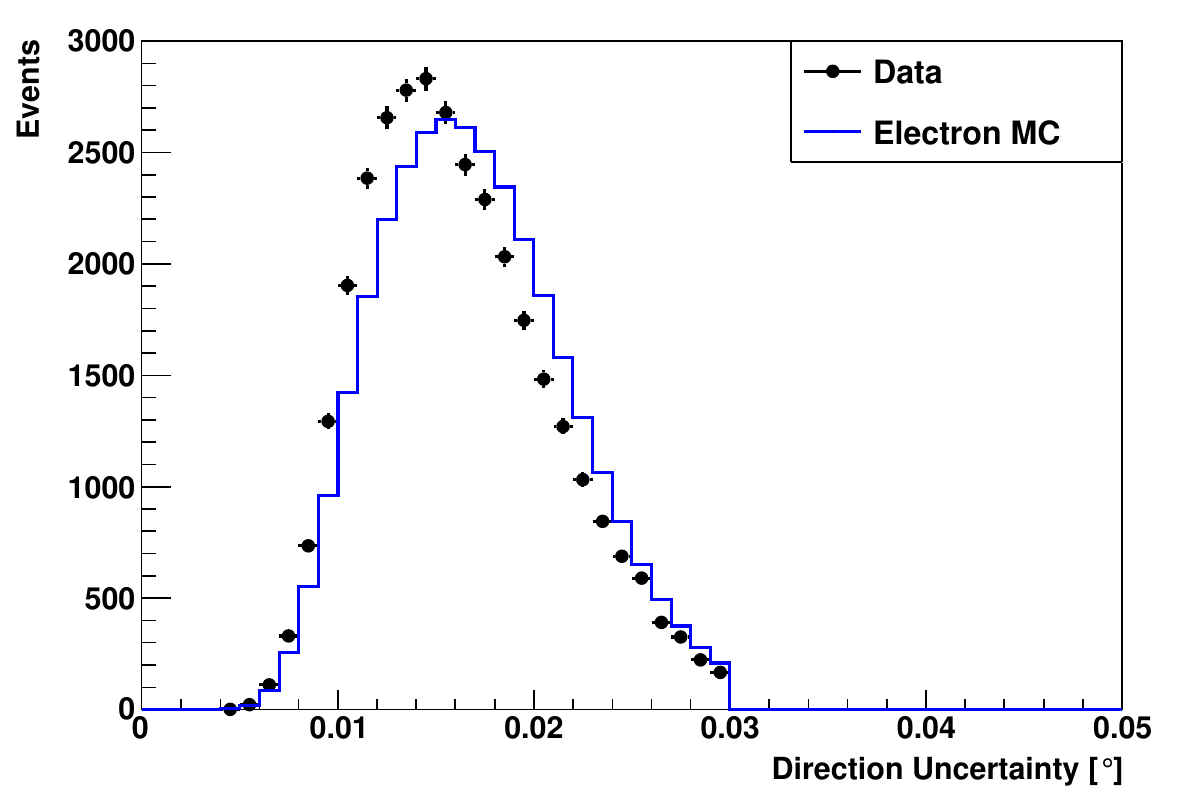}
          \includegraphics[width=0.30\textwidth]{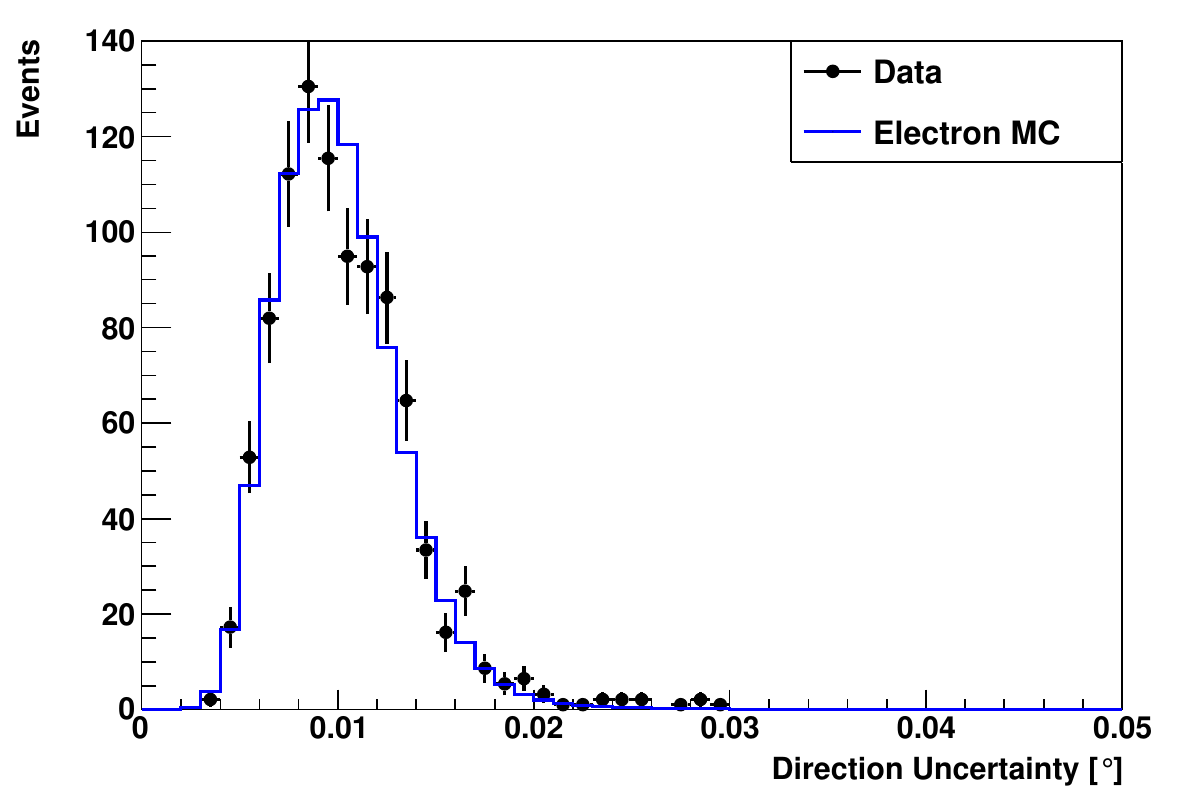}
         \caption{Distribution in direction error}
     \end{subfigure}

     \begin{subfigure}[b]{\textwidth}
         \centering
         \includegraphics[width=0.30\textwidth]{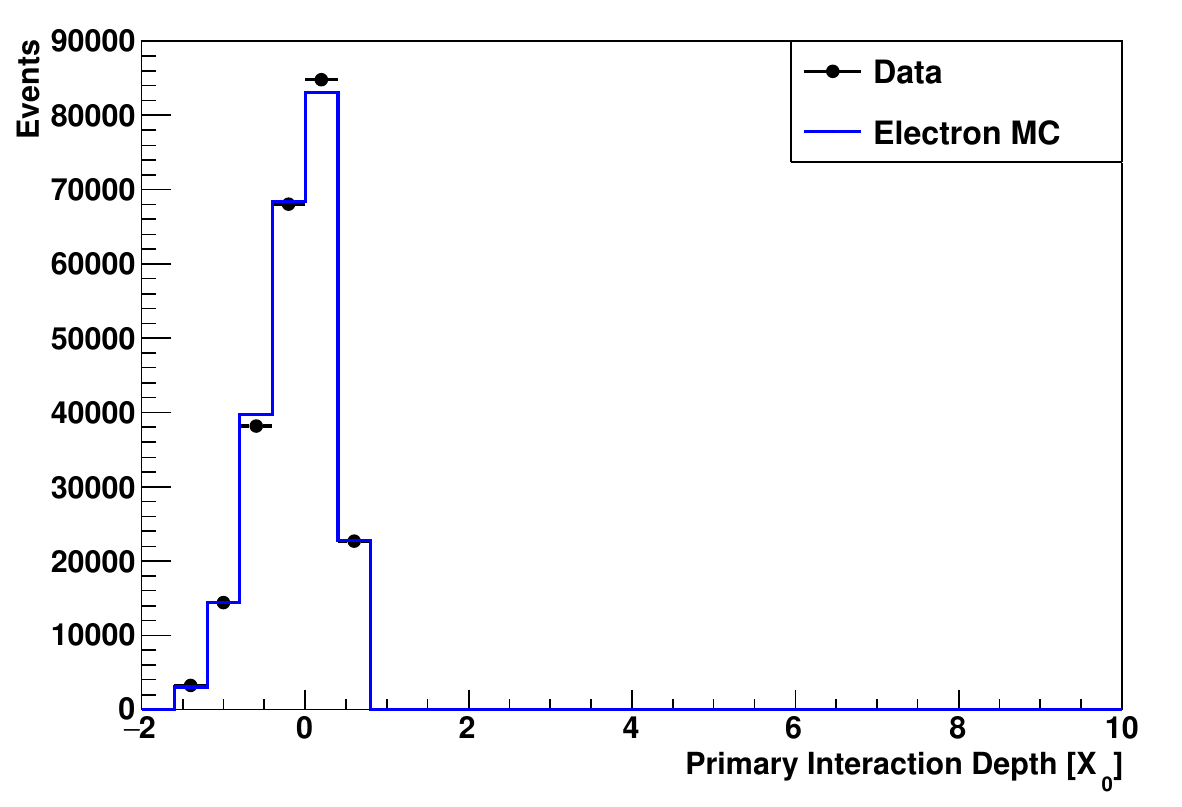}
          \includegraphics[width=0.30\textwidth]{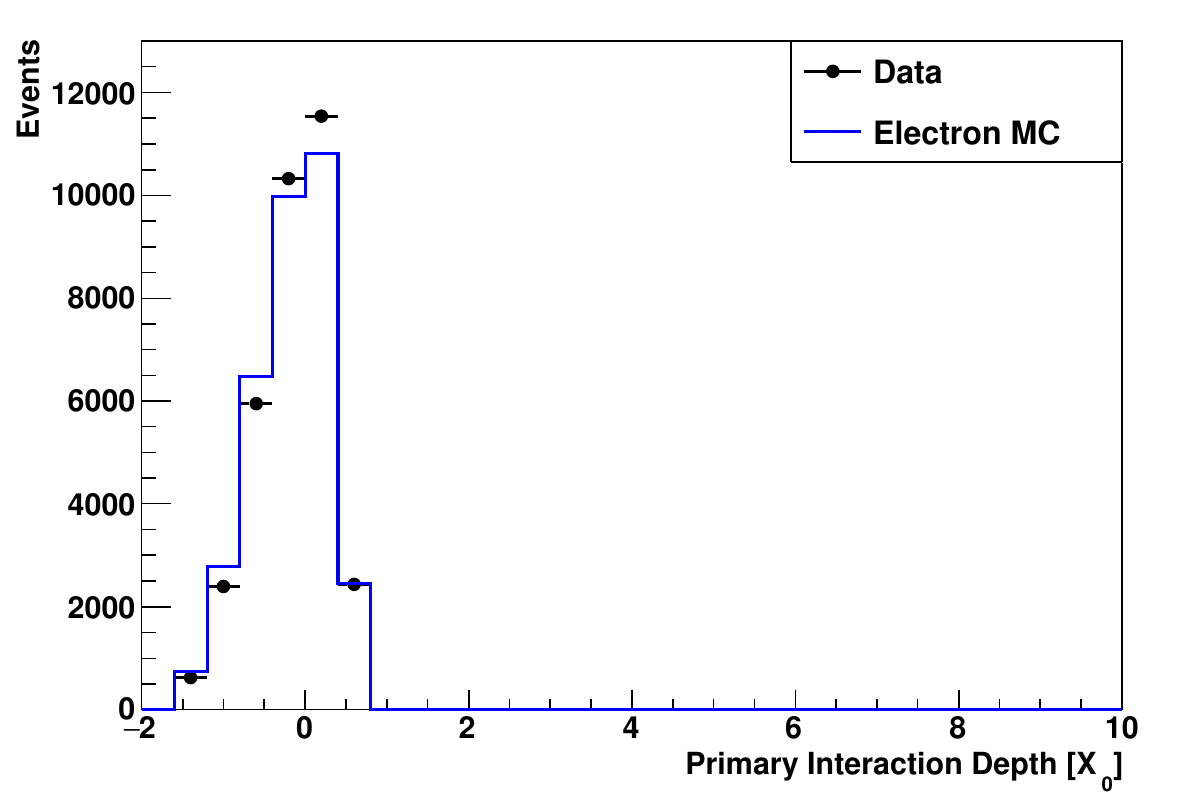}
          \includegraphics[width=0.30\textwidth]{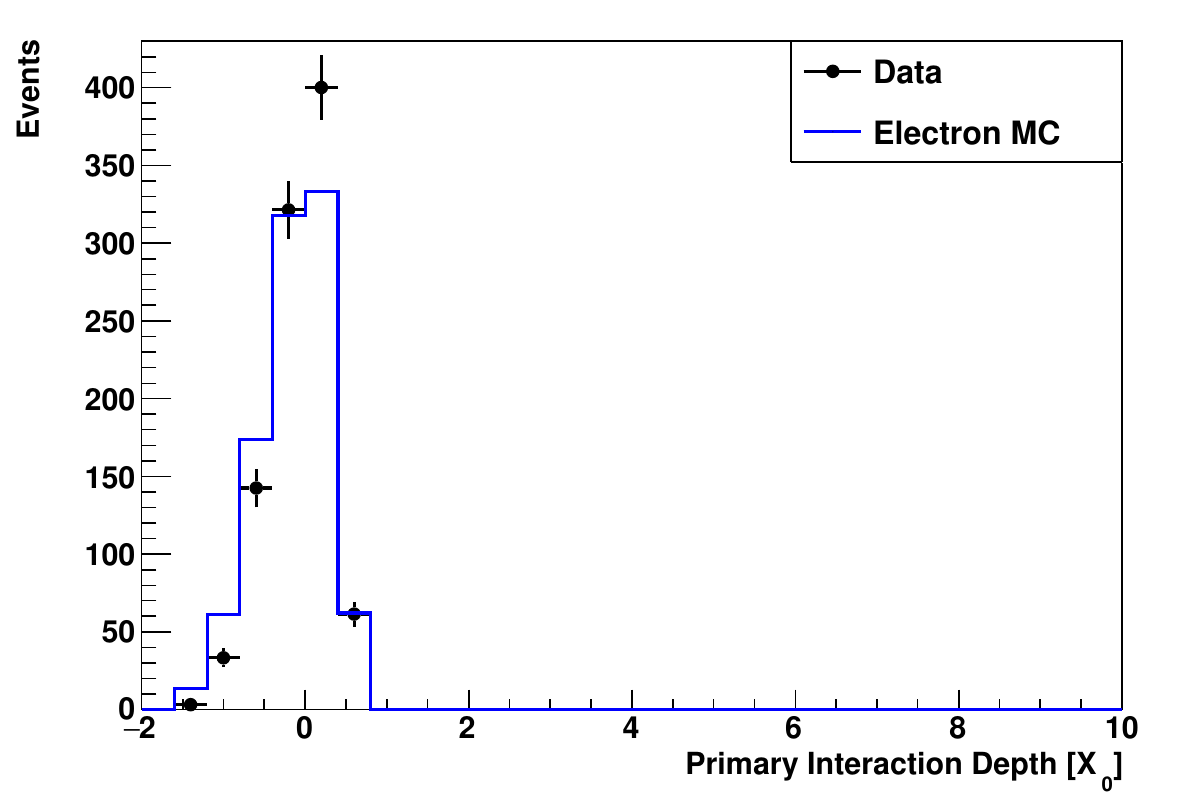}
         \caption{Distribution in reconstructed depth of 1st interaction}
     \end{subfigure}
     
     \caption{Distributions in event parameters after CRe selection (data points), compared to electron simulations (blue histogram). Left column: energy range $0.3 - 1$ TeV; middle column: $1 - 3$ eV; right column: $>$ 3 TeV. Shown are (top to bottom): (a) core distance, (b) direction error and (c) depth of $1^\mathrm{st}$ interaction. The distributions in MSSG are shown in the main body of the paper, and are not reproduced here. { Deviations towards higher energies increase due to the increased hadronic contamination.}}
     \label{fig:elecdis_2}

\end{figure*}

Figs. \ref{fig:elecdis_1} and  \ref{fig:elecdis_2} compare distributions in key event parameters between the data -- after applying CRe selection cuts -- and the simulation, using the identical cuts (see Table \ref{table_cuts}). We note that unlike for the background-subtracted $\gamma$-ray data, where perfect agreement between data and simulations can be expected for a sufficiently precise simulation, the CRe sample contains hadronic background events, that are not accounted for in the simulations. Indeed, agreement between data and simulations is reasonably
good for the energy range $0.3 - 1$ TeV (left column of Figs. \ref{fig:elecdis_1} and  \ref{fig:elecdis_2}), but for the $1 - 3$ TeV range (middle column) and the $>$ 3 TeV range (right column), deviations become visible that are expected from the estimated proton contamination. Background events seem to have larger image amplitudes, and in particular concentrate at larger core distances.

\section{Stability of the results and systematic errors}

This section serves to discuss the consistency of the results between different (temporal, zenith angle, latitude) subsets of the data, the consistency under variation of selection cuts, and the implications for systematic errors.

\subsection{Stability of flux over time}

Owing to the diffusive propagation of CRe, the CRe flux is expected to be constant on the decade time scales of H.E.S.S. data acquisition, and to a good approximation isotropic. This expected spatial and temporal stability enables further tests of the quality and stability of the results, and { allows us to probe the systematic uncertainties of the analysis}. 

\begin{figure*}[htbp] 
   \centering
   \includegraphics[width=0.95\textwidth]{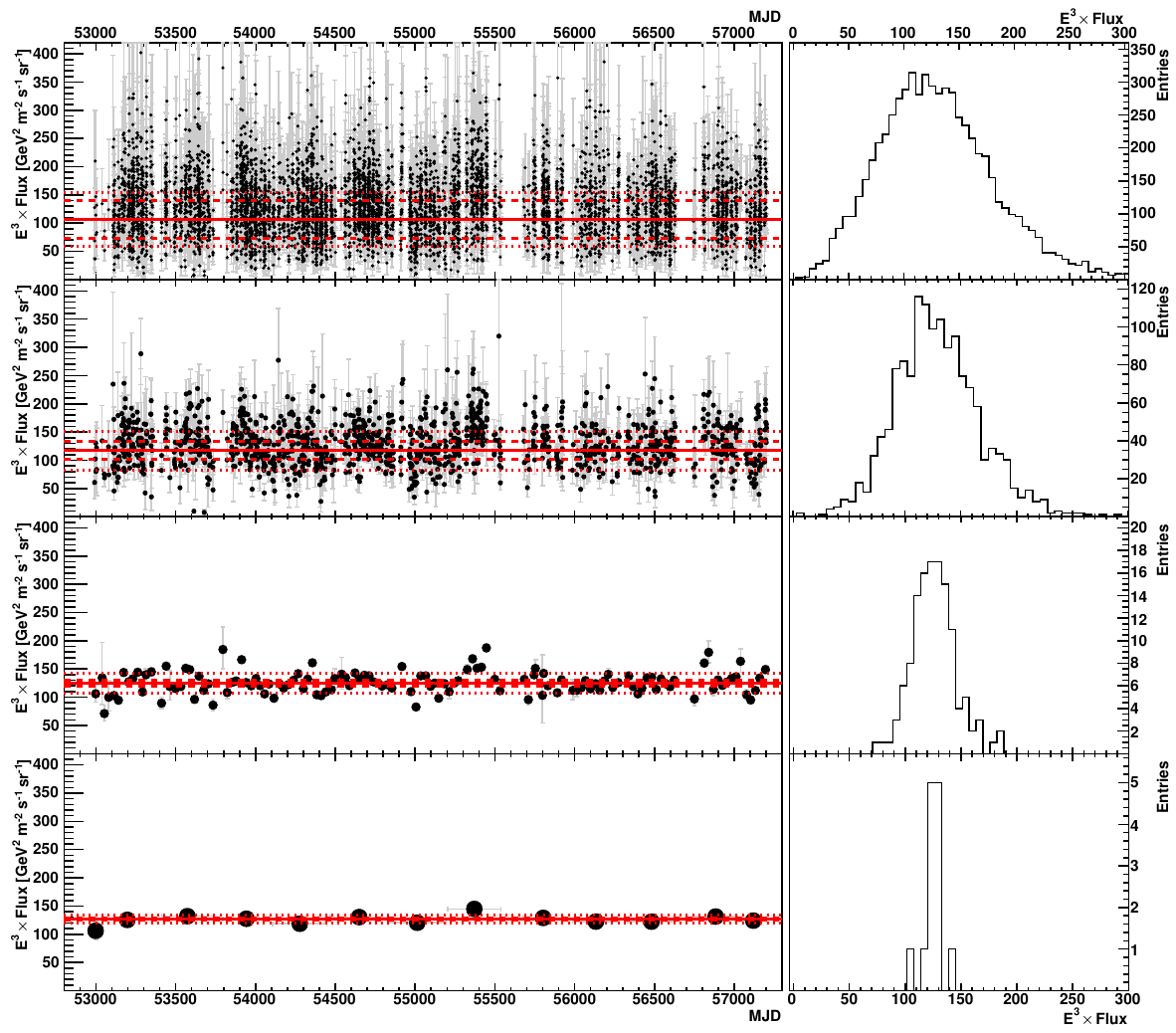}
   \caption{CRe flux at 1 TeV determined on different time scales: for 28-min runs, nights, moon periods, and years (top to bottom). The left panels show the CRe flux versus time (error bars show statistical errors), while the right panels show the distribution of flux values. Flux values are determined assuming the average spectral shape, with the flux normalization as single free parameter. The flux values are not corrected for hadronic contamination. The solid red line indicates the uncertainty-averaged flux. The dashed lines indicate the band around the average flux corresponding to the average individual flux uncertainties (mean of error bars). The dotted line correspond to the $\pm 1 \,\sigma$ spread of the individual flux points.}
   \label{fig:lightcurve}
\end{figure*}

Temporal stability is tested by investigating flux variability on different time scales: for 28-min runs, nights, 30-day moon periods (with data taking interrupted during full moon), and years (Fig. \ref{fig:lightcurve}). In generating these light curves, the shape of the electron spectrum was fixed, with the flux normalization as single free parameter. 
Despite variations in the telescope system, such as degradation of mirror reflectivity over time, or adjustments in the high voltage of the photo sensors, and despite significant variations in atmospheric transparency on annual time scales, the CRe `light curves' shown in Fig. \ref{fig:lightcurve} are free of strong systematic trends, showing that variations in instrument and atmosphere are properly taken into account in the run-wise simulations. The CRe signal is detected at the 2.5$\sigma$ level in individual 28-min runs (top panels of Fig. \ref{fig:lightcurve}), and becomes significant when runs from a night are combined (2nd row of panels). Table \ref{table_variance} shows the observed root mean square variation in the CRe flux measured on the different time scales, and the excess variance that needs to be added (in quadrature) to the statistical errors to explain the observed variation. The excess variance is attributed to systematic errors, likely related to the not fully perfect description of the state of the instrument and of the atmosphere at any given time. The flux measurement on a run-by-run basis is dominated by statistical errors; in the nightly fluxes, systematic errors start to become visible, and they dominate on moon-period and year time scales. Systematic errors are smaller for larger time scales, indicating that some of the short-term systematic deviations average out. This is to be expected, since despite the use of run-wise simulations \cite{RunWise2020} not all instrument and atmosphere parameters can be fine-tuned on run time scales.

\begin{figure}[htbp] 
   \centering
   \includegraphics[width=0.4\textwidth]{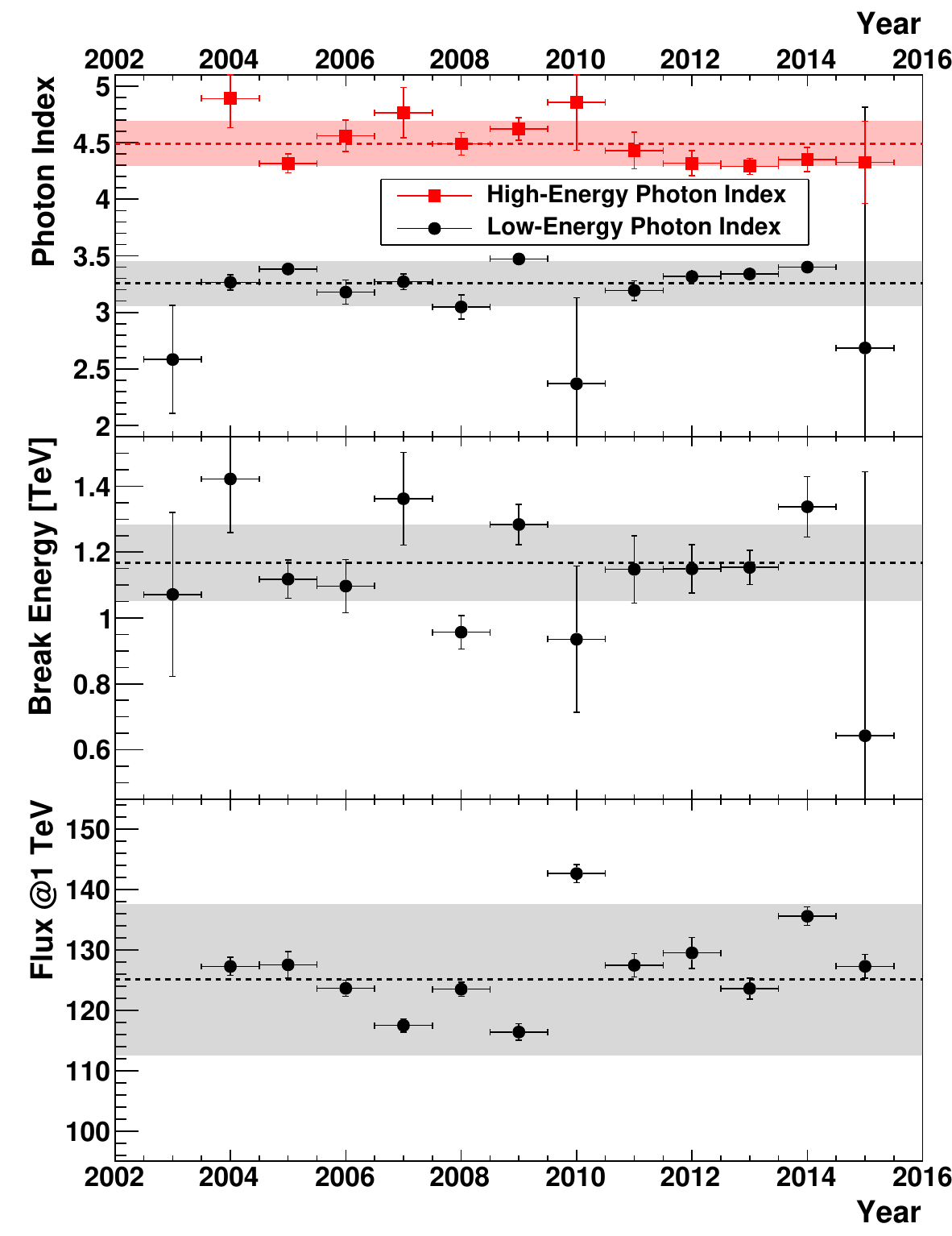}
   \caption{ Spectral parameters derived for individual years of data taking: spectral index below and above the break, break energy, and flux normalization (top to bottom). The flux values are not corrected for hadronic contamination. The shaded bands indicate the quoted systematic errors. }
   \label{fig:annual_parameters}
\end{figure}

For the annual spectra, the spectral parameters (low-energy index, high-energy index, break energy and flux normalisation) can be determined individually, rather than assuming an average spectral shape. The result is shown in Fig. \ref{fig:annual_parameters}, also compared with the quoted range of systematic uncertainties on the spectral parameters. 

\begin{table}[htp]
\caption{Variance of measured CRe flux for different time scales, and excess variance beyond statistical errors. The excess variance is added in quadrature to the statistical errors, and is attributed to systematic errors.}
\begin{center}
\begin{ruledtabular}
\begin{tabular}{l c c}
Time scale & Variance & Excess variance \\
\hline
Run &  40\% & 17\% \\
Night & 29\% & 17\% \\
Period & 16\% & 13.5\%\\
Year & 7\% & 6\%\\
\end{tabular}
\end{ruledtabular}
\end{center}
\label{table_variance}
\end{table}%

\begin{figure*}[htbp]
   \centering
   \includegraphics[height=0.5\textwidth]{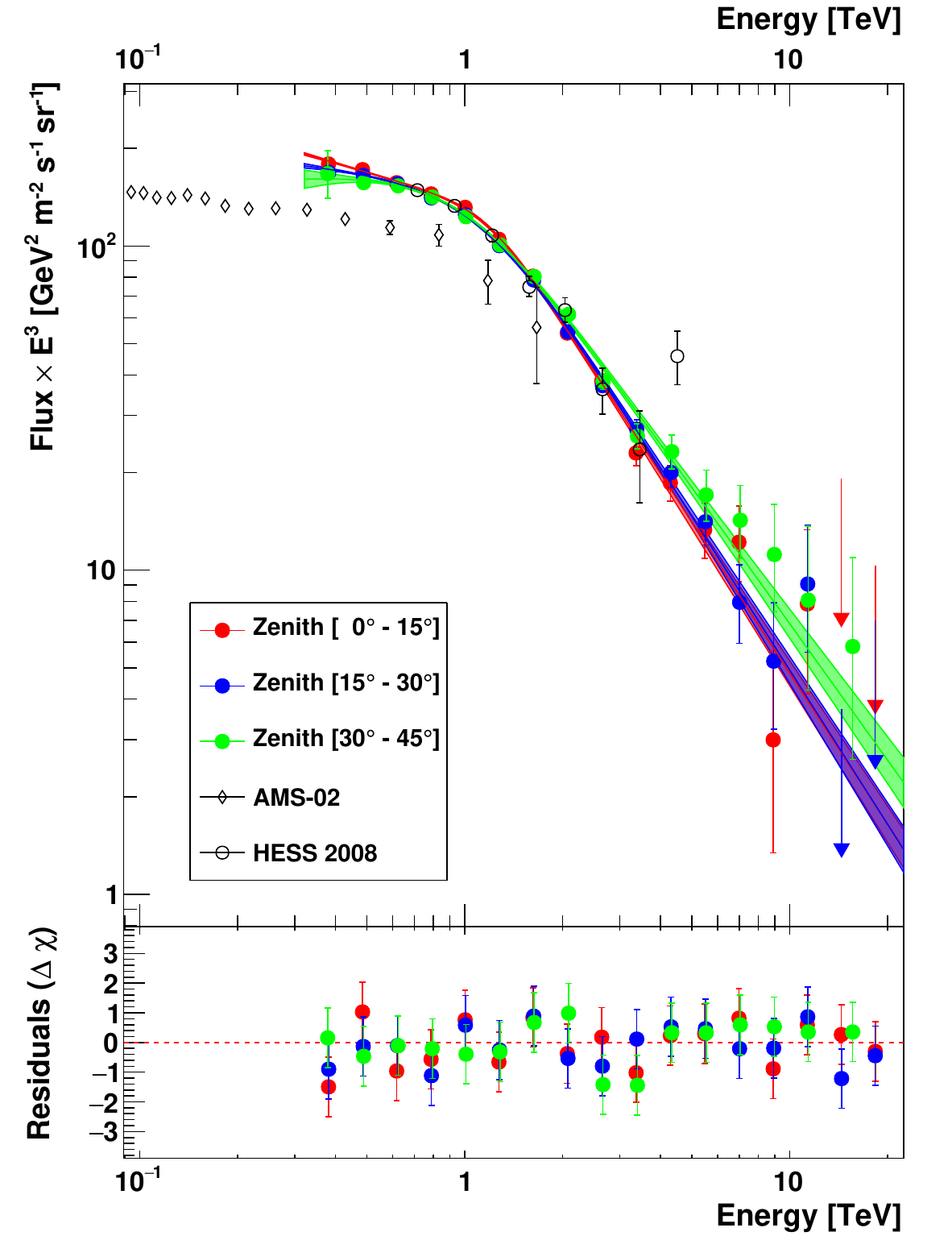}
   \includegraphics[height=0.5\textwidth]{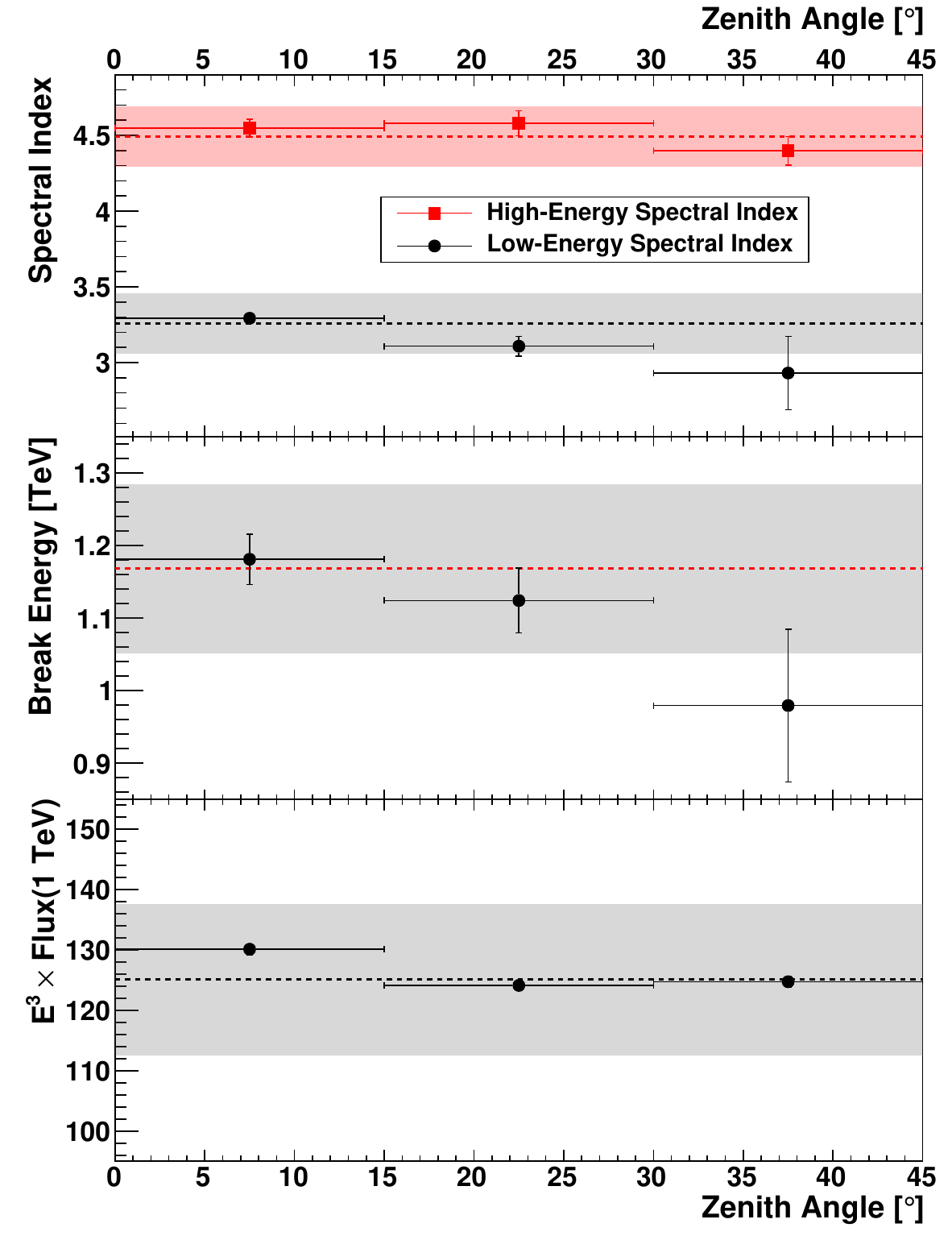}
   
   \caption{Left: CRe spectrum measured for three different zenith angles: $0^\circ$ - $15^\circ$, $15^\circ$ - $30^\circ$ and $30^\circ$ - $45^\circ$. The flux values are not corrected for hadronic contamination. Right: Fit parameters as a function of zenith angle: spectral index below and above the break, break energy, and flux normalization (top to bottom).  The shaded bands indicate the quoted systematic errors.}
   \label{fig:zenith}
\end{figure*}
\subsection{Zenith-angle dependence of CRe spectrum}

With increasing zenith angle, showers are more distant in the atmosphere and the size of the Cherenkov light pool increases. This results in a decrease of the intensity of Cherenkov light for a given shower energy, both due to the wider distribution of light and due to the increased atmospheric attenuation. This is illustrated in Fig. \ref{fig:effarea} which shows the effective detection area with CRe cuts as a function of energy, for different zenith ranges; while the effective area saturates above 1 TeV, the sub-TeV threshold behavior depends significantly on zenith angle. A comparison of CRe spectra measured at different zenith angles is hence a very efficient means to probe systematics related to energy reconstruction and effective area determination, in particular in the threshold region. Fig. \ref{fig:zenith} shows the measured CRe flux and the spectral parameters -- the flux at 1 TeV, the spectral index below and above the break, and the break energy -- for three different zenith angle ranges, up to $45^\circ$. 

The measured spectral data points at energies below a few TeV are consistent across zenith angles; above $\sim 5$\,TeV the flux at the larger zenith angles is marginally higher, possibly caused by an increasing hadronic contamination. The fit parameters show a slight variation but are consistent within the estimated systematic errors of 0.2 for spectral indices, and $10\%$ for the energy scale and flux. The stability of the low-energy spectral index is particularly relevant, since at energies below 1 TeV, the effective area is strongly dependent on energy and zenith angle.
\\
\begin{figure*}[htbp] 
   \centering
   \includegraphics[height=0.5\textwidth]{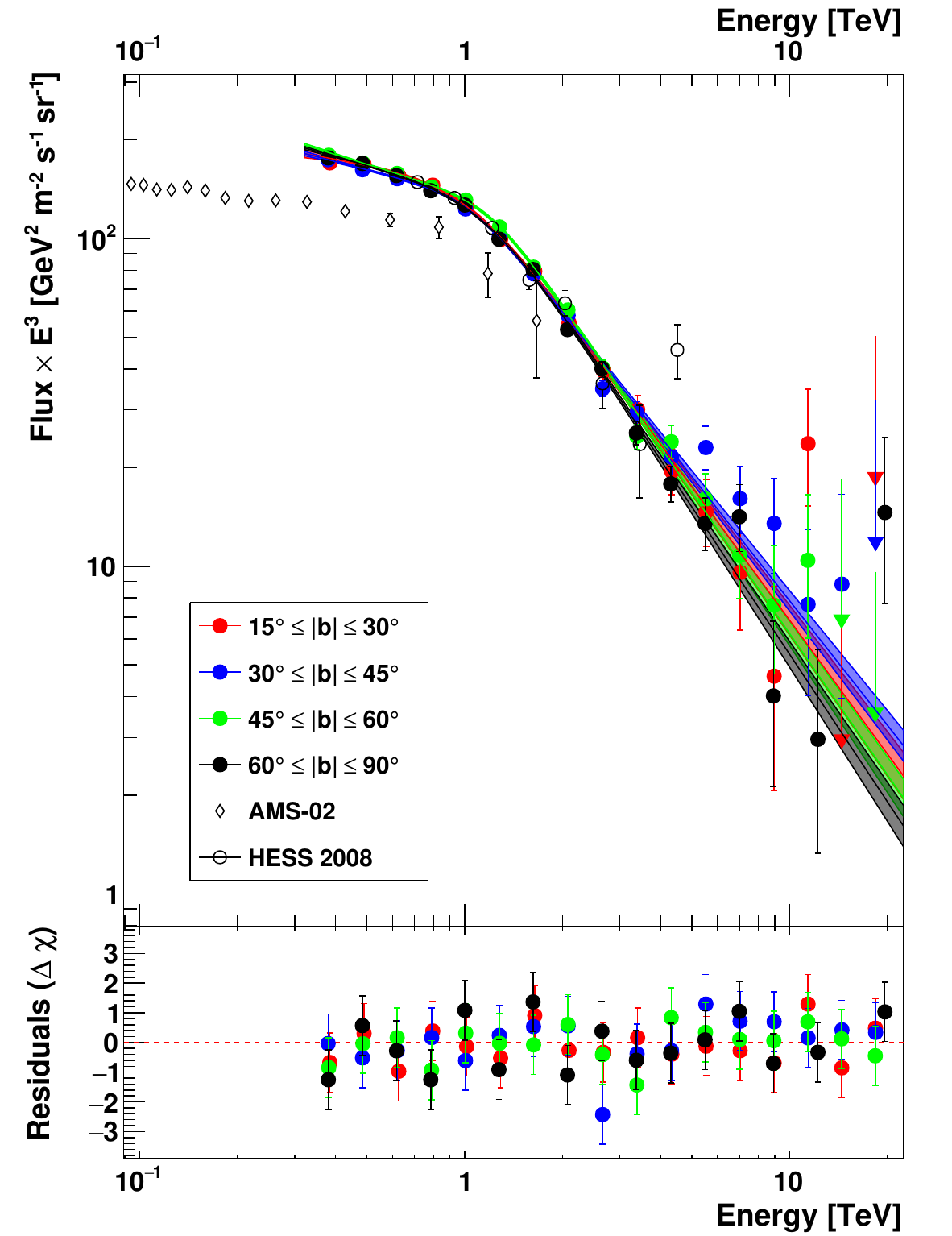} 
   \includegraphics[height=0.5\textwidth]{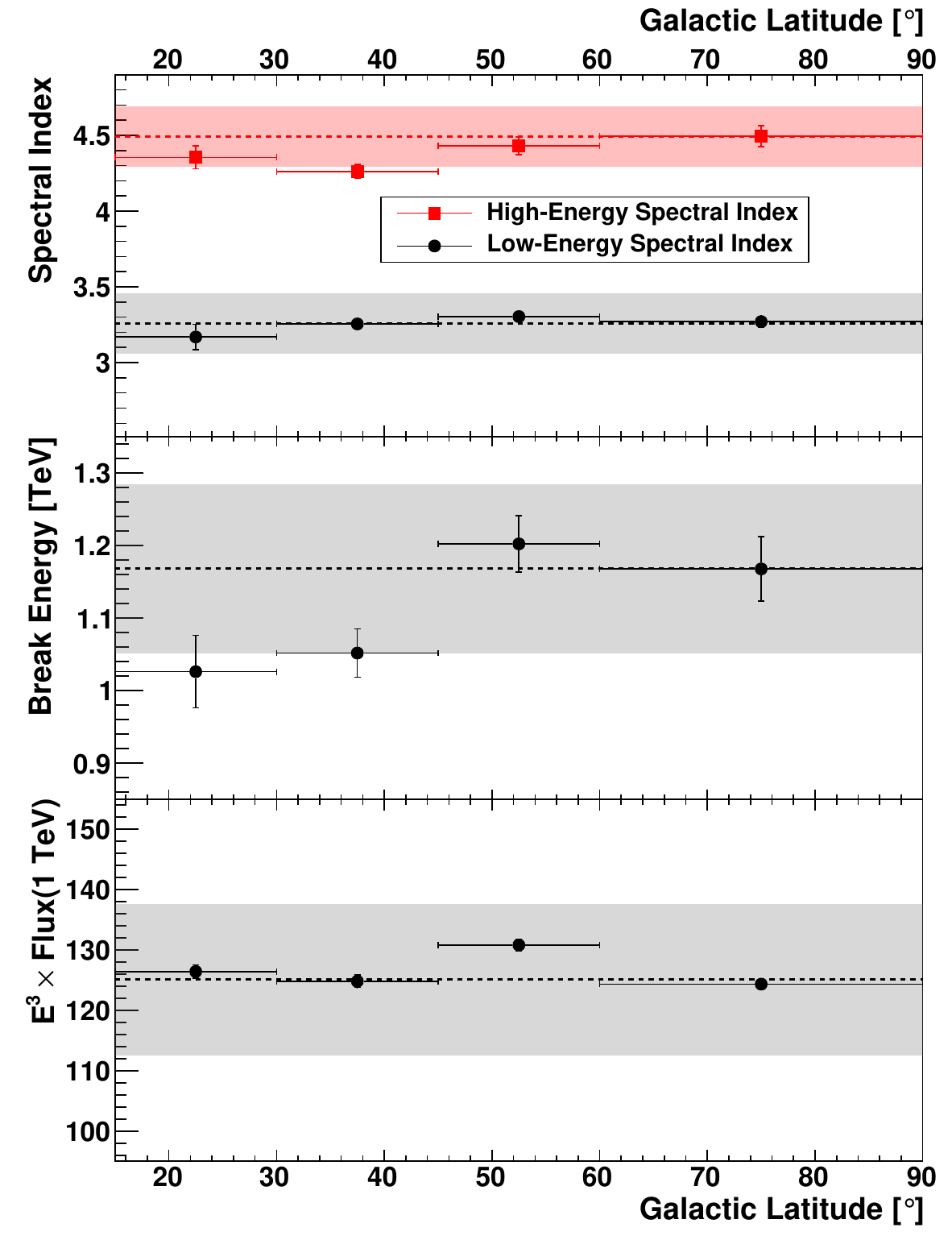} 
   
   \caption{Left: CRe spectrum measured for different bands in Galactic latitude: $15^\circ$ - $30^\circ$, $30^\circ$ - $45^\circ$, $45^\circ$ - $60^\circ$ and $60^\circ$ - $90^\circ$. The flux values are not corrected for hadronic contamination. Right: Fit parameters as a function of latitude: spectral index below and above the break, break energy, and flux normalization (top to bottom). The shaded bands indicate the quoted systematic errors. } 
   \label{fig:latitude}
\end{figure*}
\subsection{Latitude dependence of electron spectrum}

To test for any remaining contamination from Galactic diffuse $\gamma$-ray emission, CRe spectra derived for different bands in Galactic latitude are compared (Fig. \ref{fig:latitude}). Again, spectra and spectral parameters are compatible within quoted systematic uncertainties. Due to visibility constraints, latitude correlates to a certain extent with seasons during the year.\\

\begin{figure}[htbp] 
   \centering
   \includegraphics[width=0.46\textwidth]{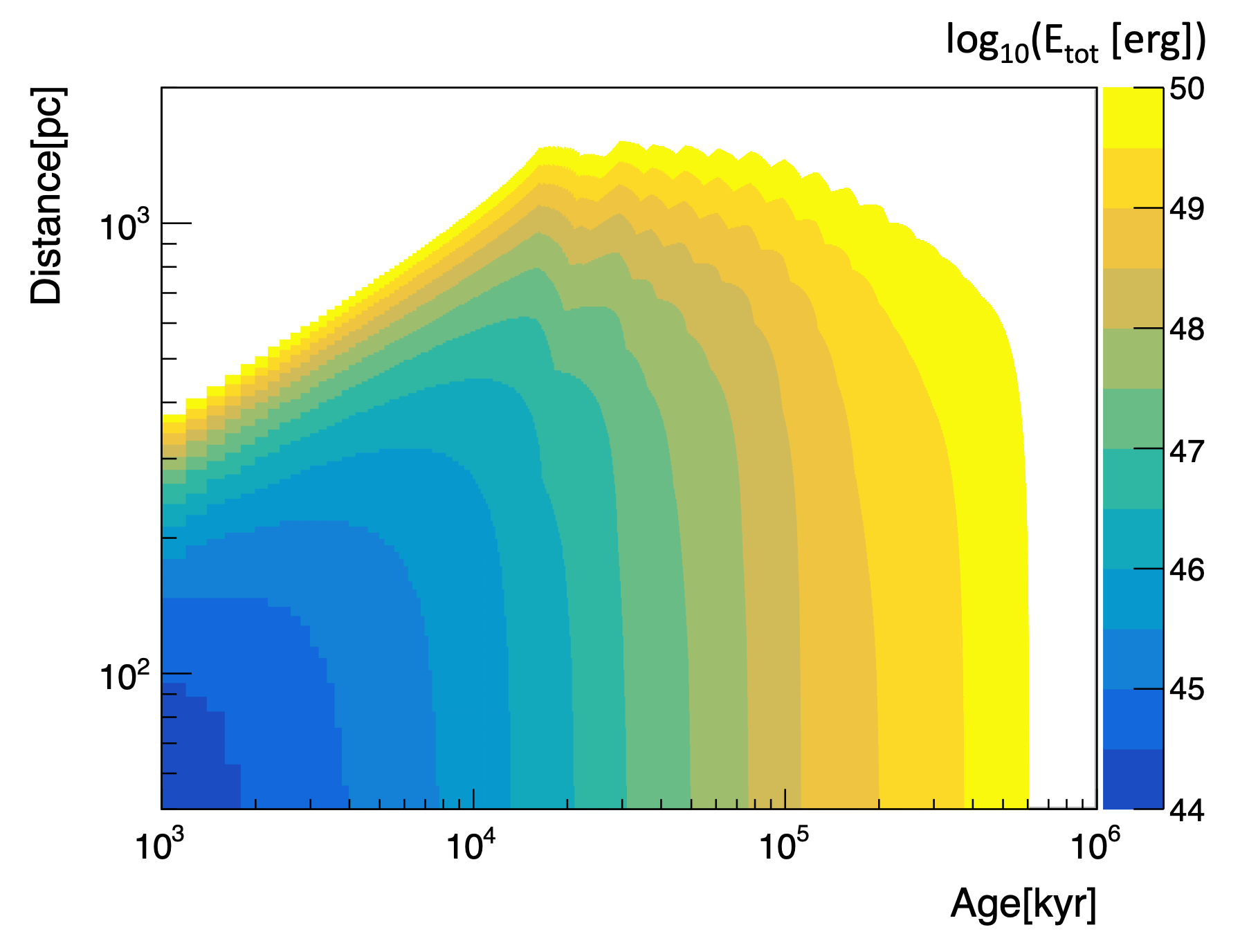}
   \caption{The maximum energy released in CRe by a source located at a certain distance and age, constrained by the H.E.S.S. CRe candidate spectrum assuming source spectra and diffusion properties as used in \cite{1995A&A...294L..41A}.
   }
   \label{fig:limit}
\end{figure}

{
\section{Implications of the measurement for local CRe accelerators}
The measured spectrum of CRe candidate events represents an upper limit on the contribution of individual local electron sources. As discussed e.g. in \cite{1995A&A...294L..41A}, a burst-like injection of electrons with spectral index around 2 will produce a flux of electrons at Earth peaking in $E^3F(E)$ at an energy that depends on the age and distance of the source, as well as on the assumed diffusion coefficient. While a general discussion of the constraints on source parameters is beyond the scope of this paper, we illustrate in Fig.~\ref{fig:limit} the limits on energy release as a function of age and distance, under the conservative assumption that the calculated flux cannot exceed the measured spectrum (i.e. allowing the spectrum to be entirely dominated by a single source), for an example parameter set as used in \cite{1995A&A...294L..41A} (injection spectral index 2.2 up to $\gamma_{max} = 10^9$, diffusion coefficient $D(10\,\mbox{GeV}) = 10^{28}$ cm$^2$/s and energy index $\delta = 0.6$).
}

\end{document}